\documentclass[aps,showpacs,preprintnumbers,amsmath,amssymb]{revtex4}
\usepackage{dcolumn}
\usepackage[dvips]{epsfig}

\begin{document}
\baselineskip=0.8 cm
\title{\bf Strong gravitational lensing by a Konoplya-Zhidenko  rotating non-Kerr compact object}

\author{Shangyun Wang$^{1,2}$, Songbai Chen$^{1,2,3}$\footnote{Corresponding author: csb3752@hunnu.edu.cn}, Jiliang
Jing$^{1,2,3}$ \footnote{jljing@hunnu.edu.cn}}

\affiliation{$^{\textit{1}}$Institute of Physics and Department of Physics, Hunan
Normal University,  Changsha, Hunan 410081, People's Republic of
China \\ $^{\textit{2}}$Key Laboratory of Low Dimensional Quantum Structures \\
and Quantum Control of Ministry of Education, Hunan Normal
University, Changsha, Hunan 410081, People's Republic of China\\
$^{\textit{3}}$Synergetic Innovation Center for Quantum Effects and Applications,
Hunan Normal University, Changsha, Hunan 410081, People's Republic
of China}

\begin{abstract}
\baselineskip=0.6 cm
\begin{center}
{\bf Abstract}
\end{center}

 Konoplya and Zhidenko have proposed recently a rotating non-Kerr black hole metric beyond General Relativity and make an estimate for the possible deviations from the Kerr solution with the data of GW 150914. We here study the strong gravitational lensing in such a rotating non-Kerr spacetime with an extra deformation parameter. We find that the condition of existence of horizons is not inconsistent with that of the marginally circular photon orbit. Moreover, the deflection angle of the light ray near the weakly naked singularity covered by the marginally circular orbit diverges logarithmically in the strong-field limit. In the case of the completely naked singularity, the deflection angle near the singularity tends to a certain finite value, whose sign depends on the rotation parameter and the deformation parameter.
 These properties of strong gravitational lensing are different from those in the Johannsen-Psaltis rotating non-Kerr spacetime and in the Janis-Newman-Winicour spacetime. Modeling the supermassive central object of the Milk Way Galaxy as a Konoplya-Zhidenko rotating non-Kerr compact object, we estimated the numerical values of observables for the strong gravitational lensing including the time delay between two relativistic images.

\end{abstract}

 \maketitle
\newpage
\section{Introduction}

It is now widely believed that there resides a supermassive black hole at the centre of every galaxy. The Sagittarius A* in our own galaxy is a compelling candidate of such a compact object \cite{t1,t2}. According to General Relativity,  a neutral rotating
black hole in asymptotically flat and matter-free spacetime is
described completely by the Kerr metric only with two parameters,
the mass $M$ and the rotation parameter $a$. Although General Relativity has successfully passed a series of observational and experimental tests \cite{t3} and then the astrophysical black holes in our Universe are expected to be Kerr black holes, there exist a lot of investigations focusing on black hole solutions in other alternative theories of gravity. The main reason is that modifying Einstein's theory of gravity \cite{t4} is one of promising ways to explain the accelerating expansion of the current Universe observed through astronomical experiments \cite{t5,t501,t51,t6,t61}. Moreover, there is no direct evidence to confirm unambiguously that black hole candidates are Kerr black holes, and the current observations including GW 150914 \cite{gw,gw1} cannot exclude the possibility that the geometry of these candidates significantly deviates from the Kerr metric.

The rotating spacetimes deviated from the Kerr metric is usually called as the rotating non-Kerr spacetimes, which have been studied extensively in astrophysics. One of important non-Kerr spacetimes is proposed by Johannsen and Psaltis \cite{TJo} to test the no-hair theorem \cite{noh,noh1,noh2,noh3,noh4}. Besides the mass $M$ and rotation parameter $a$, Johannsen-Psaltis non-Kerr spacetime \cite{TJo} possesses an extra a deformation parameter, which describes the deviation from the usual Kerr spacetime. Johannsen-Psaltis non-Kerr spacetime has the same asymptotic behaviors of Kerr spacetime in the far-field region,  but possesses qualitatively different features near the event horizon. The change of spacetime structure originating from the deformation parameter and the corresponding observable effects are studied in \cite{Cos1,Cos101,Cos102,chen1,chen12,chen2,chen201,Kraw,Ra1,Fa1,yong1,And1}. The possible constraints of the deformation parameter are made for this non-Kerr spacetime by many observational data including the continuum-fitting and iron line \cite{Test1,Test2,Test201,Test202}, quasi-periodic oscillations \cite{Test3,Test301,Test302,Test303,Test304,Test305}, and so on.

Through adding a static deformation, Konoplya and Zhidenko \cite{RA2} have proposed recently another rotating non-Kerr black hole metric beyond general relativity,  which can be regarded as a vacuum solution of a unknown alternative theory of gravity \cite{RA20}. The deformation changes the relation between the black hole mass and position of the event horizon, but preserves asymptotic properties of the Kerr spacetime. With the frequencies of the black hole ringing from the gravitational waves detection by the LIGO and VIRGO collaborations \cite{gw,gw1}, Konoplya and Zhidenko \cite{RA2} made an estimate for the possible deviations from the Kerr solution by using this rotating non-Kerr metric, and found that some non-negligible deformations of the Kerr spacetime can lead to the same frequencies of the black-hole ringing. Moreover, the constraints from quasi-periodic oscillations \cite{Cos3} and the iron line \cite{Cos30} show that the gravity of a real celestial objects in Universe could be described by Konoplya-Zhidenko rotating non-Kerr mertic. Thus, it is necessary to investigate furtherly the properties and observational effects in this Konoplya-Zhidenko rotating non-Kerr spacetime, especially in the strong-field limit.

In this paper, we will study the strong gravitational
lensing by such a Konoplya-Zhidenko rotating non-Kerr compact object \cite{RA2}. It is well known that the deflection angles in strong gravitational lensing become so large that an observer would detect two infinite sets of faint relativistic images on each side of the compact object. These relativistic images are produced by photons which travelled complete loops around the compact object before reaching the observer. It is shown that these relativistic images carry some essential signatures about the central celestial objects and could provide the profound verification of alternative theories of gravity in their strong field regime \cite{Ein1,Darwin,KS1,KS2,KS3,KS4, VB1,VB2,VB201,VB202,Gyulchev,Gyulchev1,Fritt,Bozza1,Eirc1,whisk,
Bhad1,Song1,Song2,Song201,TSa1,AnAv,agl1,agl101,gr1,gr101,gr2,gr201,gr3,gr4,gr401}. The main purpose is to probe whether  the signature of the deformation parameter resides in the
deflection angle and the observables for gravitational
lensing including the time delay between two relativistic images in the strong field limit. Moreover, we will explore how it
differs from that in the Johannsen-Psaltis rotating non-Kerr lensing.

The paper is organized as follows: in the following section we will
review briefly the rotating no-Kerr black hole metric proposed by Konoplya and Zhidenko, and then study the deflection  for
light rays propagating in this background. In Sec.III, we study the
physical properties of the strong gravitational lensing by the
Konoplya-Zhidenko rotating non-Kerr compact object and probe the effects of the deformation parameter on the deflection angle and the coefficients in strong gravitational lensing. In Sec.IV, we study furtherly how the deformation parameter affects the observable in strong gravitational lensing and the time delay between relativistic images. We end the paper with a summary.

\section{Konoplya-Zhidenko rotating non-Kerr spacetime and the deflection angle for light ray}

Let us now review briefly the Konoplya-Zhidenko rotating non-Kerr spacetime. As a usual rotating non-Kerr metric, this metric can be obtained by deforming the Kerr metric. Konoplya and Zhidenko modify the metric as \cite{RA2}
\begin{eqnarray}
ds^2&=&-\frac{N^2(r,\theta)-W^2(r,\theta)\sin^2\theta}{K^2(r,\theta)}dt^2
-{2rW(r,\theta)\sin^2\theta}dtd\phi+{K^2(r,\theta)r^2\sin^2\theta}d\phi^2
\nonumber\\
&+&\Sigma(r,\theta)\bigg[{\frac{B^2(r,\theta)}{N^2(r,\theta)}+r^2d\theta^2}\bigg],\label{metric1}
\end{eqnarray}
with
\begin{eqnarray}
N^2(r,\theta)&=&\frac{r^2-2Mr+a^2}{r^2}-\frac{\eta}{r^3},\;\;\;\;
B(r,\theta)=1,\;\;\;\;
\Sigma(r,\theta)=\frac{r^2+a^2\cos^2\theta}{r^2},\nonumber\\
K^2(r,\theta)&=&\frac{(r^2+a^2)^2-a^2\sin^2\theta(r^2-2Mr+a^2)}{r^2(r^2+a^2\cos^2\theta)}
+\frac{a^2\eta\sin^2\theta}{r^3(r^2+a^2\cos^2\theta)},\nonumber\\
W(r,\theta)&=&\frac{2Ma}{r^2+a^2\cos^2\theta}
+\frac{\eta a}{r^2(r^2+a^2\cos^2\theta)},
\end{eqnarray}
where $M$ is the mass of black hole and $a$ is the rotation parameter. The quantity $\eta$ is a deformation parameter, which describes the deviations from the Kerr spacetime. If $\eta=0$, one can find that the metric reduces to that of usual Kerr black hole. It seems that the choice of the metric in Eq. (\ref{metric1}) is somewhat arbitrary, but it possesses the similar asymptotic behavior as that of Kerr one, which means that it is difficult to be distinguished from Kerr black hole in the examination in the weak field such as the solar system experiments. However, the presence of deformation parameter $\eta$ changes the properties of spacetime near the event horizon, which the test in the strong field region can provide some signals to estimate the possible deviations from the Kerr one. With data of GW150914 \cite{gw,gw1}, Konoplya and Zhidenko find that it is allowed for some non-negligible deformations of the Kerr spacetime resulting in the same frequencies of the black-hole ringing \cite{RA2}.

The position of event horizon of the black hole
is defined by
\begin{eqnarray}
\Delta=r^2-2Mr+a^2-\frac{\eta}{r}=0,\label{metric5}
\end{eqnarray}
which is different from that of usual Kerr one. Solving Eq.(\ref{metric5}),
we find that there exist two critical values for the existence of horizon in the spacetime (\ref{metric1})
\begin{eqnarray}\label{etam0}
\eta_1&=&\frac{2}{27}\bigg(\sqrt{4M^2-3a^2}-2M\bigg)^2
\bigg(\sqrt{4M^2-3a^2}+M\bigg),\nonumber\\ \eta_2&=&-\frac{2}{27}\bigg(\sqrt{4M^2-3a^2}+2M\bigg)^2
\bigg(\sqrt{4M^2-3a^2}-M\bigg).
\end{eqnarray}
The dependence of the existence of horizon on the deformation parameter $\eta$ and the rotation parameter $a$ is plotted in Fig.(1). As $\eta<\eta_2<0$, there is no horizon and the metric describes the geometry of a naked singularity. As $\eta_2\leq\eta\leq0$ or $\eta=\eta_1$, there exist two horizons which like in the Kerr black hole spacetime. However, the black hole spacetime (\ref{metric1}) has three horizons if $0<\eta<\eta_1$ in the case with the negative $\eta_2$ or $\eta_2<\eta<\eta_1$ for the positive $\eta_2$. Moreover, we also find that the black hole possesses a single horizon as $\eta>\eta_1$ or $0<\eta<\eta_2$.  As the rotation parameter $|a|>\frac{2\sqrt{3}M}{3}$, it is easy to find that both the critical values $\eta_1$ and $\eta_2$ become imaginary, which means that $\eta_1$ and $ \eta_2$ are not again the threshold values of existence of horizon. However, we find that in the case $|a|>\frac{2\sqrt{3}M}{3}$ there is a single horizon as $\eta>0$ and no any horizon as $\eta<0$. Thus,  comparing with the Kerr black hole, we find that for the Konoplya-Zhidenko rotating non-Kerr black hole the presence of the deformation parameter extends the allowed range of the rotation parameter $a$ and  changes the spacetime structure in the strong field region.
\begin{figure}[ht]
\begin{center}
\includegraphics[width=7cm]{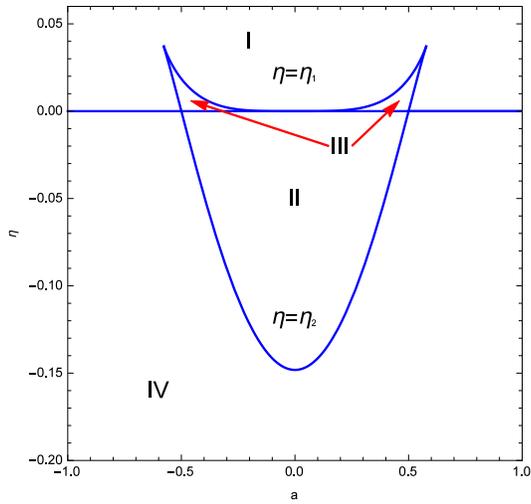}
\caption{The dependence of the existence of horizon on the deformation parameter $\eta$ and the rotation parameter $a$. The regions $I$,  $II$,  $III$ and $IV$ separated by curves $\eta=\eta_1$, $\eta=0$ and $\eta=\eta_2$ in the panel are corresponded to the cases of Konoplya-Zhidenko non-Kerr spacetime with a single horizon, two horizons, three horizons and no any horizon, respectively. The presence of the deformation parameter
extends the allowed range of the rotation parameter $a$ for the Konoplya-Zhidenko rotating non-Kerr  black hole. Here, we set $2M=1$.}
\end{center}
\end{figure}

We are now in position to study the strong gravitational lensing in the background of a Konoplya-Zhidenko rotating non-Kerr spacetime. For simplicity, we here focus on the case in which both the source and the observer lie in the equatorial plane in the Konoplya-Zhidenko rotating non-Kerr spacetime (\ref{metric1}) and the whole trajectory of the
photon is limited on the same plane. With this condition $\theta=\pi/2$, the metric (\ref{metric1}) reduced in the equatorial plane can be expressed as
\begin{eqnarray}
ds^2=-A(x)dt^2+B(x)dx^2+C(x)d\phi^2-2D(x)dtd\phi. \label{metric2}
\end{eqnarray}
Here $x$ is related to the radial coordinate by $x=r/2M$, and then the corresponding metric coefficients become
\begin{eqnarray}
A(x)&=&1-\frac{1}{x}-\frac{\eta}{x^3},\\
B(x)&=&\frac{x^3}{a^2x-x^2+x^3-\eta},\\
C(x)&=&\frac{x^5+a^2(x^2+x^3+\eta)}{x^3},\\
D(x)&=&\frac{a(x^2+\eta)}{x^3},
\end{eqnarray}
where we measure the parameters $a$ and $\eta$ in the units of $2M$.
The null geodesics for the metric (\ref{metric2}) obey
\begin{eqnarray}
\frac{dt}{d\lambda}&=&\frac{C(x)-JD(x)}{D(x)^2+A(x)C(x)},\label{u3}\\
\frac{d\phi}{d\lambda}&=&\frac{D(x)+JA(x)}{D(x)^2+A(x)C(x)},\label{u4}\\
\bigg(\frac{dx}{d\lambda}\bigg)^2&=&\frac{C(x)-2JD(x)-J^2A(x)}{B(x)[D(x)^2+A(x)C(x)]}.\label{cedi}
\end{eqnarray}
where $J$ is the angular momentum
of the photon and $\lambda$ is an affine parameter along the null geodesics. In the background of a Konoplya-Zhidenko rotating non-Kerr black hole, it is easy to obtain the relation
between the impact parameter $u(x_0)$ and the distance of the closest approach of the light ray $x_0$ by the conservation of the
angular momentum along the null geodesics
\begin{eqnarray}
u(x_0)=J(x_0)=\frac{-x^{5/2}_0\sqrt{a^2x_0-x^2_0+x^3_0-\eta}+a(x^2_0+\eta)}{x^2_0-x^3_0+\eta}.
\end{eqnarray}
It is well known that as the closest distance of approach $x_0$ tends to the marginally circular orbit radius $x_{ps}$ of photon,  the deflect angle of the light becomes unlimited large. The equation of circular photon orbits in a stationary axially-symmetric spacetime can be given by
\begin{eqnarray}
A(x)C'(x)-A'(x)C(x)+2J[A'(x)D(x)-A(x)D'(x)]=0.\label{root}
\end{eqnarray}
The marginally circular radius of photon
$x_{ps}$ is defined by the biggest real root outside the horizon of this equation. For a Konoplya-Zhidenko rotating non-Kerr metric (\ref{metric1}),
the equation of circular photon orbits takes a form
\begin{eqnarray}
(2x^3-3x^2-5\eta)^2-8a^2x(x^2+3\eta)=0.\label{root0}
\end{eqnarray}
As expected, it depends on both the deformed parameter $\eta$ and the
rotation parameter $a$ of the black hole. However, the appearance of $\eta$ makes the equation so complicated that it is impossible to obtain an analytical form for the marginally circular photon orbit radius in this case. In Fig. (2), we present the variety of the marginally circular photon orbit radius $x_{ps}$ with the deformed parameter $\eta$ and the rotation parameter $a$ by solving Eq. (\ref{root0}) numerically.
It is show that the marginally circular photon orbit radius $x_{ps}$ increases with the deformation parameter $\eta$. This is different from that in the Johannsen-Psaltis rotating non-Kerr spacetime \cite{TJo} in which the radius $x_{ps}$ decreases with the deformation parameter. Moreover, we find that $x_{ps}$
decreases with the rotation parameter $a$, which is similar to that in the Kerr black hole spacetime. This implies that the decrease of $x_{ps}$ with rotation parameter is a common feature of photon propagating in the background of a rotating black hole.
\begin{figure}[ht]
\begin{center}
\includegraphics[width=5cm]{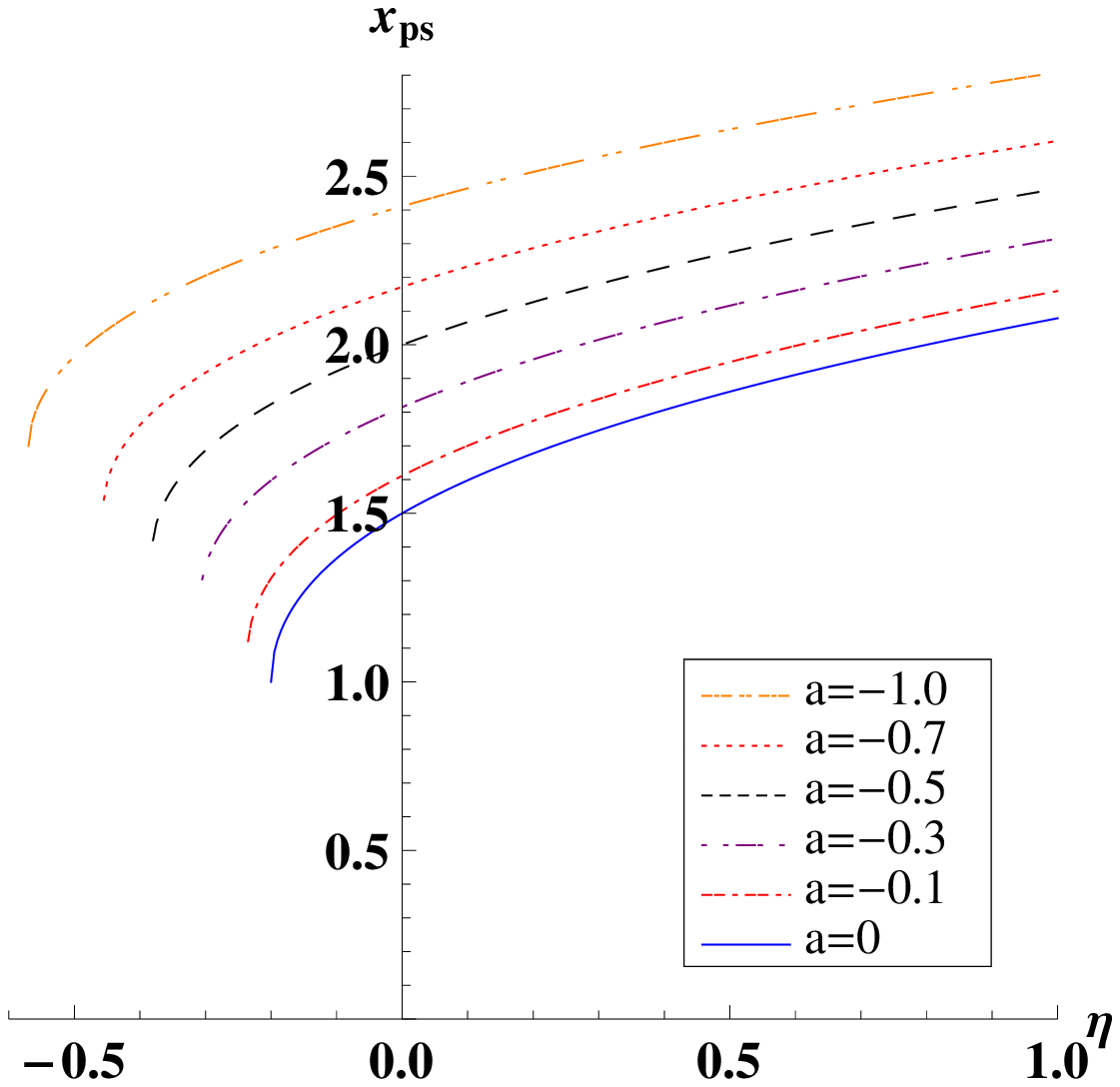}
\;\;\;\;\includegraphics[width=5cm]{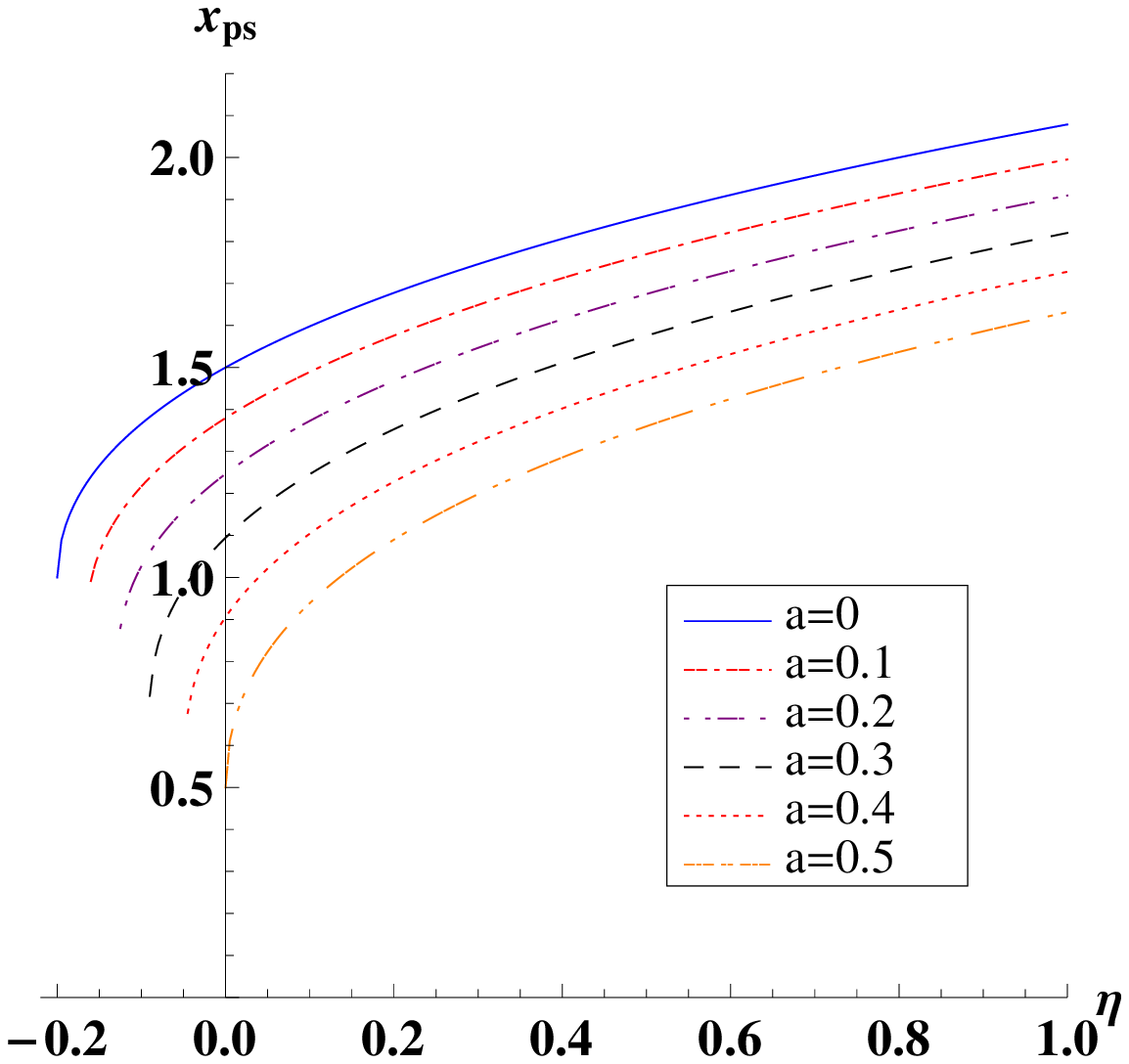}
\;\;\;\;\includegraphics[width=5cm]{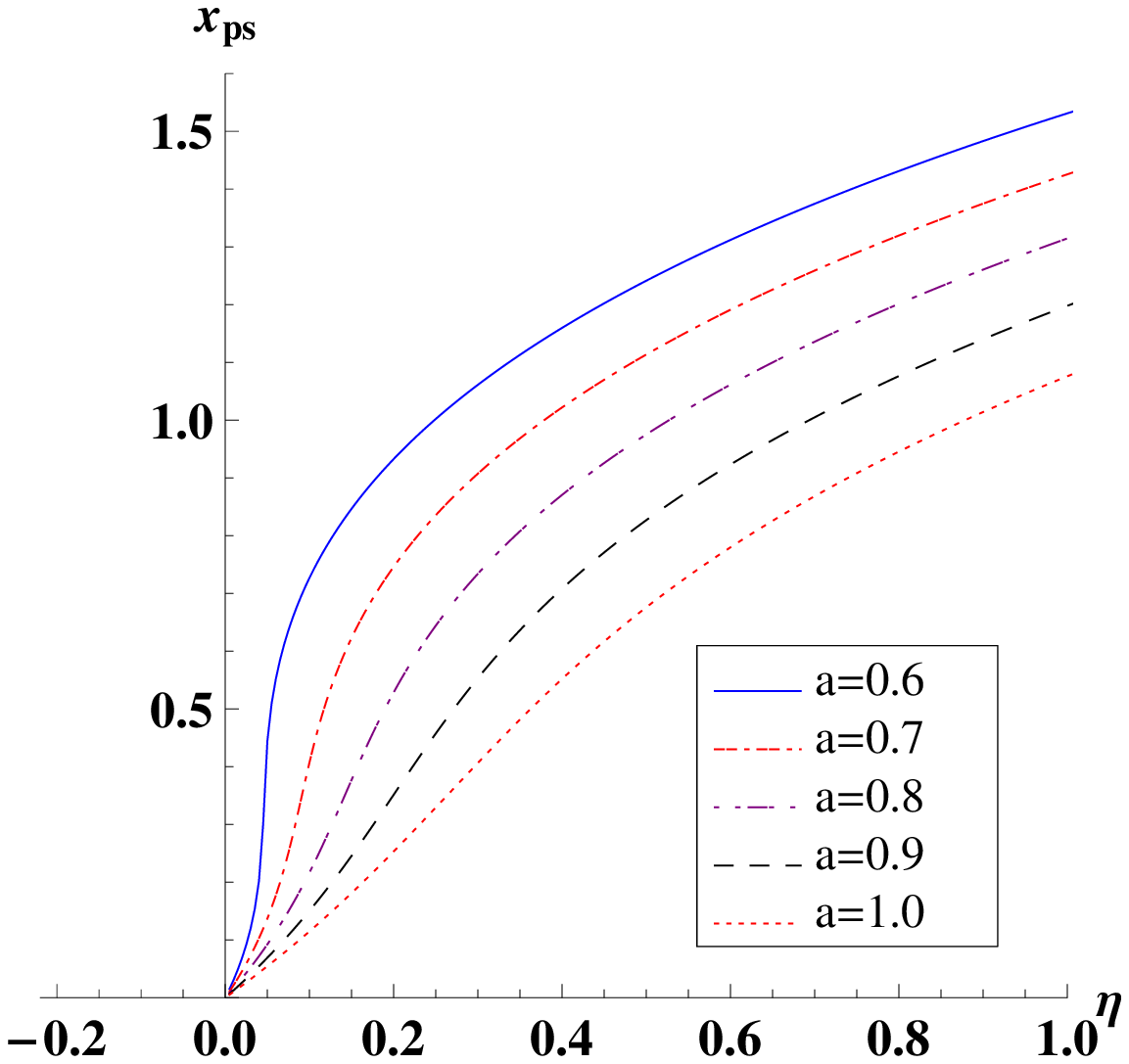}
\caption{Variety of the marginally circular orbit radius of photon
with the deformed parameter $\eta$ for different $a$. Here, we
set $2M=1$.}
\end{center}
\end{figure}
Moreover, we also find that the marginally circular photon orbit radius  $x_{ps}$  always exists for the case with the positive deformation parameter.
For the negative $\eta$, the radius $x_{ps}$ exists only in the regime $\eta\geq\eta_{min}$ for $a\leq0.5$. The value of the lower limit $\eta_{min}$ depends on the rotation rotation parameter and the direction of photon motion,  and its form can be expressed as
\begin{eqnarray}\label{etam}
\eta_{min}=\left\{
\begin{array}{cl}
\eta_3=\frac{[5^{2/3}(\sqrt{3}+i)\mathcal{A}^{2/3}+20i\mathcal{A}^{1/3}-
5^{4/3}(\sqrt{3}-i)]^2}{4500\mathcal{A}^{2/3}}, & a<0,\\
&\\
\eta_4=\frac{[-5^{2/3}(\sqrt{3}-i)\mathcal{A}^{2/3}+20i\mathcal{A}^{1/3}+
5^{4/3}(\sqrt{3}+i)]^2}{4500\mathcal{A}^{2/3}}, & 0\leq a<0.3464,\\ & \\
\eta_2=-\frac{1}{27}(1+\sqrt{1-3a^2})^2(2\sqrt{1-3a^2}-1), & 0.3464\leq a\leq 0.5,
\end{array}\right.
\end{eqnarray}
with $\mathcal{A}=6\sqrt{3a^2 (27 a^2-5)}-54 a^2+5$. We also note that in the cases with $a>0.5$ the marginally circular photon orbit radius $x_{ps}$ exists only in the regime $\eta>\eta_{min}=0$ and $x_{ps}$ is very close to zero as the value of $\eta$ is near the threshold value $\eta_{min}$, which implies that in the region near $\eta_{min}$ the properties of the strong gravitational lensing for $a>0.5$ could differ from those of the cases $a\leq0.5$.
In the parameter panel $(a, \eta)$, the whole region is split into two region $I$ and $II$ by the curves in Eq.(\ref{etam}) as shown in Fig.(3). Here the regions $I$ and $II$  denote the cases with the existence and non-existence of $x_{ps}$, respectively, in the Konoplya-Zhidenko non-Kerr spacetime.
\begin{figure}[ht]
\begin{center}
\includegraphics[width=7cm]{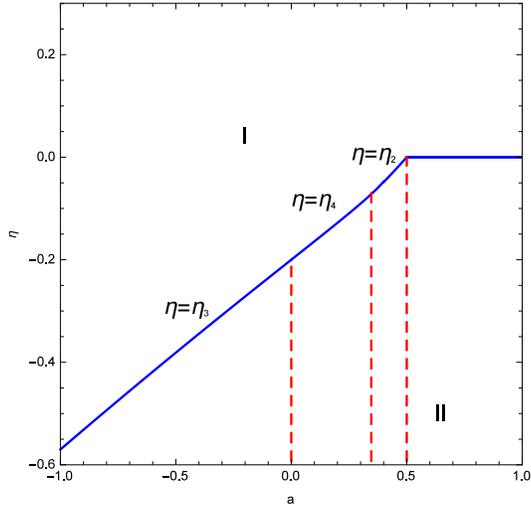}
\caption{The dependence of the existence of the radius of marginally circular orbit of photon $x_{ps}$ on the deformation parameter $\eta$ and the rotation parameter $a$. The regions $I$ and $II$ denote the cases with the existence and non-existence of $x_{ps}$, respectively, in the Konoplya-Zhidenko non-Kerr spacetime.  Here, we set $2M=1$.}
\end{center}
\end{figure}
\begin{figure}
\begin{center}
\includegraphics[width=7cm]{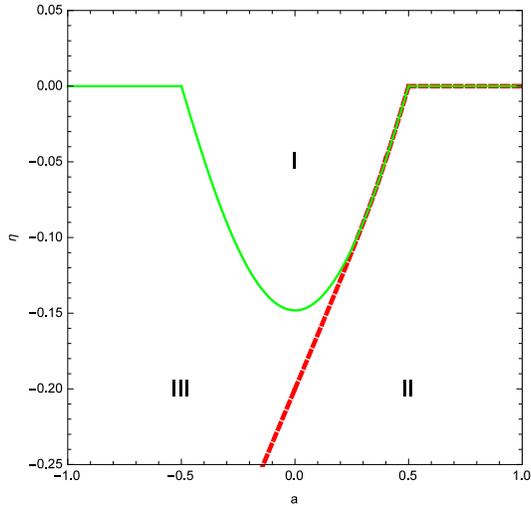}
\caption{ The boundary of the existence of horizon (green line) and of the marginally circular orbit radius for photon (red dashed line) in the Konoplya-Zhidenko non-Kerr spacetime.  Here, we set $2M=1$.}
\end{center}
\end{figure}
Comparing Eq.(\ref{etam0}) with Eq.(\ref{etam}), we find that the condition of existence of horizons is not inconsistent with that of the marginally circular photon orbit, which is different from those in the Johannsen-Psaltis  rotating non-Kerr spacetime. Thus, one can find that in Fig.(4), there exist both horizon and  marginally circular orbit radius $x_{ps}$ when the parameters ($a, \eta$) lie in the region $I$. This situation is similar to that of in the Kerr black hole spacetime with $-M\leq a\leq M$ where both the horizon and the marginally circular orbit radius appear.
Moreover, as ($a, \eta$) lies in the region $II$, there is no horizon and no  marginally circular orbit radius, which means that the singularity is completely naked in this case. However, when ($a, \eta$) is located in region $III$, there is no horizon but the marginally circular orbit radius appears. In other words, the singularity is covered by the marginally circular orbit radius. The later two situations are correspond to the cases of strong naked singularity (SNS) and weakly naked singularity (WNS), respectively, named firstly by Virbhadra and Ellis in the study of strong gravitational lensing in the Janis-Newman-Winicour spacetime \cite{KS4,Gyulchev1}. The weakly naked singularity does not appear in the Johannsen-Psaltis rotating non-Kerr spacetime \cite{TJo,chen12}. This implies that strong gravitational lensing could provide a way to distinguish  two kind of rotating non-Kerr compact objects in the strong field regime.

Let us now discuss the
behavior of the deflection angle for the lens described by a Konoplya-Zhidenko
rotating non-Kerr metric (\ref{metric1}). In a stationary, axially-symmetric background with the metric
(\ref{metric2}), the deflection angle for the photon coming from infinite can be expressed as \cite{Ein1}
\begin{eqnarray}
\alpha(x_{0})=I(x_{0})-\pi,
\end{eqnarray}
where $I(x_0)$ is given by
\begin{eqnarray}
I(x_0)=2\int^{\infty}_{x_0}\frac{\sqrt{B(x)|A(x_0)|}[D(x)+JA(x)]dx}{\sqrt{D^2(x)+A(x)C(x)}
\sqrt{A(x_0)C(x)-A(x)C(x_0)+2J[A(x)D(x_0)-A(x_0)D(x)]}}.\label{int0}
\end{eqnarray}
\begin{figure}[ht]\label{pas1}
\begin{center}
\includegraphics[width=5cm]{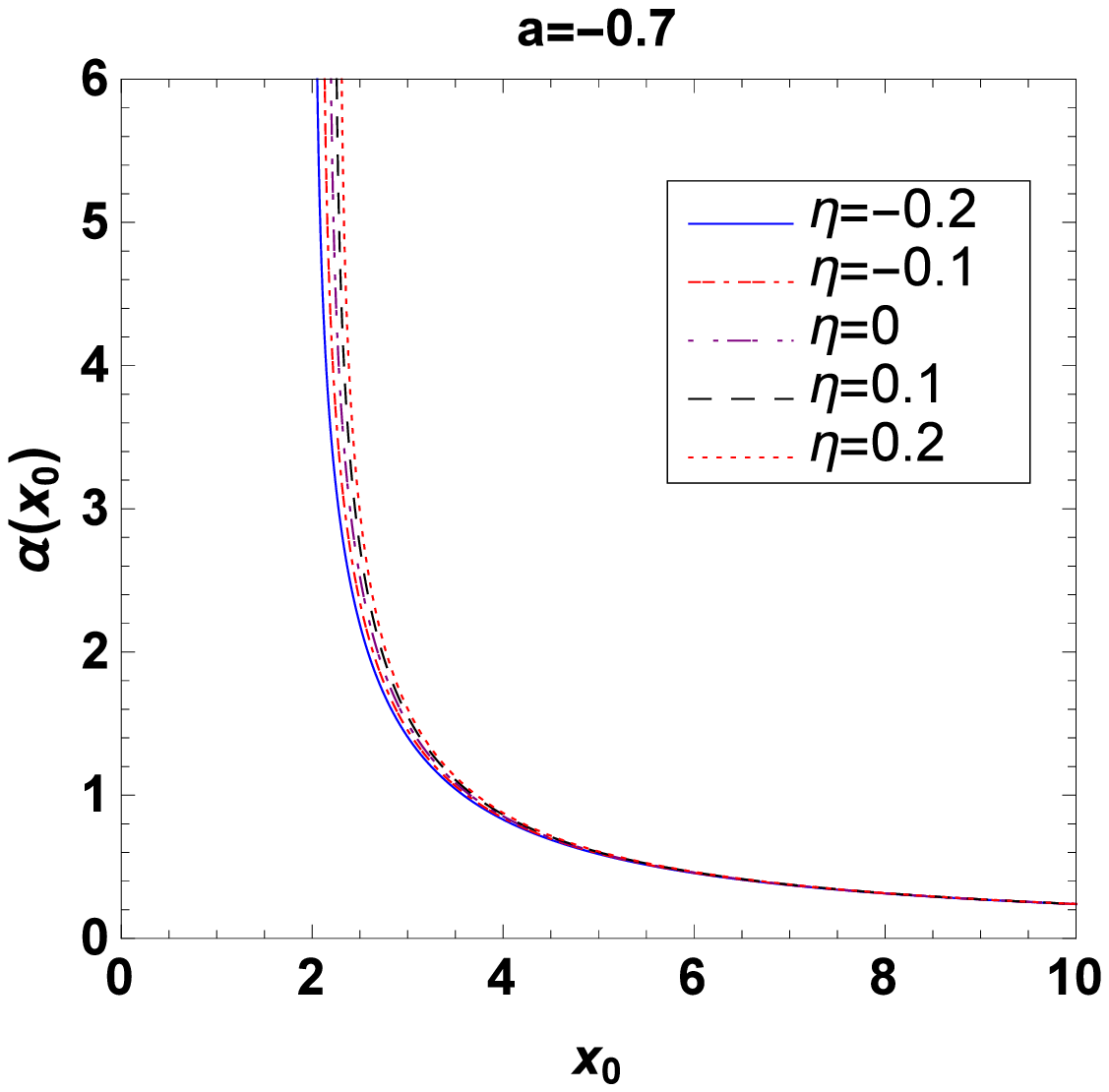}\;\includegraphics[width=5cm]{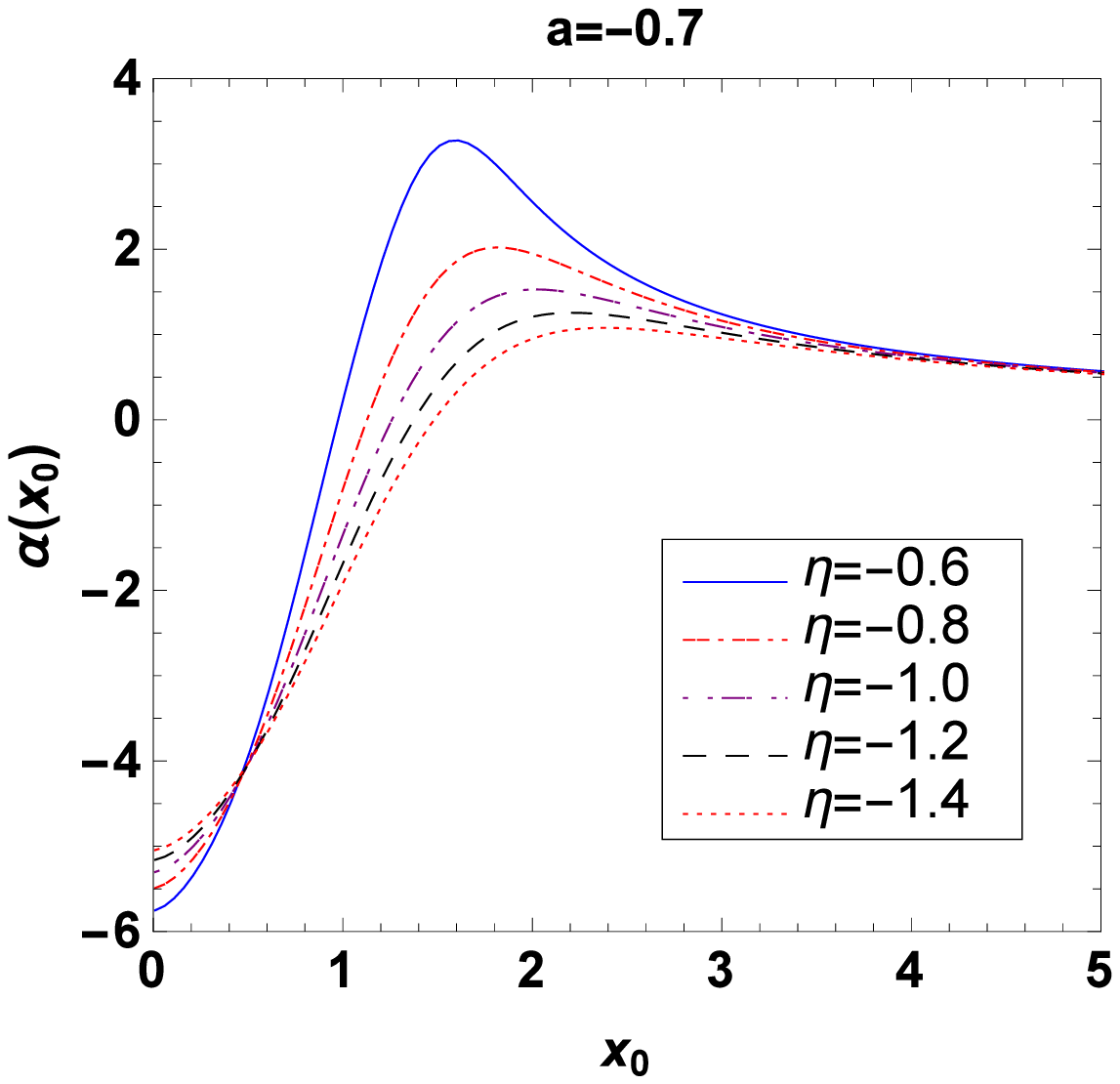}\;
\includegraphics[width=5cm]{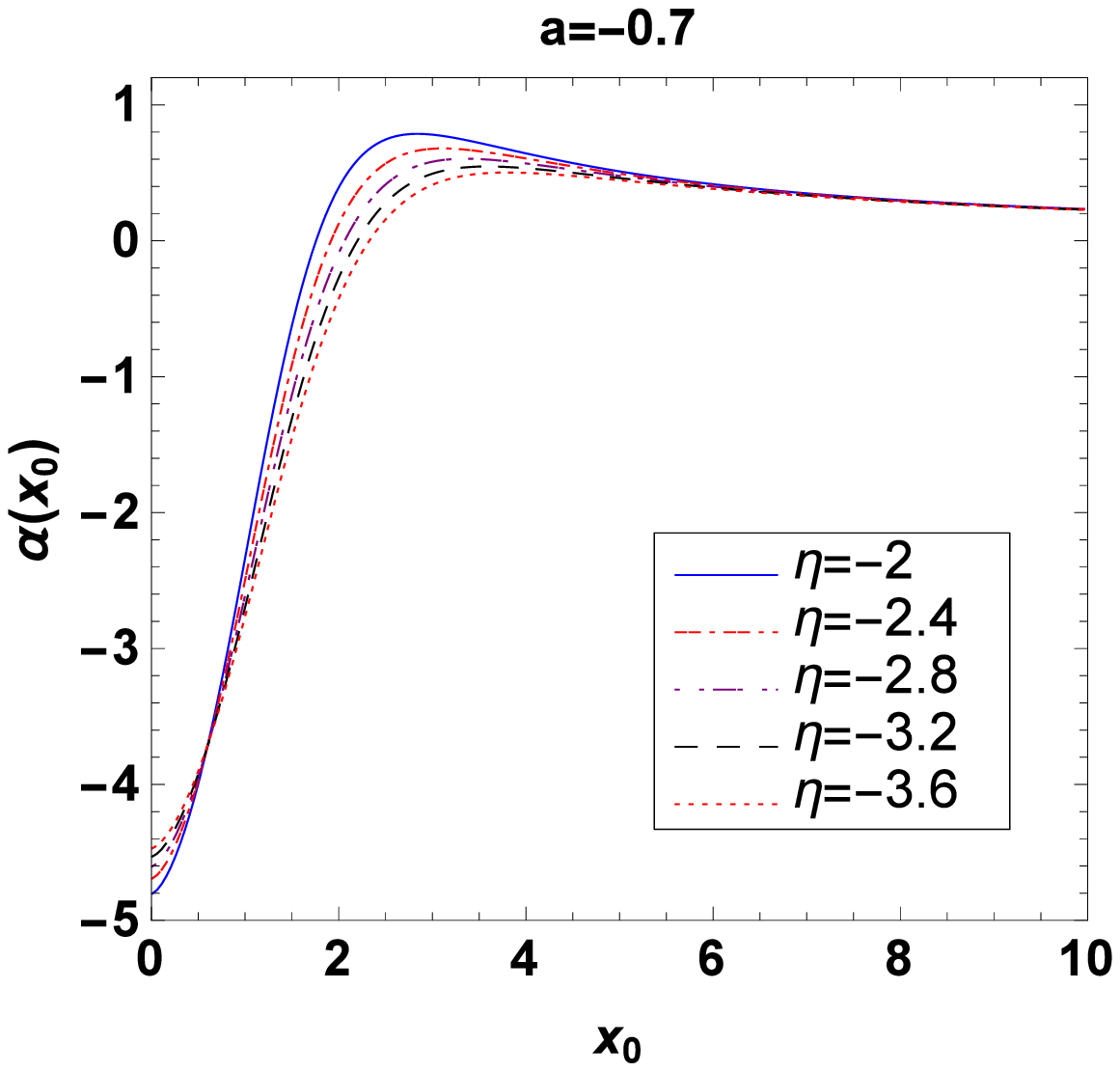}\\
\includegraphics[width=5cm]{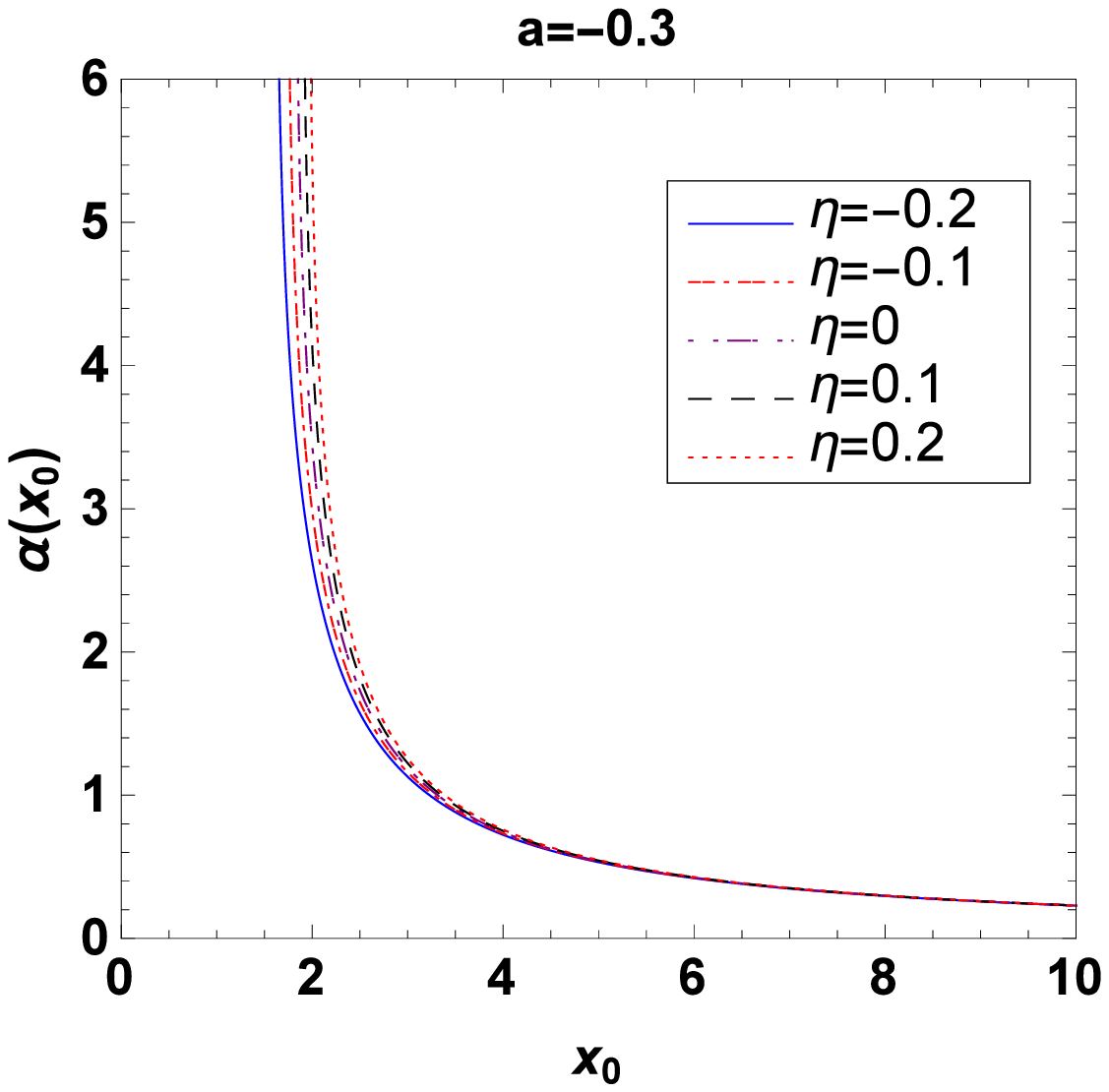}\;\includegraphics[width=5cm]{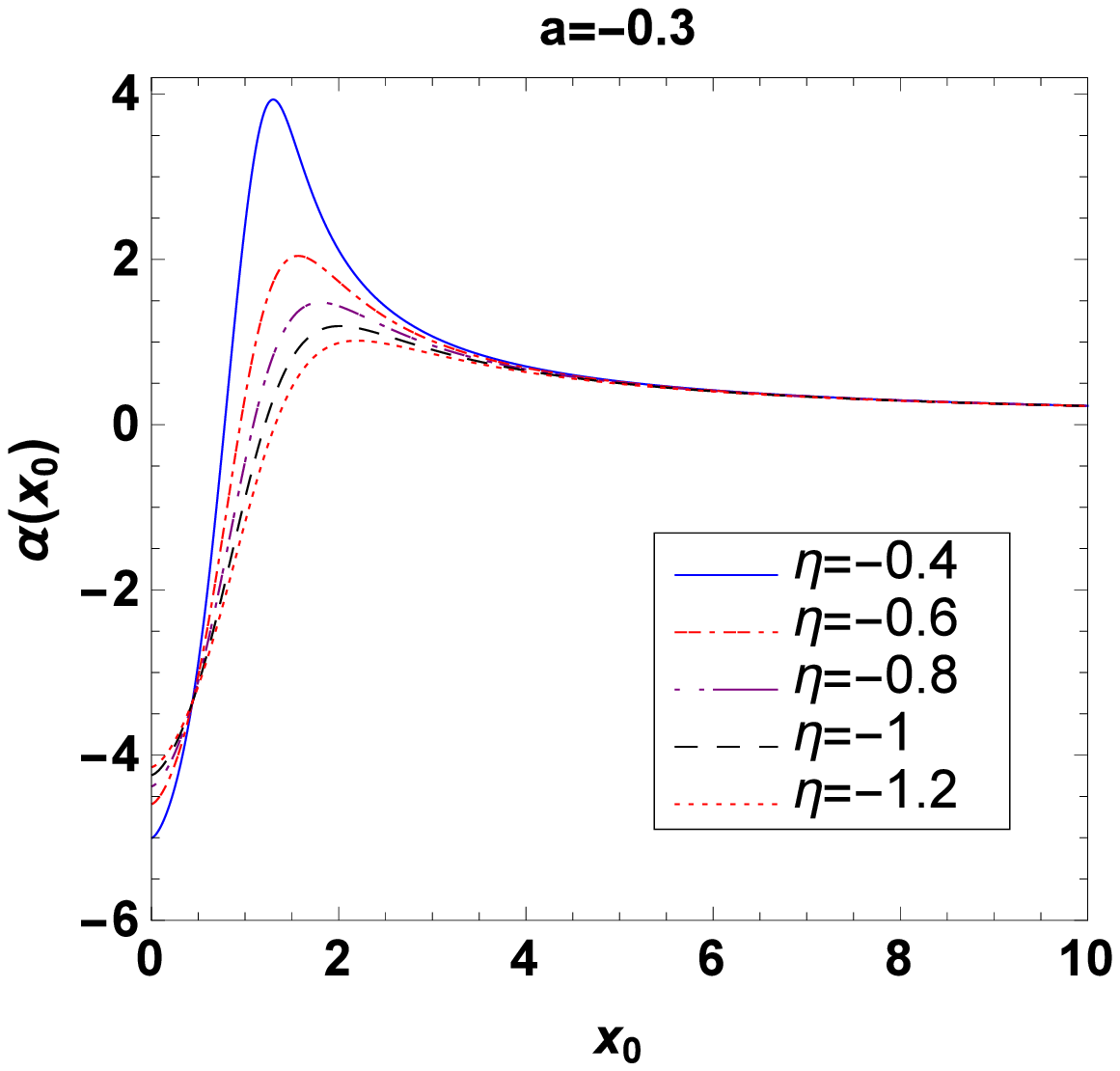}\;
\includegraphics[width=5cm]{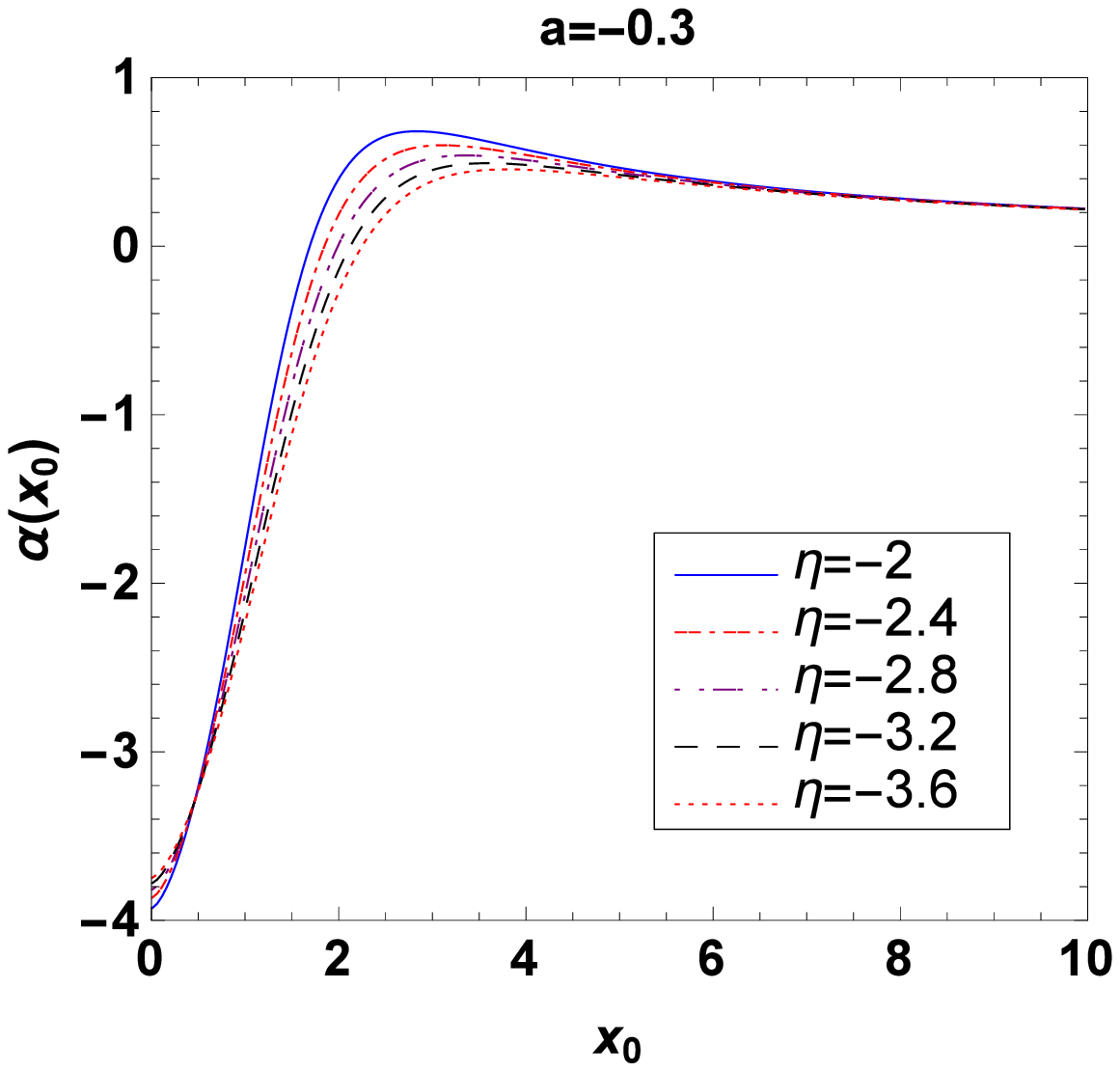}
\caption{Deflection angle $\alpha(x_0)$ as a function of the closest
distance of approach $x_0$ for the negative angular momentum $a$ in the Konoplya-Zhidenko rotating non-Kerr spacetime. Here, we set $2M=1$.}
\end{center}
\end{figure}
\begin{figure}[ht]\label{pas2}
\begin{center}
\includegraphics[width=5cm]{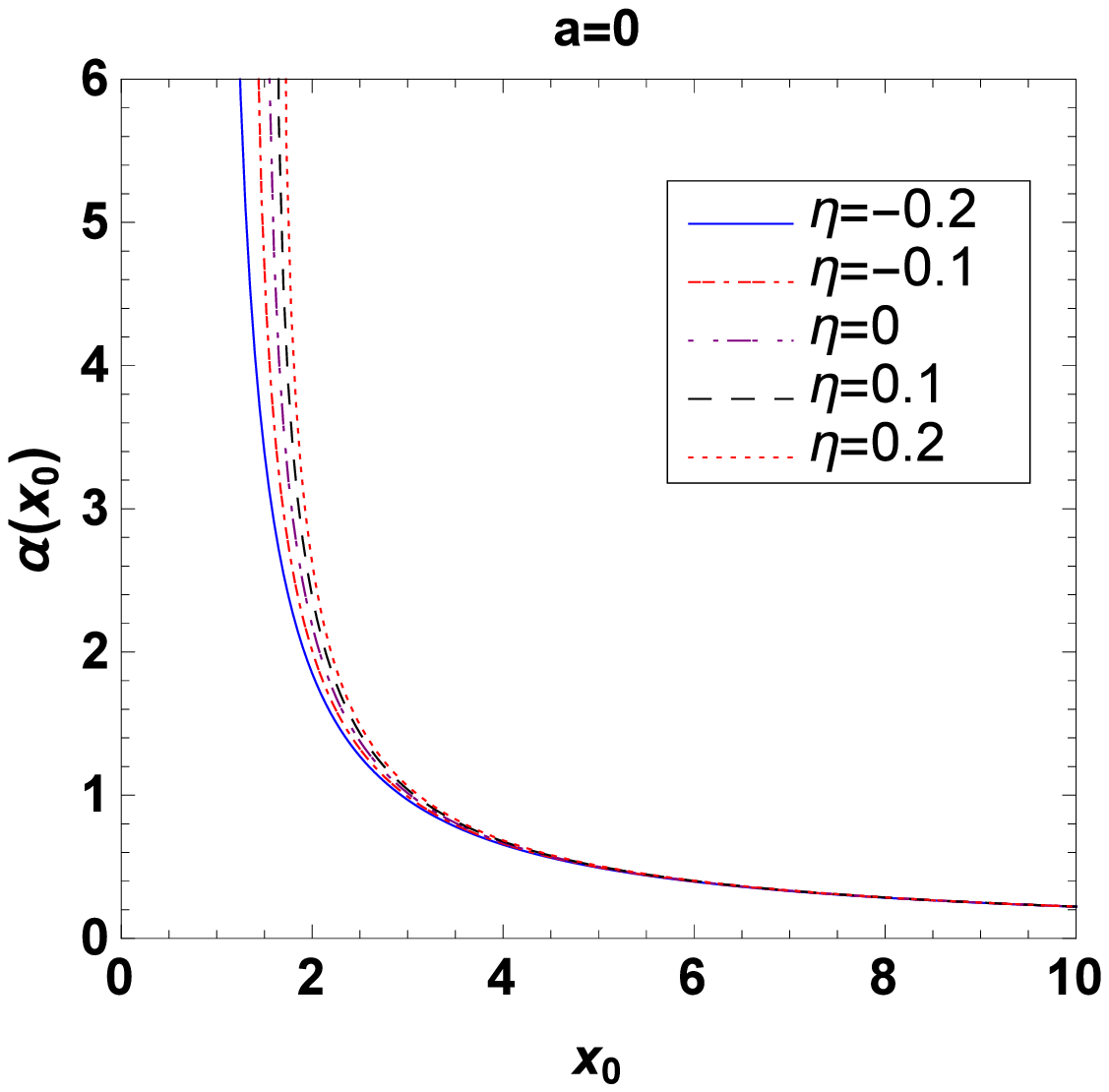}\;\includegraphics[width=5cm]{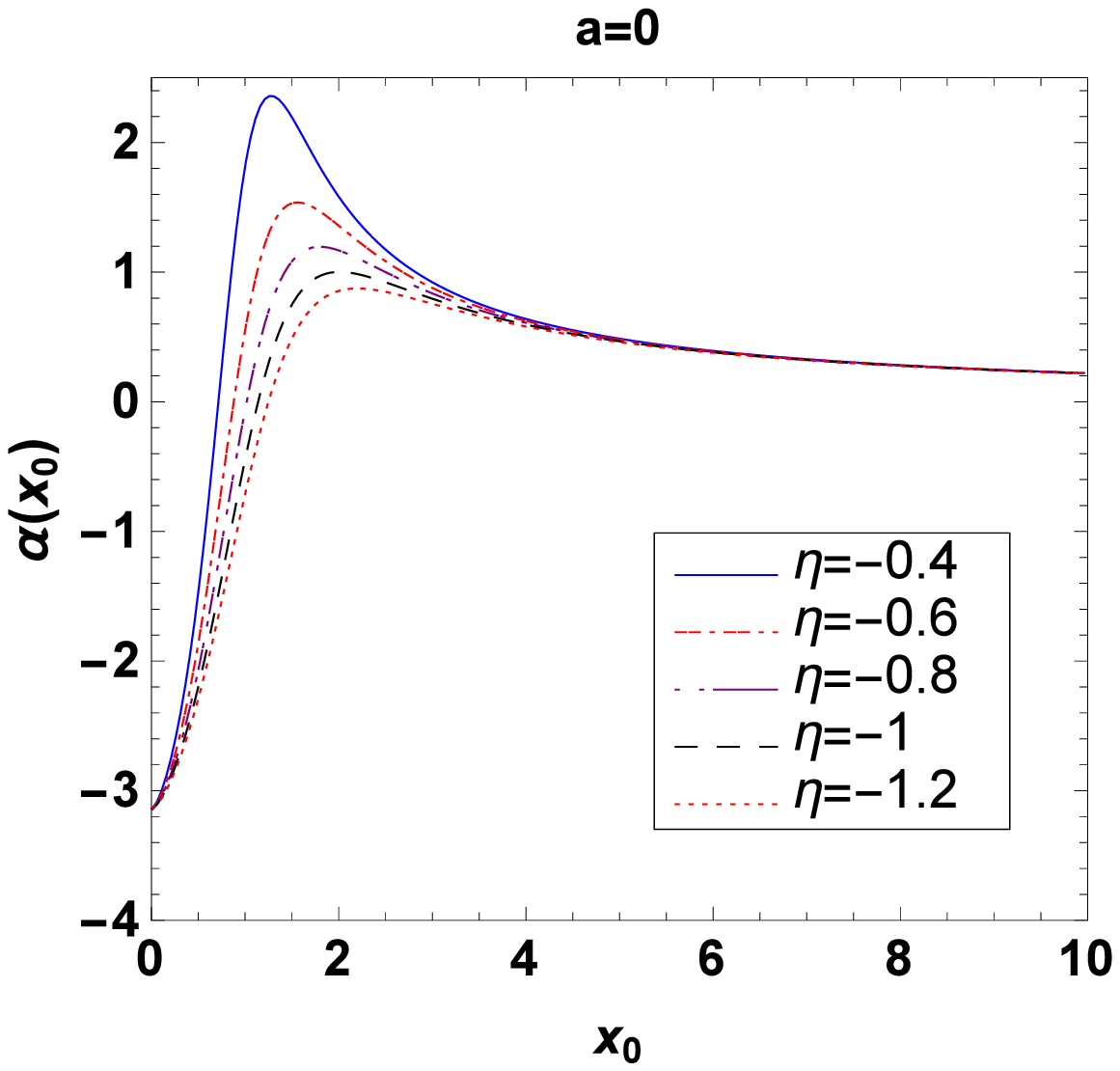}\;
\includegraphics[width=5cm]{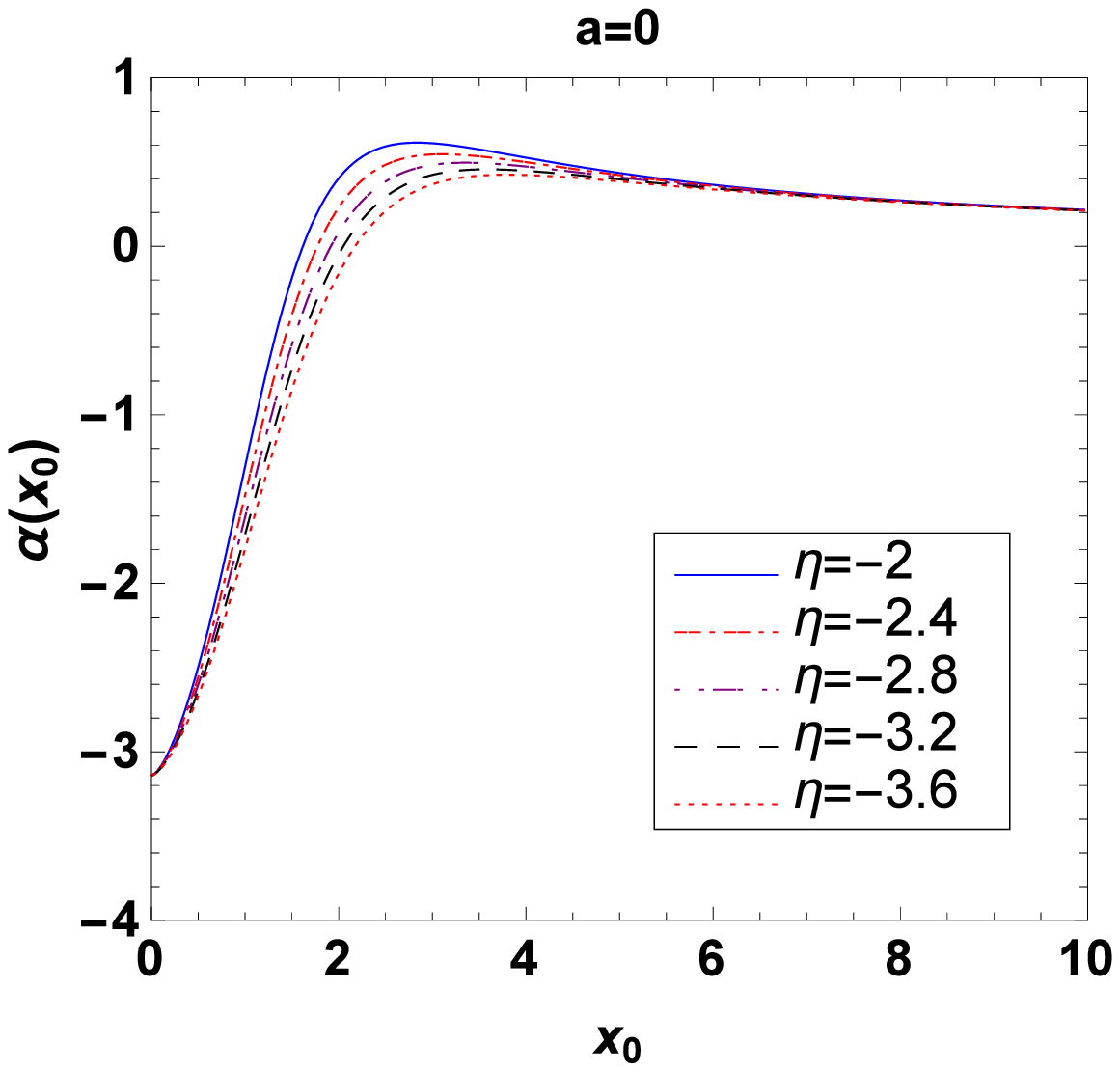}\\
\includegraphics[width=5cm]{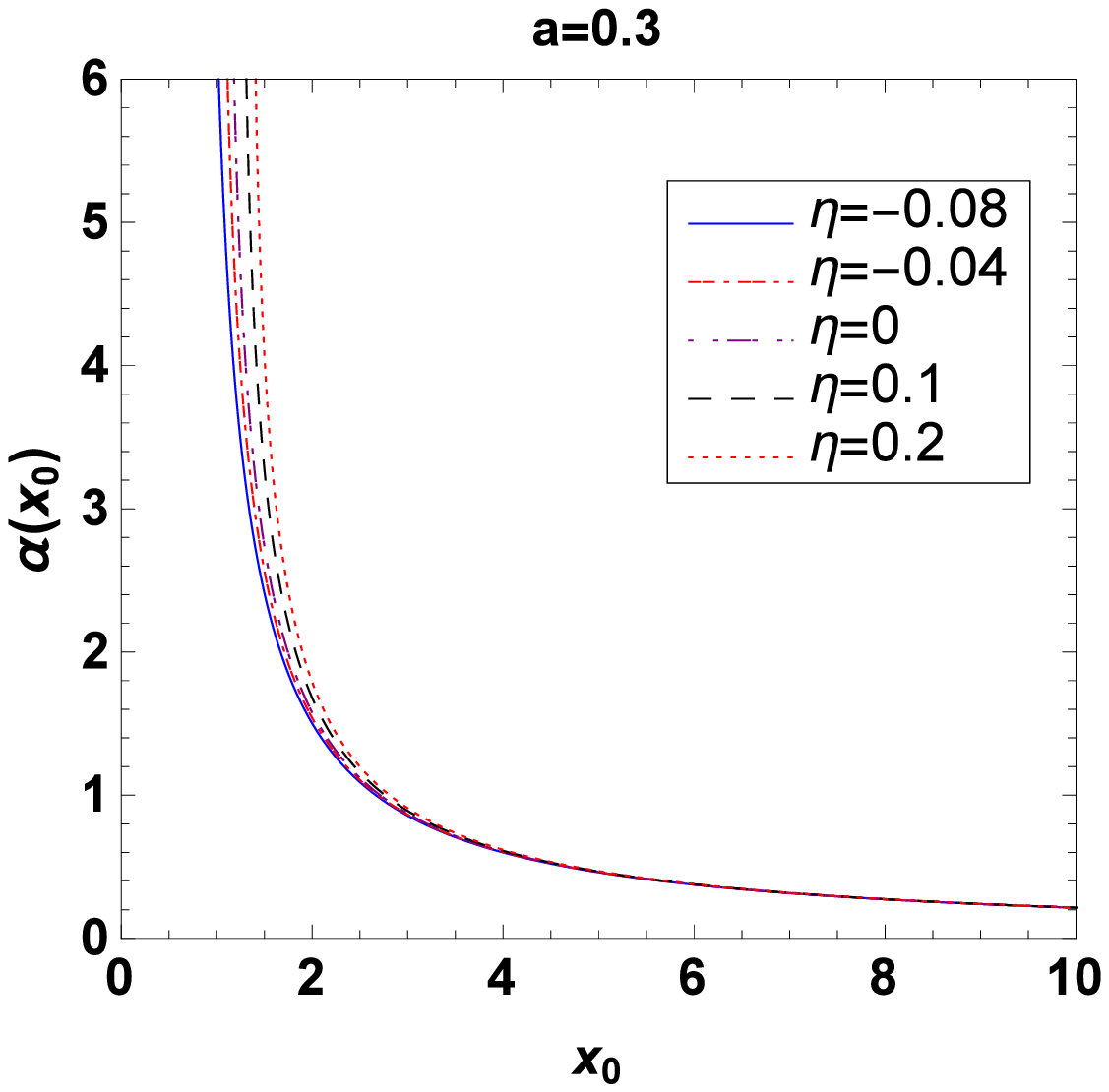}\;\includegraphics[width=5cm]{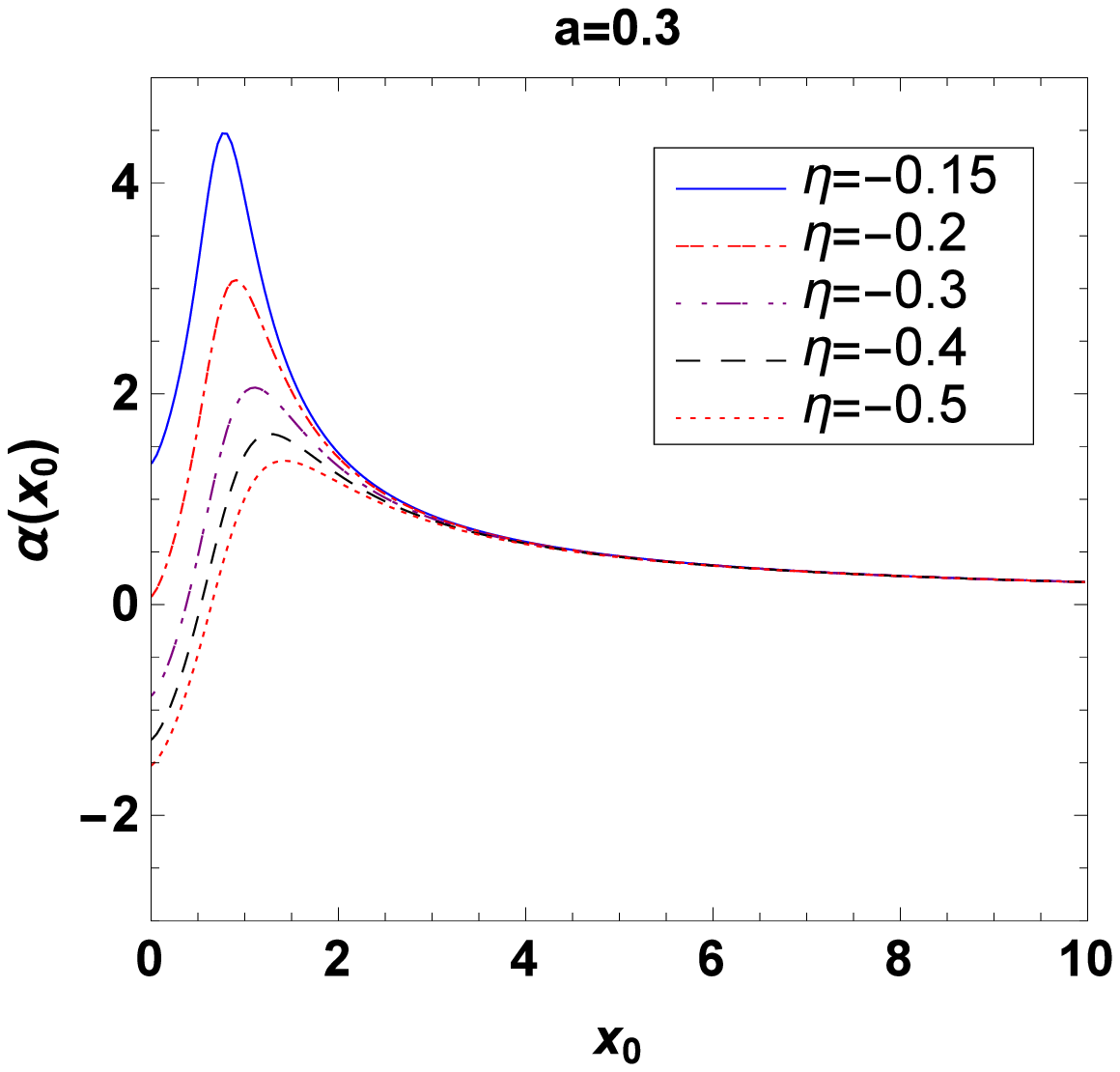}\;
\includegraphics[width=5cm]{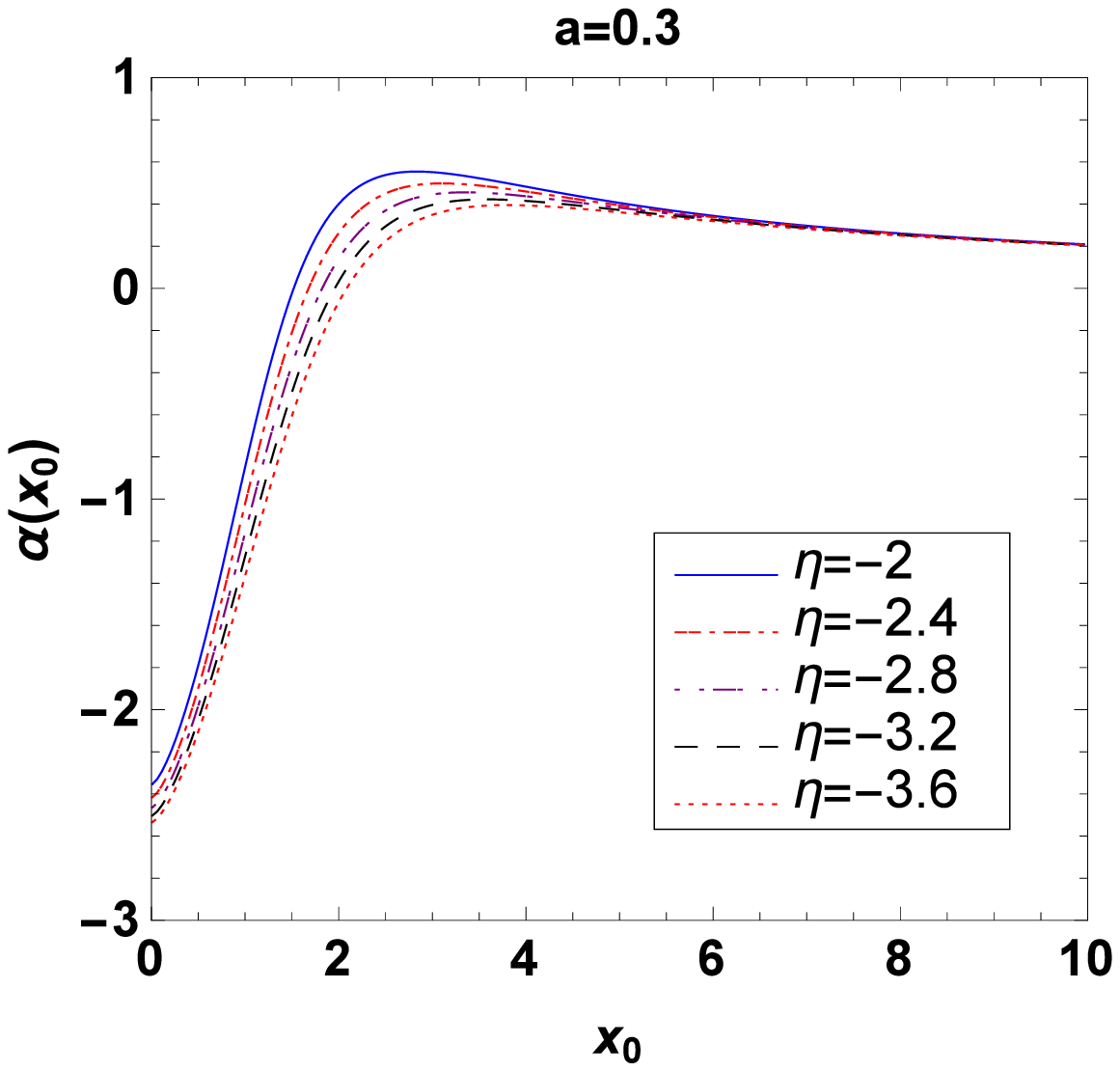}\\
\includegraphics[width=5cm]{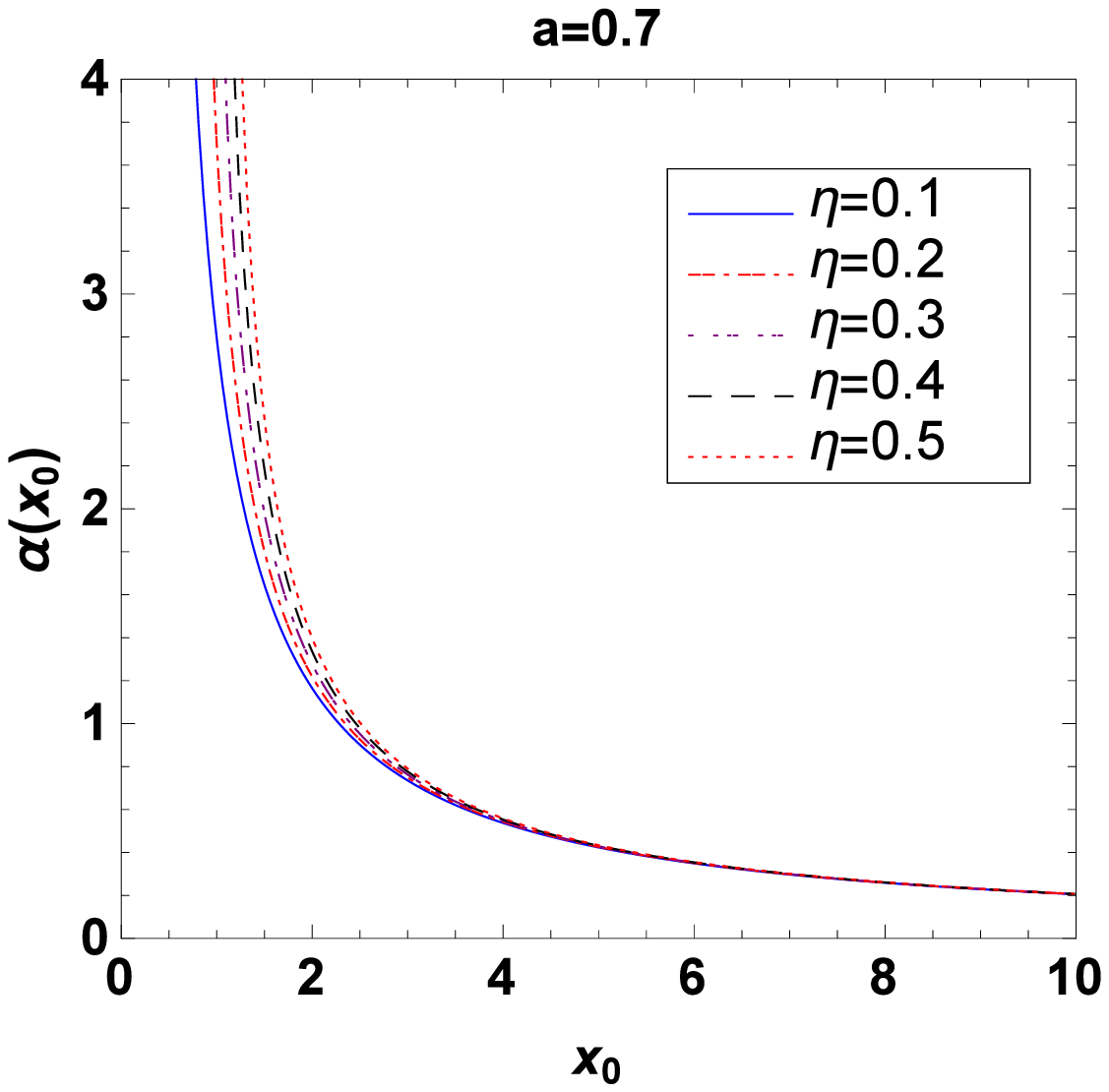}\;\includegraphics[width=5cm]{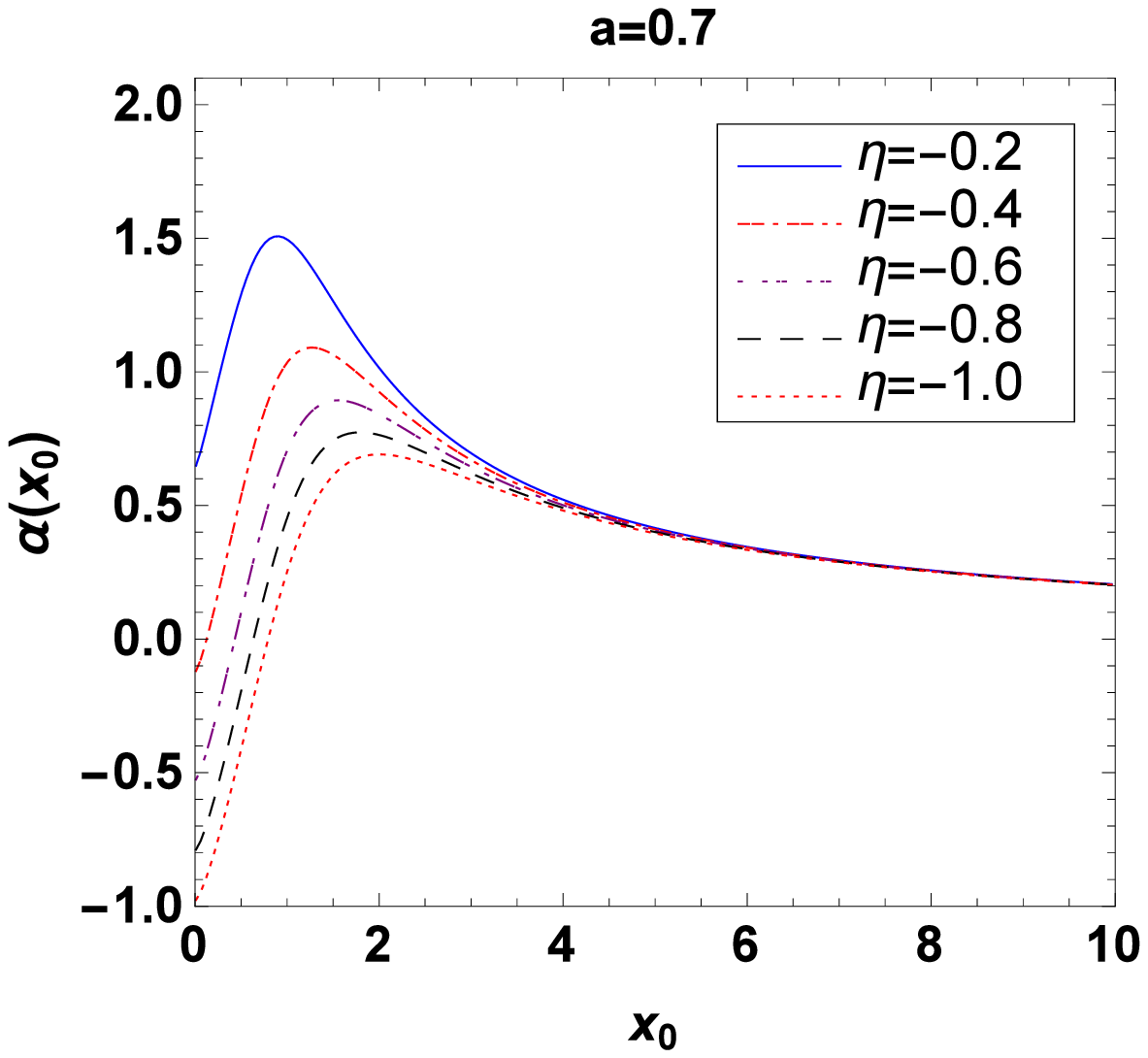}\;
\includegraphics[width=5cm]{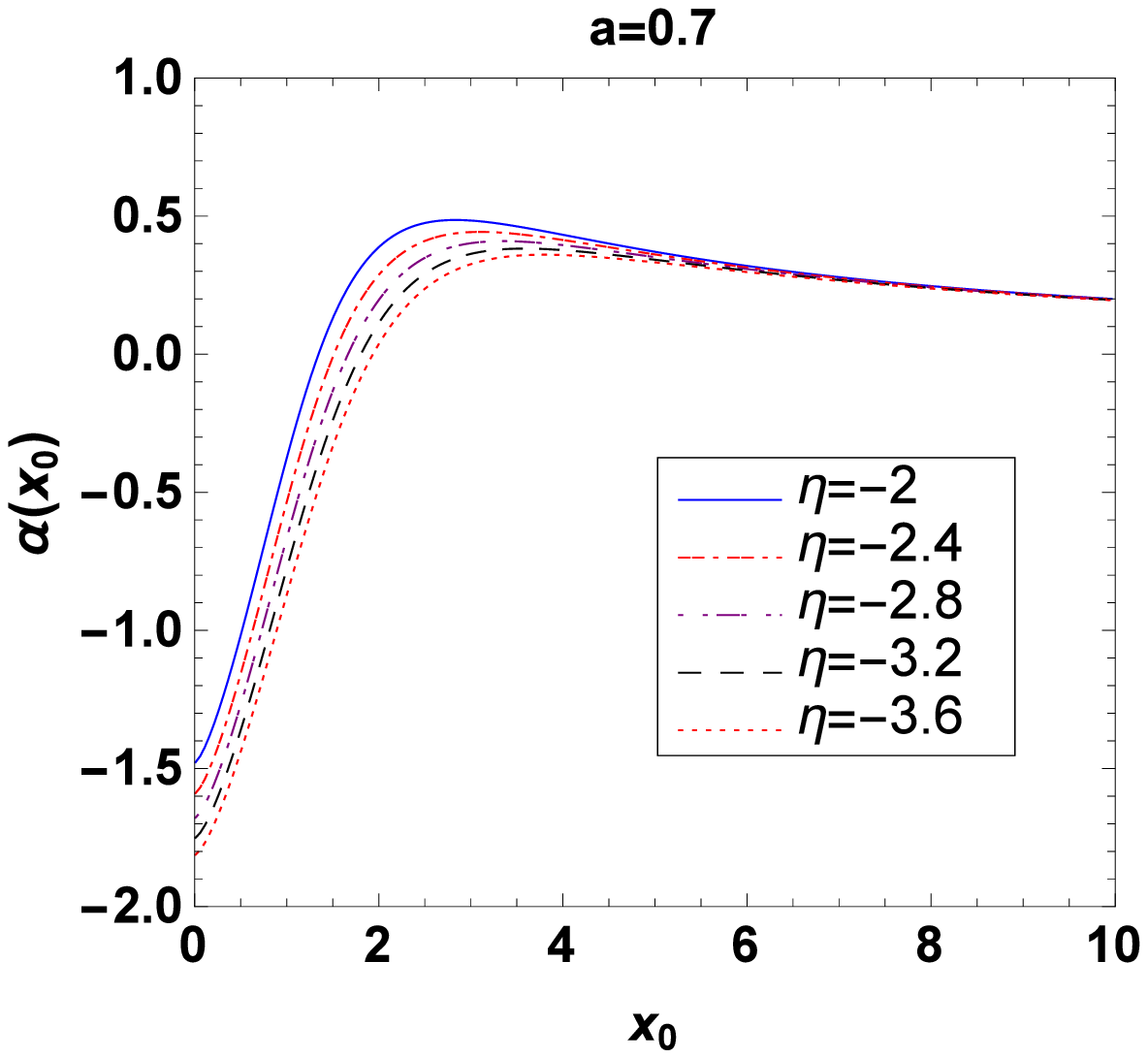}
\caption{Deflection angle $\alpha(x_0)$ as a function of the closest
distance of approach $x_0$ for the non-negative angular momentum $a$ in the Konoplya-Zhidenko rotating non-Kerr spacetime. Here, we set $2M=1$.}
\end{center}
\end{figure}
In Figs.(5) and (6), we present the dependence of the deflection angle $\alpha(x_0)$ on the distance of approach $x_0$ for different angular momentum $a$ and deformation parameter $\eta$ in the Konoplya-Zhidenko rotating non-Kerr spacetime. It is shown that $\text{lim}_{x_0\rightarrow\infty}\alpha(x_{0})=0$
for all values of parameters $\eta$ and $a$, which can be understandable since in the far-field limit the effect of gravity is negligible. When there exists a marginally circular photon orbit for the
compact object (i.e., the parameters ($a, \eta$) lie in the regions $I$ and $III$ in Fig.(4) ),  we find that the deflection angle possesses
similar qualitative properties for the different deformed
parameter $\eta$, and it strictly increases with the decreases
of the the closest distance of approach $x_{0}$ and finally becomes unlimited large as $x_{0}$ tends to the respective marginally circular orbit radius
$x_{ps}$, i.e., $\text{lim}_{x_0\rightarrow x_{ps}}\alpha(x_{0})=\infty$ in this case. When the parameters ($a, \eta$) lie in the region $II$ in Fig.(4),
the marginally circular photon orbit vanishes and then the singularity is naked completely, we find that the deflection angle of the light
ray closing to the singularity tends to a certain finite value $\alpha_s$, i.e., $\text{lim}_{x_0\rightarrow 0}\alpha(x_{0})=\alpha_s$.
This means that the photon could not be captured by the Konoplya-Zhidenko rotating  non-Kerr compact object in this situation. We also find that the sign of $\alpha_s$ depends on the rotation parameter $a$ and the deformation parameter $\eta$, which is shown in Fig.(7). It tells us that the deflection angle $\alpha_s$ is positive as ($a, \eta$) lie in the region $I$ and is negative as ($a, \eta$) are in the region $II$. This means that in the Konoplya-Zhidenko rotating  non-Kerr spacetime the
behavior of the deflection angle near the singularity differs from those in Janis-Newman-Winicour spacetime with the naked singularity \cite{KS4,Gyulchev1} and  in the Johannsen-Psaltis rotating non-Kerr spacetime \cite{chen12} since the values of $\alpha_s$ are always negative in the Janis-Newman-Winicour spacetime  and are always positive in the Johannsen-Psaltis one.
\begin{figure}
\begin{center}
\includegraphics[width=7cm]{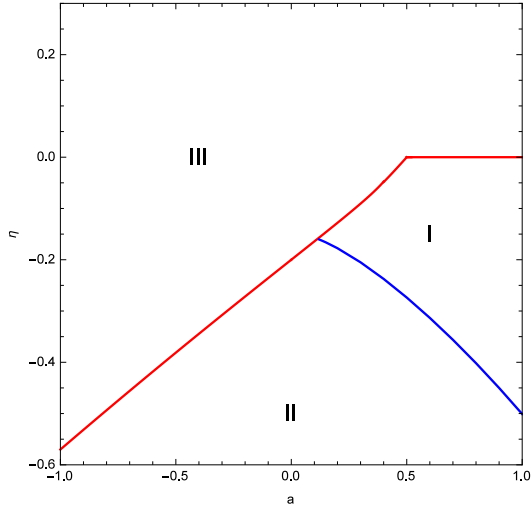}
\caption{ The sign distribution of $\alpha_s$ in the panel of $(a,~\eta)$ for the Konoplya-Zhidenko non-Kerr spacetime. The regions $I$ and $II$ denote the cases with the positive and negative of $\alpha_s$, respectively. The region $III$ denotes the cases with the marginally circular orbit radius
$x_{ps}$ and with the non-existence of $\alpha_s$. Here, we set $2M=1$.}
\end{center}
\end{figure}
\begin{figure}[ht]
\begin{center}
\includegraphics[width=5cm]{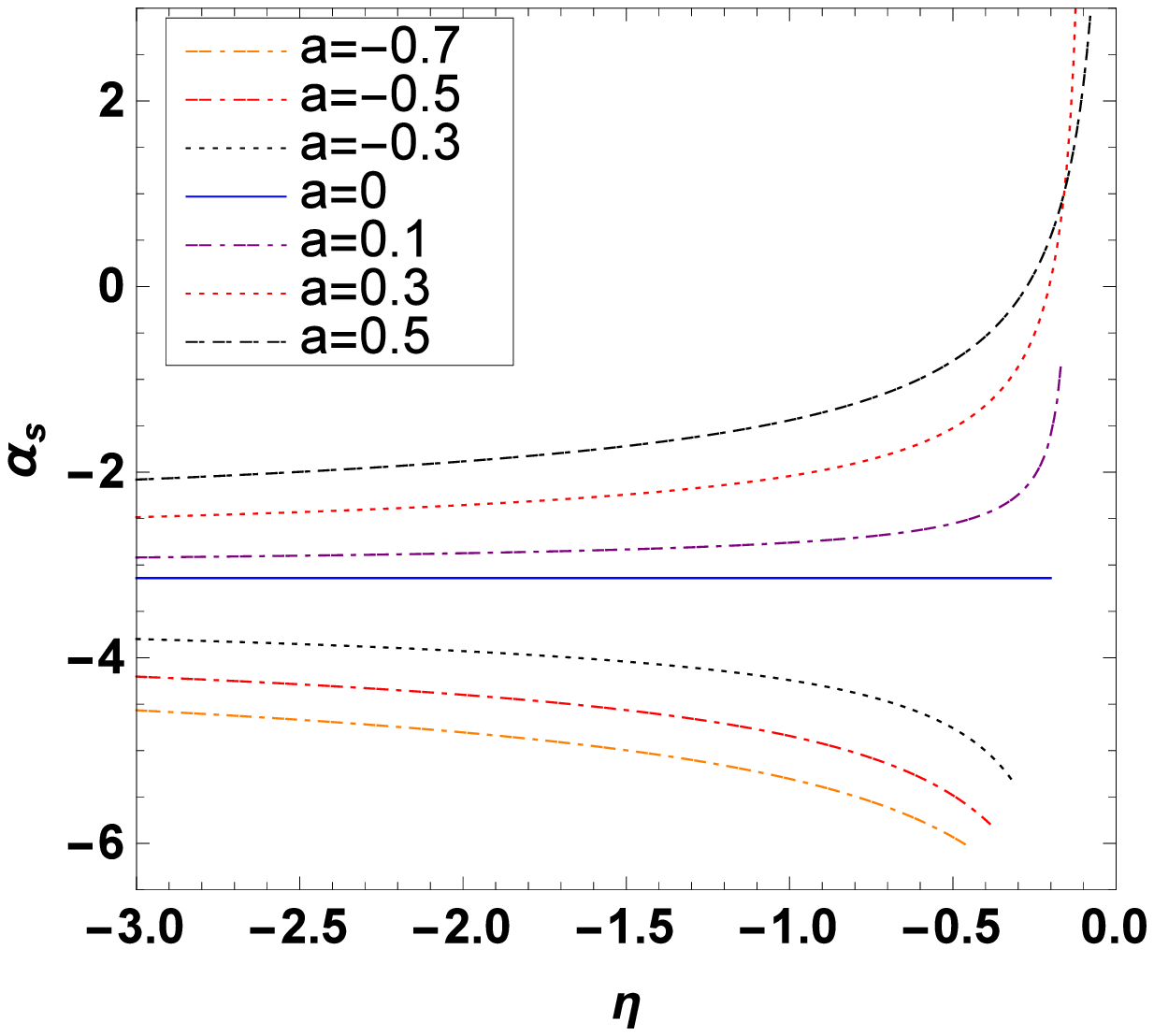}
\;\;\;\;\includegraphics[width=5cm]{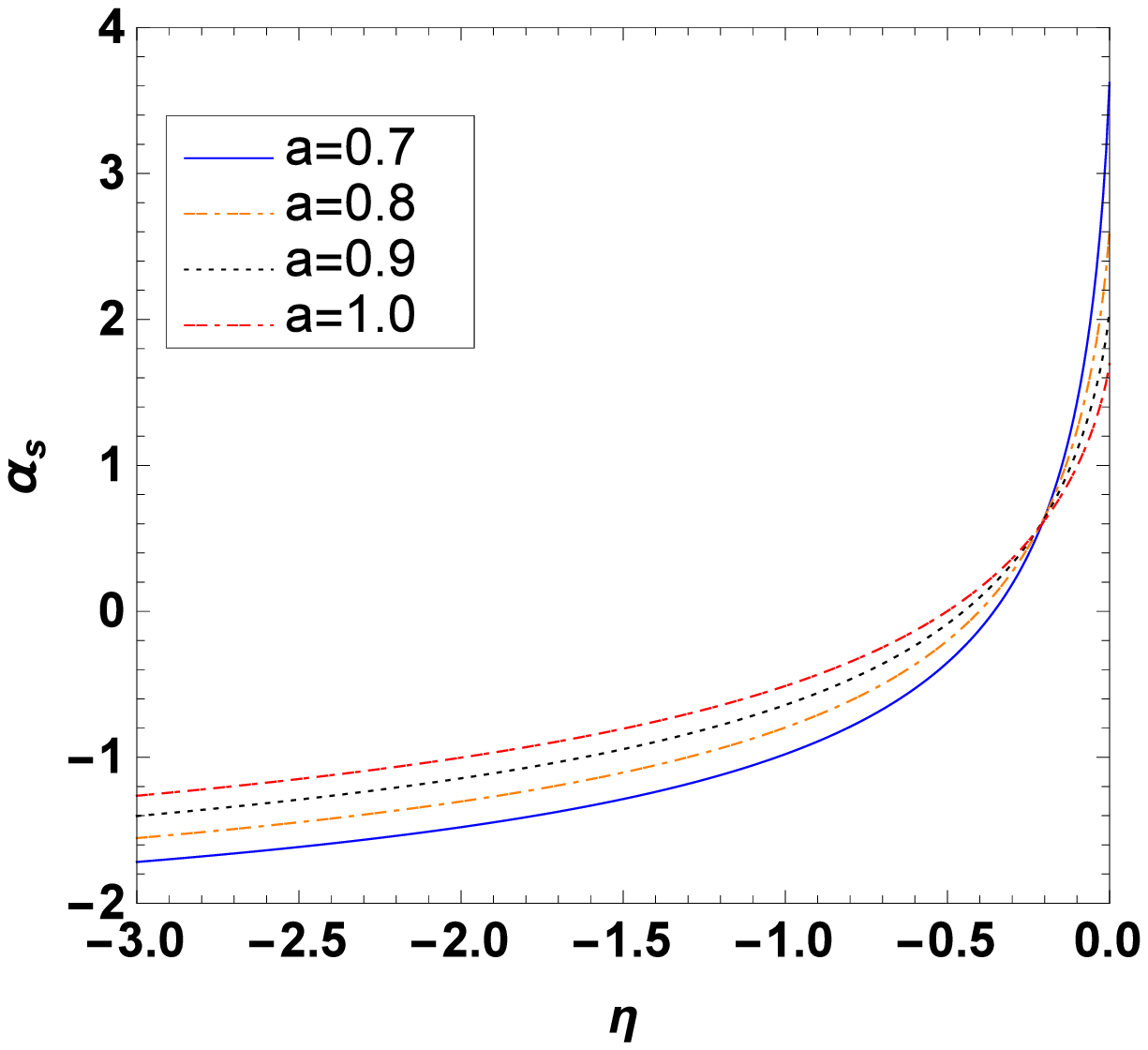}
\;\;\;\;\includegraphics[width=5cm]{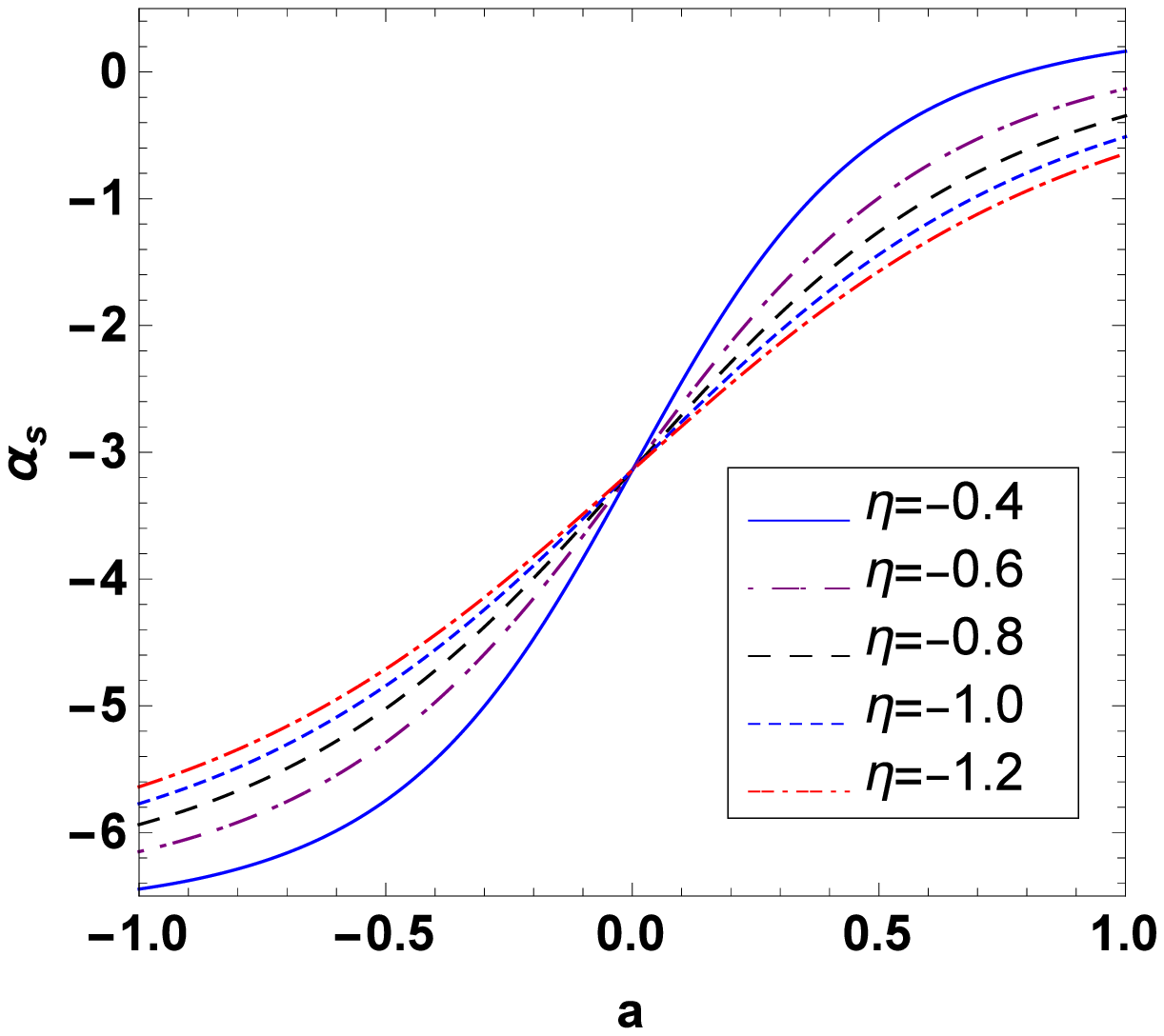}
\caption{Variety of the deflection angle $\alpha_s$ with the
deformed parameter $\eta$ and the angular momentum $a$ as the
light ray is very close to the singularity. Here, we set $2M=1$.}
\end{center}
\end{figure}
 Moreover, from Fig. (8), one can find that in the case $a\leq0$ the value of $\alpha_s$ increases monotonically with $a$ for fixed deformation parameter $\eta$, which is converse to that in the Johannsen-Psaltis non-Kerr spacetime. For the case $a>0$, $\alpha_s$ increases with $a$ as the value of $\eta$ lies in the region far from the threshold value $\eta_{min}$ and decreases with $a$ as the value of $\eta$ near $\eta_{min}$. With increase of the deformation parameter $\eta$, the deflection angle $\alpha_s$ increases monotonically for the prograde photon (i.e.,$a>0$) and decreases monotonically for the retrograde photon (i.e.,$a<0$). It is different from those in the Johannsen-Psaltis non-Kerr spacetime in which $\alpha_s$ first decreases and then increases with the deformed parameter for $a>0$, while for $a<0$ there does not exist such kind of $\alpha_s$ since  because of the existence of black hole horizon and  marginally circular orbit radius for the retrograde photon in the Johannsen-Psaltis spacetime.
 In the Konoplya-Zhidenko rotating  non-Kerr spacetime, there exists a upper limit of the deformation parameter $\eta_{sin}$ for the existence of $\alpha_s$, which is actually the lower limit $\eta_{min}$ for the existence of the marginally circular photon orbit radius. From the previous discussion, it is easy to obtain that the value of $\eta_{sin}$ increases with $a$ as $a<0.5$ and tends to zero as $a>0.5$.
 From Fig.(8), we find $\alpha_s$ does not exist as $\eta$ is larger than the upper limit $\eta_{sin}$, which means that the light ray can not close to the singularity since the marginally circular orbit radius for photon recovers, and then the behavior of deflection angle reduces to that in the Kerr black hole spacetime. Especially, in the limit case with $a=0.5$, we find that $\alpha_s$ diverges as $\eta$ tends to $\eta_{sin}=0$. This could be explained by a fact that the impact parameter for the light ray near the singularity is $J(x_0)|_{x_0\rightarrow 0}=a$, which means that  the deflection angle $\alpha_s$ can be calculated directly by
 \begin{eqnarray}\label{asqf}
\alpha_s=2\int^{\infty}_{0}\frac{\sqrt{B(x)}[D(x)+JA(x)]dx}{\sqrt{D^2(x)+A(x)C(x)}
\sqrt{C(x)-2JD(x)-J^2A(x)]}}-\pi=2\int^{\infty}_{0}\frac{ax}{x^3-x^2+ax-\eta}-\pi.
\end{eqnarray}
 The quantity $x^3-x^2+ax-\eta$ in the denominator in Eq.(\ref{asqf})
 $x^3-x^2+ax-\eta>0$ as the parameter $(a,\eta)$ lies in the region where the singularity is completely naked. When $(a,\eta)$ lies in the region near the critical curve of existing horizon overlaping the critical curve of existing the marginally circular orbit radius ( i.e., the region where the green line  overlaps the red line in Fig.(4)), there exists a point $x_c$ in the integral region $(0, +\infty)$ at which the deviations of the quantity $x^3-x^2+ax-\eta$ from zero is very small, which leads to the large deflection angle $\alpha_s$.  Especially, in the case with $a=0.5$ and $\eta$ tends to $\eta_{sin}=0$, the value of $x^3-x^2+ax-\eta$ is close to zero at $x=0.5$, which results in the divergent deflection angle $\alpha_s$. Actually, in this limit case (.i.e, $a=0.5$ and $\eta=0$), both the marginally circular orbit and horizon are recovered and the divergent deflection angle $\alpha_s$ also means that the light ray can not close to the singularity  due to the reemergence of the horizon and the marginally circular orbit. Thus, both the horizon and the marginally circular orbit play very important roles in the propagation of photon in a spacetime.

\section{Strong gravitational lensing in the Konoplya-Zhidenko rotating non-Kerr spacetime}

In this section we will study the gravitational lensing by a
Konoplya-Zhidenko rotating non-Kerr compact object with the marginally circular photon orbit and then probe how the deformed parameter
$\eta$ affects the coefficients and the lensing observables in the
strong field limit.

In order to obtain analytically the properties of the deflection angle near
the marginally circular photon orbit, we adopt to the approximation
method for the integral (\ref{int0}) proposed firstly by Bozza \cite{VB1}. Defining a variable
\begin{eqnarray}
z=1-\frac{x_0}{x},
\end{eqnarray}
we can rewrite the Eq.(\ref{int0}) as
\begin{eqnarray}
I(x_0)=\int^{1}_{0}R(z,x_0)f(z,x_0)dz,\label{in0}
\end{eqnarray}
with
\begin{eqnarray}
R(z,x_0)&=&\frac{2x^2}{x_0\sqrt{C(z)}}\frac{\sqrt{B(z)|A(x_0)|}
[D(z)+JA(z)]}{\sqrt{D^2(z)+A(z)C(z)}}, \label{R10}
\end{eqnarray}
\begin{eqnarray}
f(z,x_0)&=&\frac{1}{\sqrt{A(x_0)-A(z)\frac{C(x_0)}{C(z)}
+\frac{2J}{C(z)}[A(z)D(x_0)-A(x_0)D(z)]}}.
\end{eqnarray}
The function $R(z, x_0)$ is regular for all values of $z$
and $x_0$, but the function $f(z, x_0)$ is divergent as $z$ tends to zero. Actually, the divergence of $f(z, x_0)$ near $z=0$ reflects  that the deflection angle becomes the unbound large as the photon approaches the marginally circular photon orbit. Thus, one can split the integral (\ref{in0}) into the divergent part $I_D(x_0)$ and the regular part $I_R(x_0)$
\begin{eqnarray}
I_D(x_0)&=&\int^{1}_{0}R(0,x_{ps})f_0(z,x_0)dz, \nonumber\\
I_R(x_0)&=&\int^{1}_{0}[R(z,x_0)f(z,x_0)-R(0,x_{ps})f_0(z,x_0)]dz
\label{intbr}.
\end{eqnarray}
\begin{figure}[ht]
\begin{center}
\includegraphics[width=5cm]{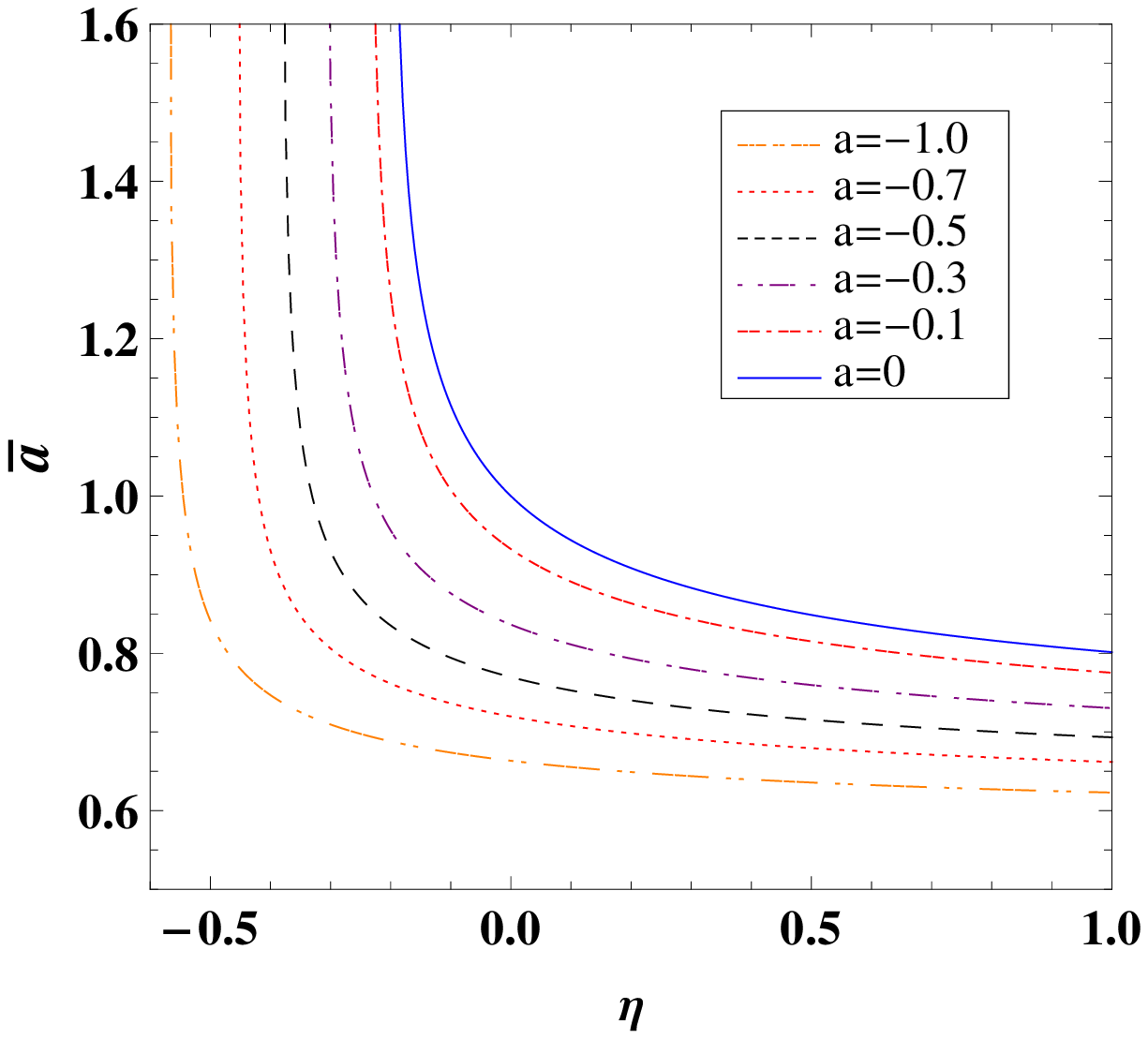}\;\;\;\;\includegraphics[width=5cm]{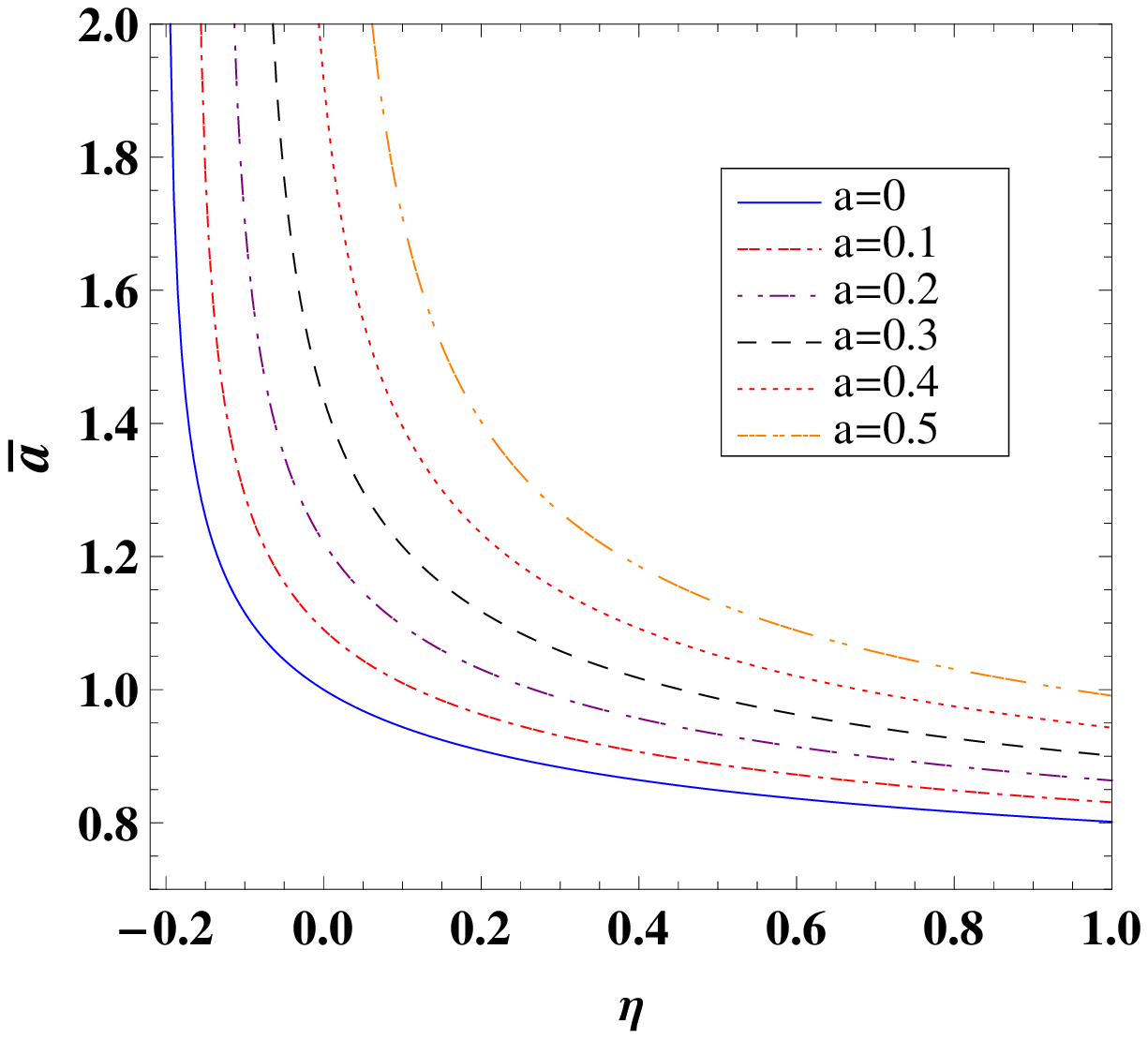}
\;\;\;\;\includegraphics[width=5cm]{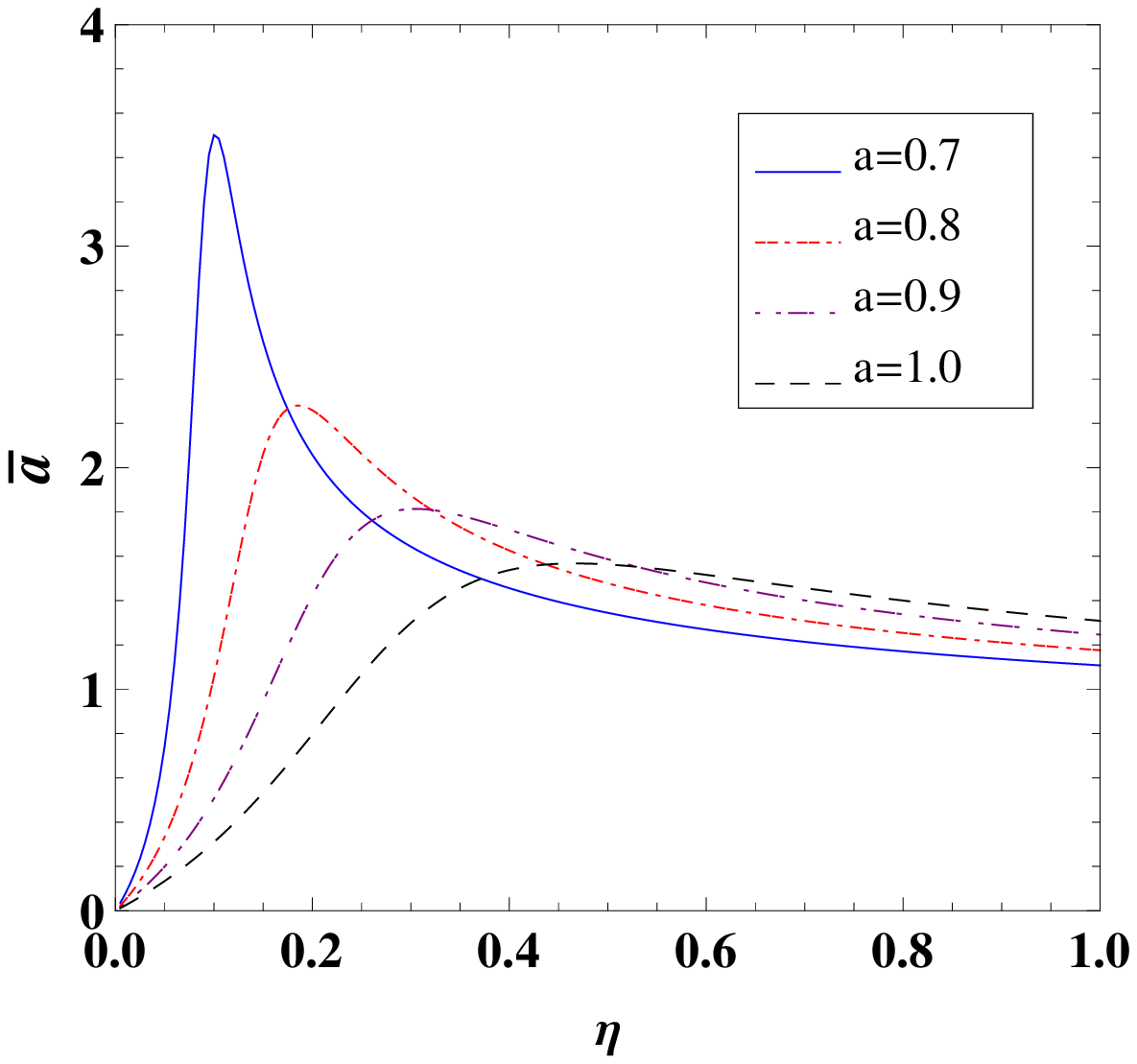}\\
\includegraphics[width=5cm]{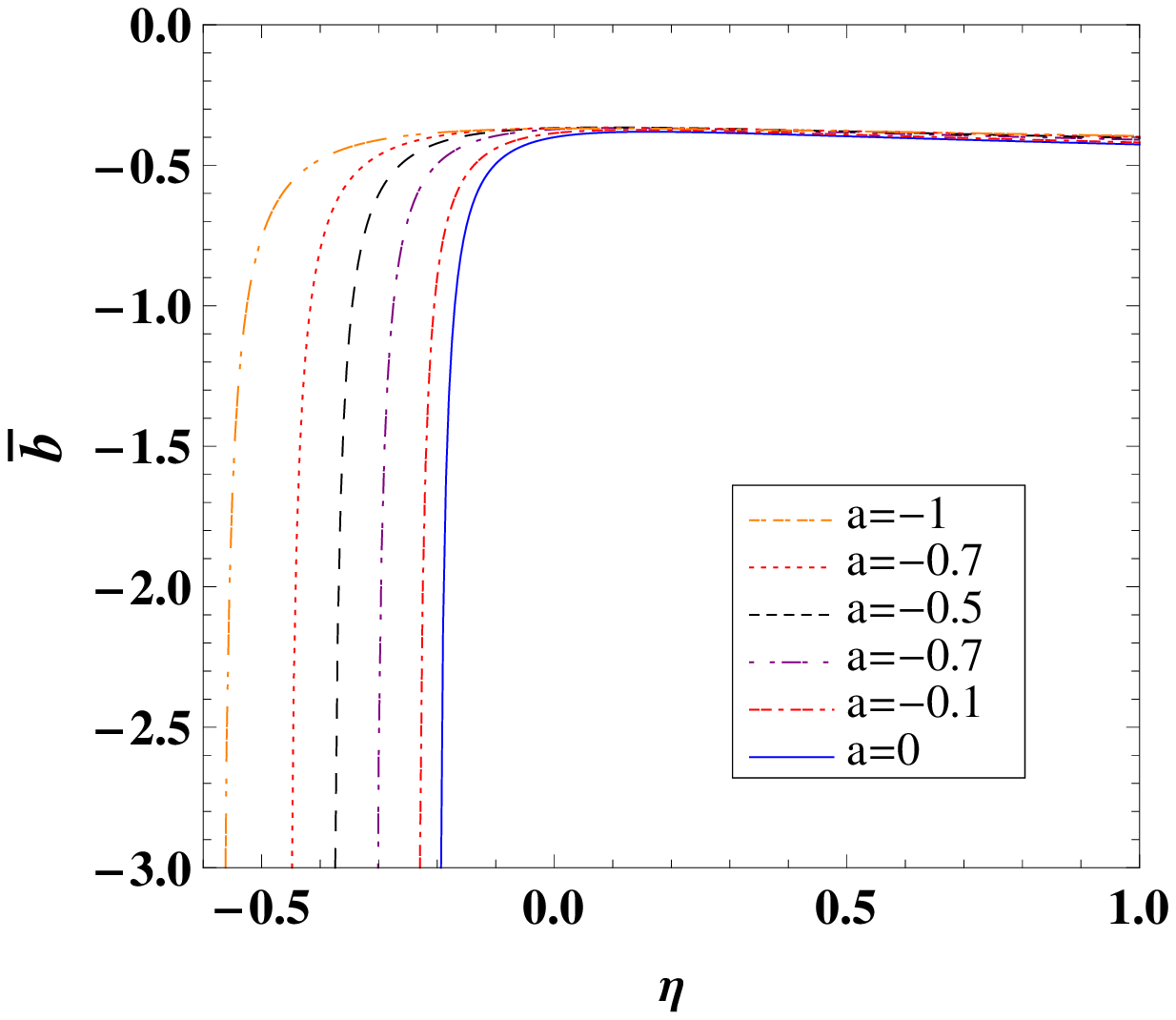}\;\;\;\;\includegraphics[width=5cm]{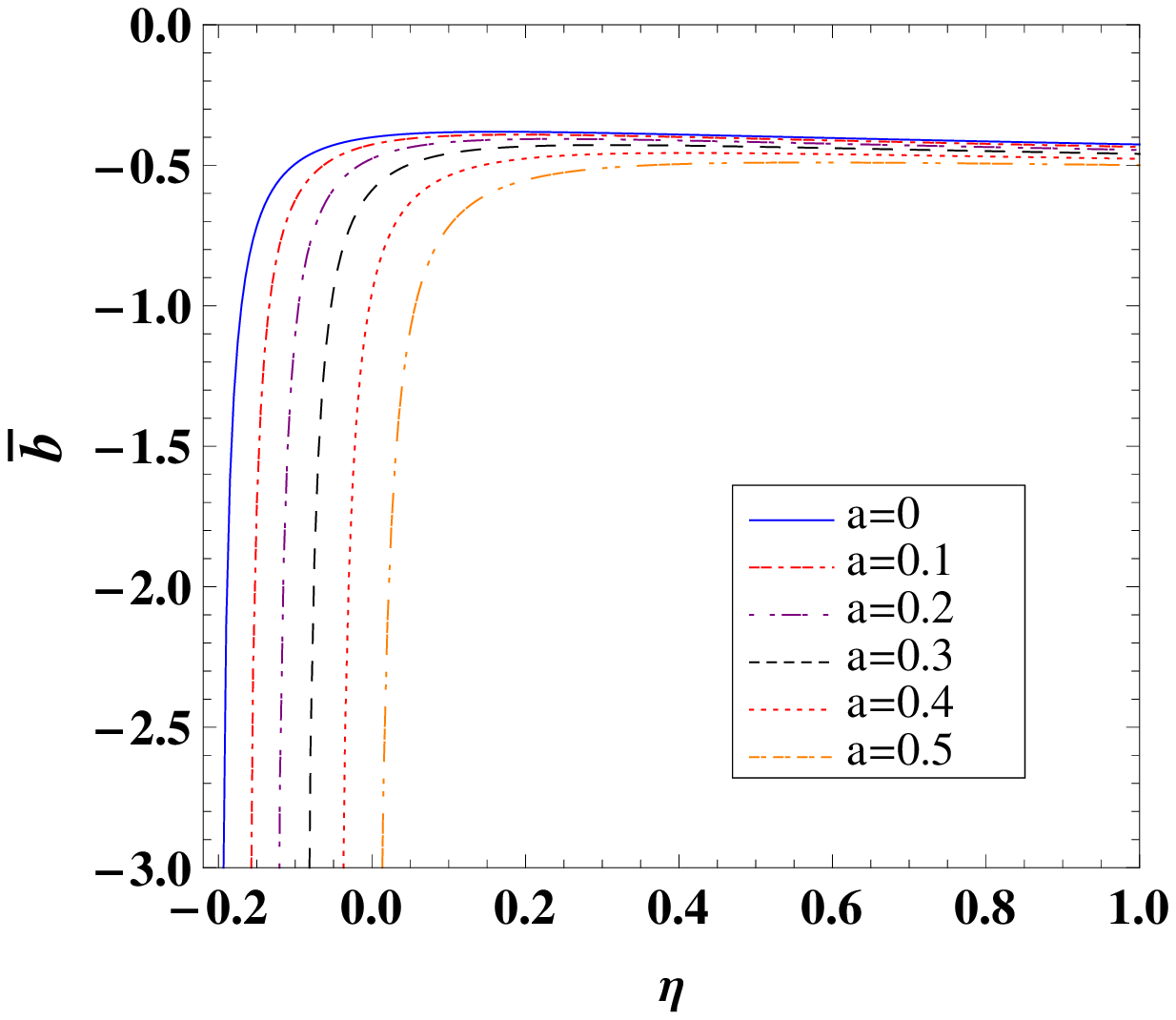}
\;\;\;\;\includegraphics[width=5cm]{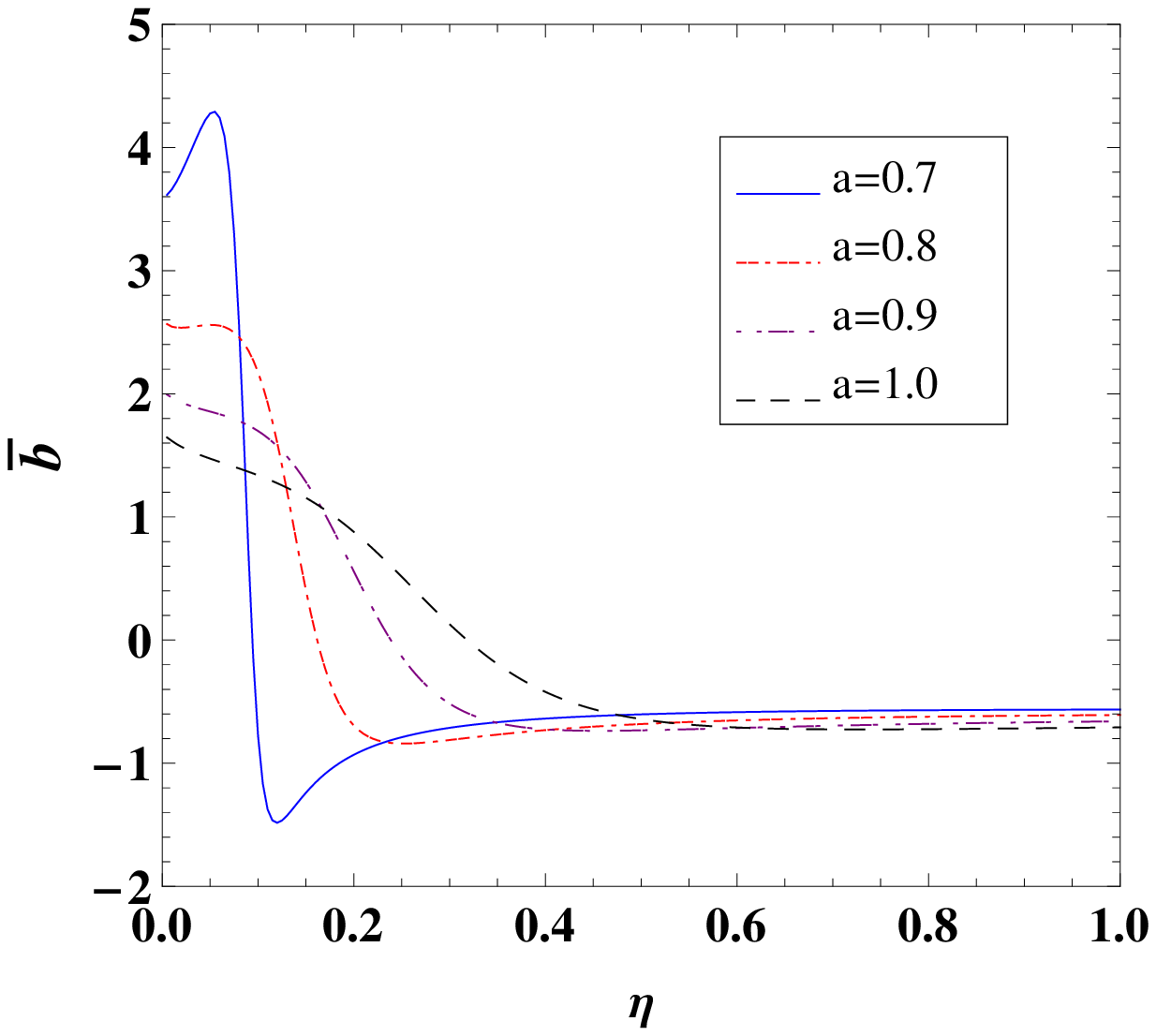}
\caption{Change of the strong deflection limit coefficients $\bar{a}$ and $\bar{b}$ with the deformation parameter $\eta$ for different $a$ in the Konoplya-Zhidenko rotating non-Kerr spacetime. Here, we set $2M=1$.}
\end{center}
\end{figure}
The function $f_0(z,x_{0})$ in Eq.(\ref{R10}) is obtained by expanding the argument of the square root in $f(z,x_{0})$ to the second order in $z$
\begin{eqnarray}
f_0(z,x_0)=\frac{1}{\sqrt{p(x_0)z+q(x_0)z^2}},\label{f0z}
\end{eqnarray}
where
\begin{eqnarray}
p(x_0)&=&\frac{x_0}{C(x_0)}\bigg\{A(x_0)C'(x_0)-A'(x_0)C(x_0)+2J[A'(x_0)D(x_0)-A(x_0)D'(x_0)]\bigg\},  \nonumber\\
q(x_0)&=&\frac{x_0}{2C^2(x_0)}\bigg\{2\bigg(C(x_0)-x_0C'(x_0)\bigg)\bigg([A(x_0)C'(x_0)-A'(x_0)C(x_0)]
+2J[A'(x_0)D(x_0)-A(x_0)D'(x_0)]\bigg)\nonumber\\&&+x_0C(x_0)
\bigg([A(x_0)C''(x_0)-A''(x_0)C(x_0)]+2J[A''(x_0)D(x_0)-A(x_0)D''(x_0)]\bigg)\bigg\}.\label{al0}
\end{eqnarray}
\begin{figure}[ht]
\begin{center}
\includegraphics[width=5cm]{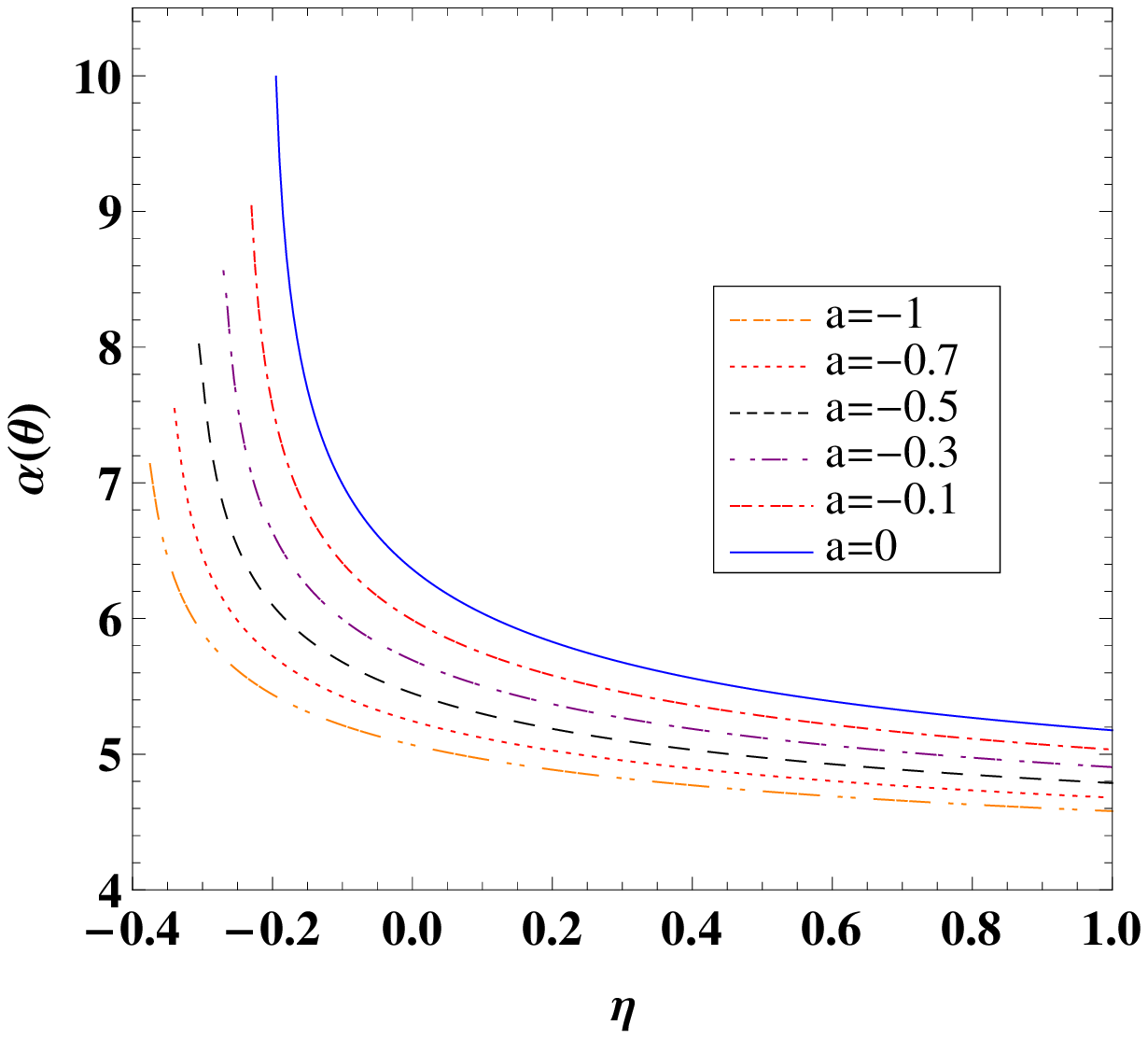}
\;\;\;\;\includegraphics[width=5cm]{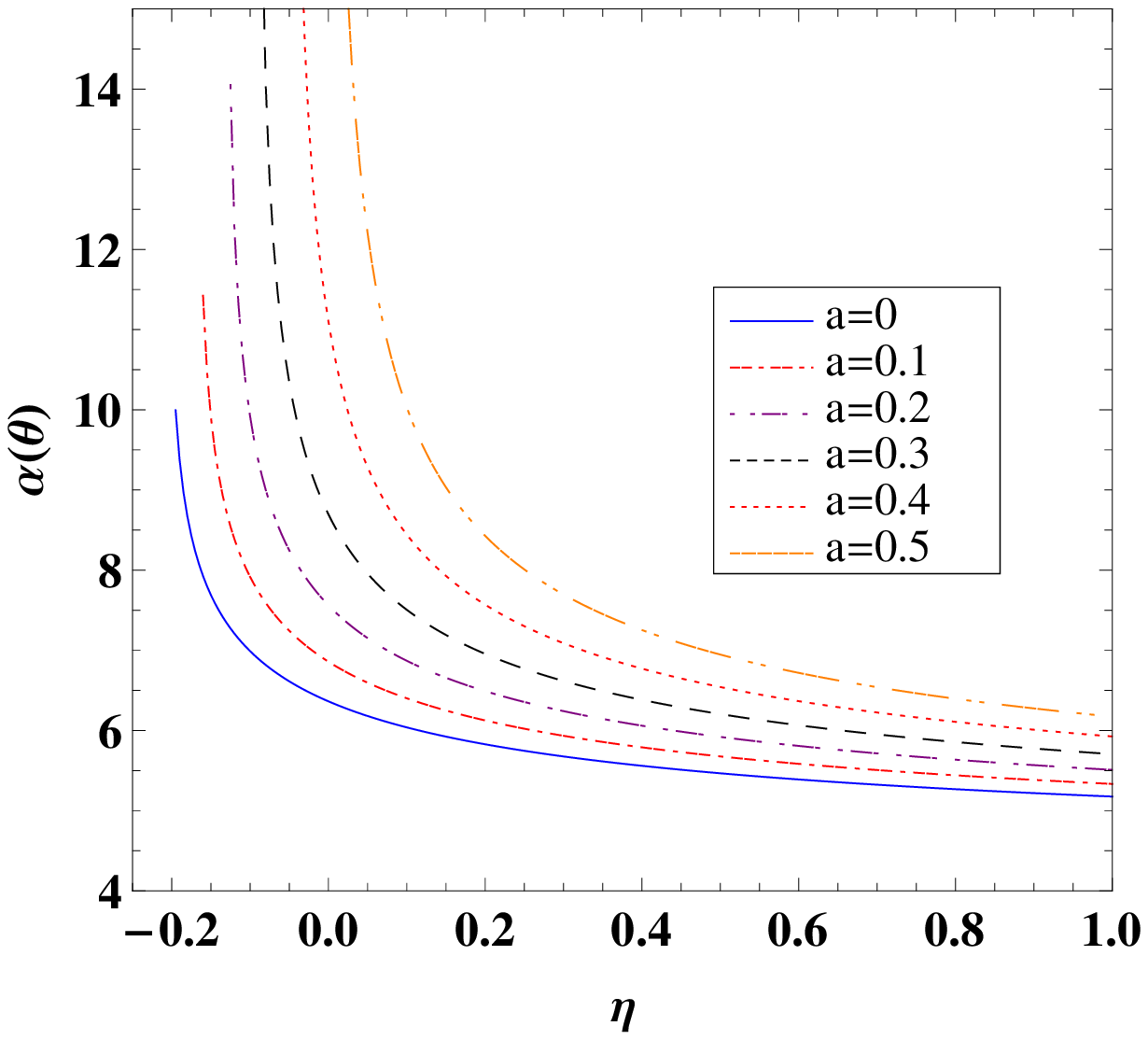}
\;\;\;\;\includegraphics[width=5cm]{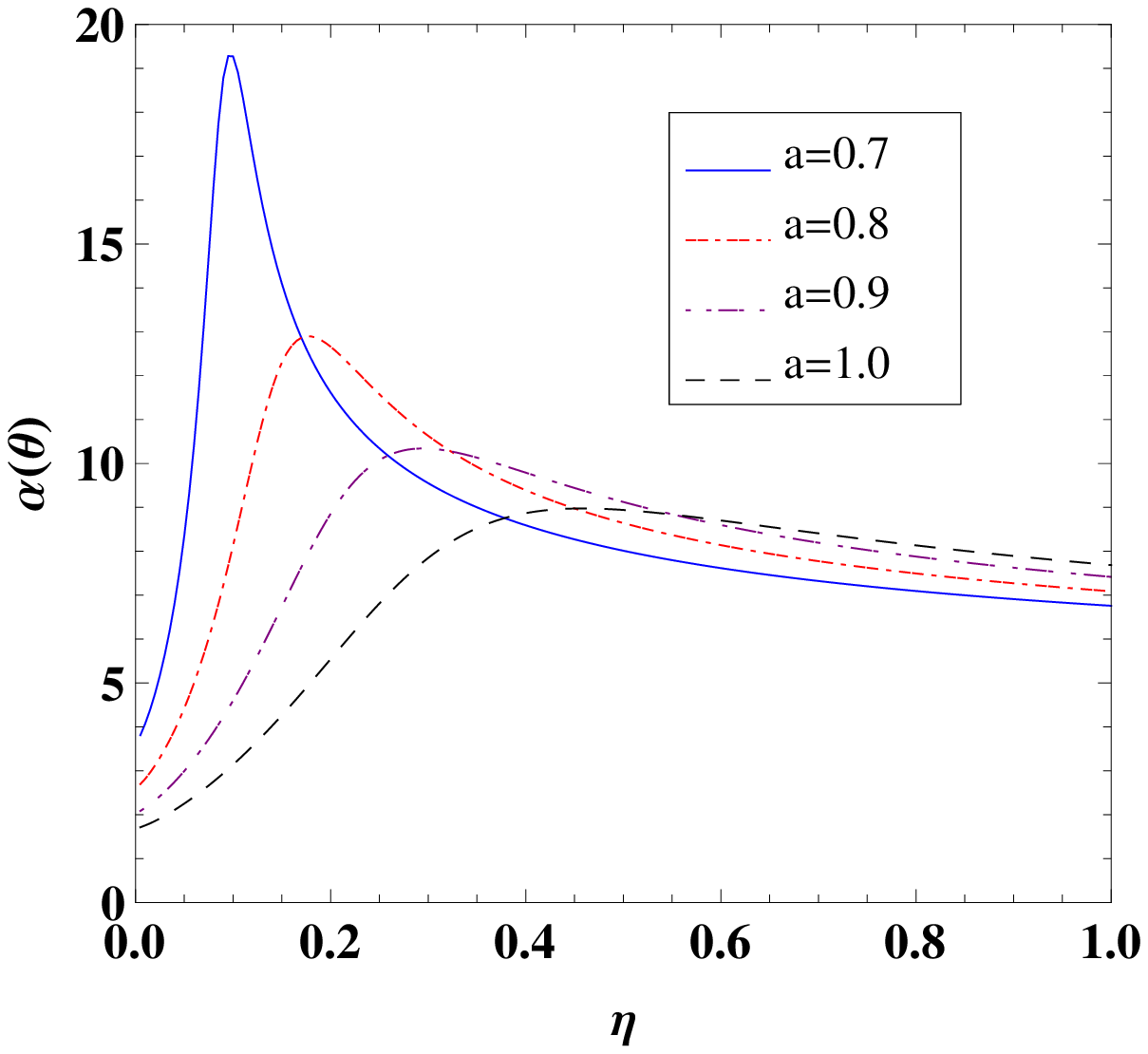}
\caption{Deflection angles evaluated at $u=u_{ps}+0.003$ is a function of the deformed parameter $\eta$ for different $a$ in the Konoplya-Zhidenko rotating non-Kerr spacetime. Here, we set $2M=1$.}
\end{center}
\end{figure}
If $x_0$ tends to the radius of the marginally circular photon
orbit $x_{ps}$, one can find that $p(x_{0})$ in Eq.(\ref{f0z}) approaches zero and then the integral (\ref{in0}) diverges logarithmically since the leading term in $f_0(z,x_{0})$ has the form of $z^{-1}$ in this limit.
Therefore,  when the photon is very close to the marginally circular photon
orbit, the deflection angle can be approximated very well as \cite{VB1}
\begin{eqnarray}
\alpha(\theta)=-\bar{a}\log{\bigg(\frac{\theta
D_{OL}}{u_{ps}}-1\bigg)}+\bar{b}+\mathcal{O}(u-u_{ps}), \label{alf1}
\end{eqnarray}
with
\begin{eqnarray}
&\bar{a}&=\frac{R(0,x_{ps})}{\sqrt{q(x_{ps})}}, \nonumber\\
&\bar{b}&= -\pi+b_R+\bar{a}\log{\bigg\{\frac{2q(x_{ps})C(x_{ps})}{u_{ps}A(x_{ps})[D(x_{ps})+JA(x_{ps})]}\bigg\}}, \nonumber\\
&b_R&=I_R(x_{ps}), \nonumber\\
&u_{ps}&=\frac{-D(x_{ps})+\sqrt{A(x_{ps})C(x_{ps})+D^2(x_{ps})}}{A(x_{ps})}.\label{coa1}
\end{eqnarray}
Here $D_{OL}$ denotes the distance between observer and
gravitational lens. Making use of Eqs.(\ref{alf1}) and (\ref{coa1}),
we can study the properties of strong gravitational lensing in the
Konoplya-Zhidenko rotating non-Kerr spacetime (\ref{metric1}). The changes of the coefficients ($\bar{a}$ and $\bar{b}$ ) with the deformed parameter $\eta$ for different rotation parameter $a$ is shown in Fig.(9). When $a<0.5$, we find that the coefficient $\bar{a}$ decreases with the deformation parameter $\eta$ and increases with the rotation parameter $a$. The coefficient $\bar{b}$ first increases and then decreases with the deformation parameter $\eta$, but it decreases monotonously with the rotation parameter $a$. When $a>0.5$, one can find that the coefficient $\bar{a}$ increases with $\eta$ and decreases with $a$ as $\eta$ is near the limit value $\eta_{min}$, but as $\eta$ is far from $\eta_{min}$, the change of $\bar{a}$ with $\eta$ and $a$ is similar to that in the cases with $a<0.5$.
The change of $\bar{b}$ with $\eta$ and $a$ becomes more complicated in the cases with $a>0.5$. From Fig.(9), we find that  the coefficient $\bar{b}$ first increases and then decreases, and finally increases with increase of $\eta$ for the case with $a=0.7$ or $a=0.8$, but it decreases monotonously with $\eta$ for the case with $a=0.9$ or $a=1$. The changes of $\bar{b}$ with the rotation parameter $a$ depends on the value of $\eta$. With increase of $\eta$, the change of $\bar{b}$ with $a$ undergoes a
process from decreasing to increasing and then to decreasing. Furthermore, as the deformed parameter $\eta$ tends to the lower limit which still hold up the marginally circular photon orbit, we also noted that  both of the coefficients $\bar{a}$ and $\bar{b}$ diverge as $a<0.5$, but as $a>0.5$ $\bar{a}$ tends to zero and $\bar{b}$ is close to a certain finite value, which implies that the deflection angle diverged logarithmically
in the strong deflection limit (\ref{alf1}) is not valid again in the case without the marginally circular photon orbit. In Fig. (10), we plotted the change of the deflection angles evaluated at $u=u_{ps}+0.003$ with the deformation parameter $\eta$ for different $a$ as in the case with the marginally circular photon orbit, which indicates that the change of the deflection angle with $\eta$ is similar to that of the coefficient $\bar{a}$. This means that in this case the deflection angles of the light rays are dominated by the logarithmic term in the strong field limit.

\section{observable in strong gravitational lensing and time delay between relativistic images }

In this section, we will estimate the numerical values for the observables of gravitational lensing and the time delay between relativistic images in the strong field limit by assuming that the spacetime of the supermassive black
hole at the Galactic center of Milky Way can be described by Konoplya-Zhidenko rotating non-Kerr metric (\ref{metric1}).

As source and observer are far enough from the lens, the lens equation can be approximated as  \cite{VB2}
\begin{eqnarray}
\gamma=\frac{D_{OL}+D_{LS}}{D_{LS}}\theta-\alpha(\theta) \; mod
\;2\pi
\end{eqnarray}
where $D_{LS}$ is the lens-source distance and $D_{OL}$ is the
observer-lens distance. $\gamma$ is the angle between the optical axis and the direction of the source. $\theta=u/D_{OL}$ is the angular
separation between the lens and the image. Following
ref.\cite{VB2}, we here focus only on the simplest situation where the source, lens and observer are highly aligned. Under this condition,  the angular separation between the lens and the $n$-th relativistic image can be simplified as
\begin{eqnarray}
\theta_n\simeq\theta^0_n\bigg(1-\frac{u_{ps}e_n(D_{OL}+D_{LS})
}{\bar{a}D_{OL}D_{LS}}\bigg),\label{nth0}
\end{eqnarray}
with
\begin{eqnarray}
\theta^0_n=\frac{u_{ps}}{D_{OL}}(1+e_n),\;\;\;\;\;\;
e_{n}=e^{\frac{\bar{b}+|\gamma|-2\pi
n}{\bar{a}}},\label{nth1}
\end{eqnarray}
where $\theta^0_n$ is the image positions corresponding to
$\alpha=2n\pi$, and $n$ is an integer. As $n\rightarrow\infty$, from Eqs.(\ref{nth0}) and (\ref{nth1}), it is easy to find that
$e_n$ tends to zero, which means that the asymptotic position of a
set of images $\theta_{\infty}$ is related to the
minimum impact parameter $u_{ps}$ by a simpler form
\begin{figure}
\begin{center}
\includegraphics[width=5cm]{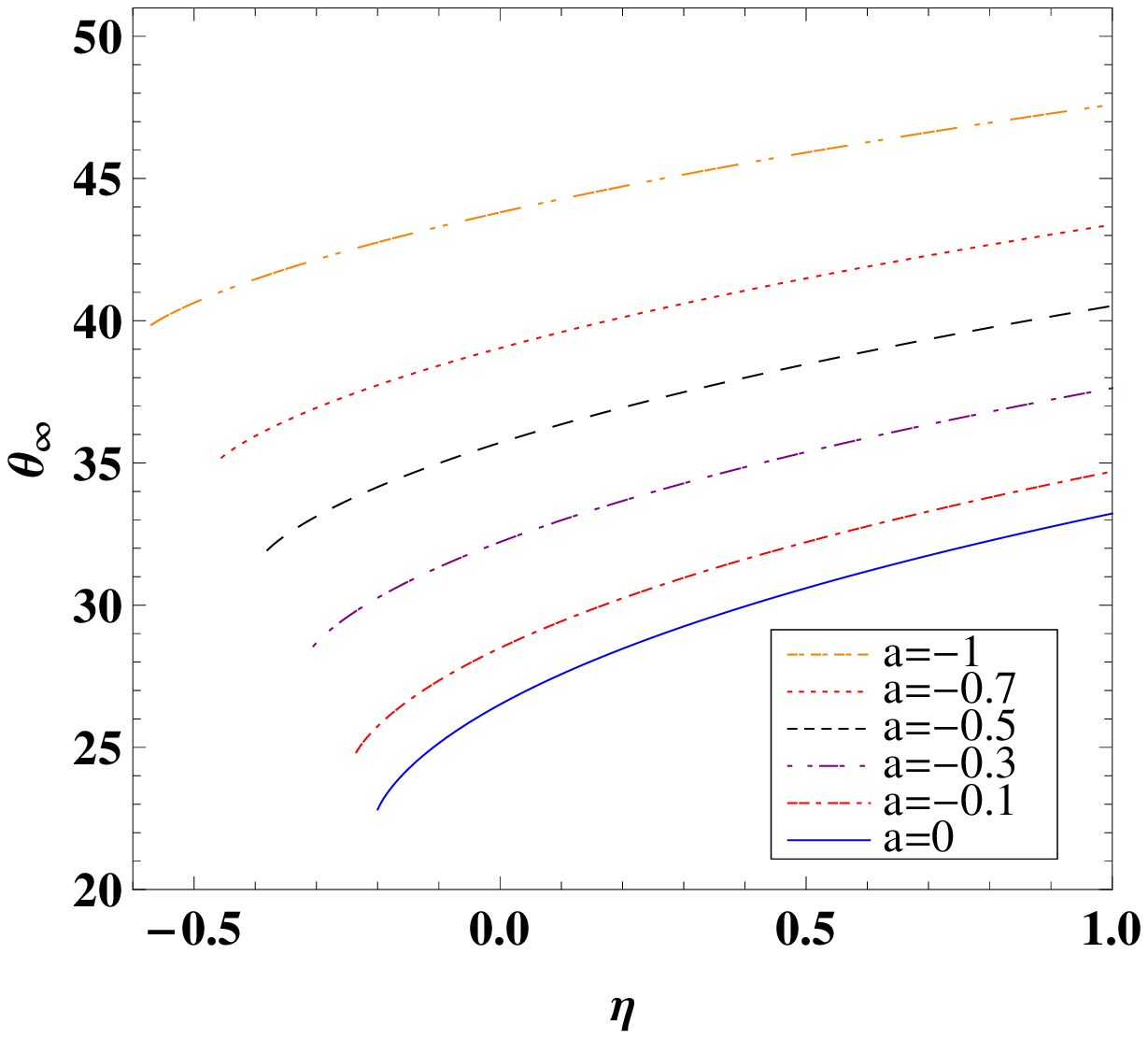}
\;\;\;\;\includegraphics[width=5cm]{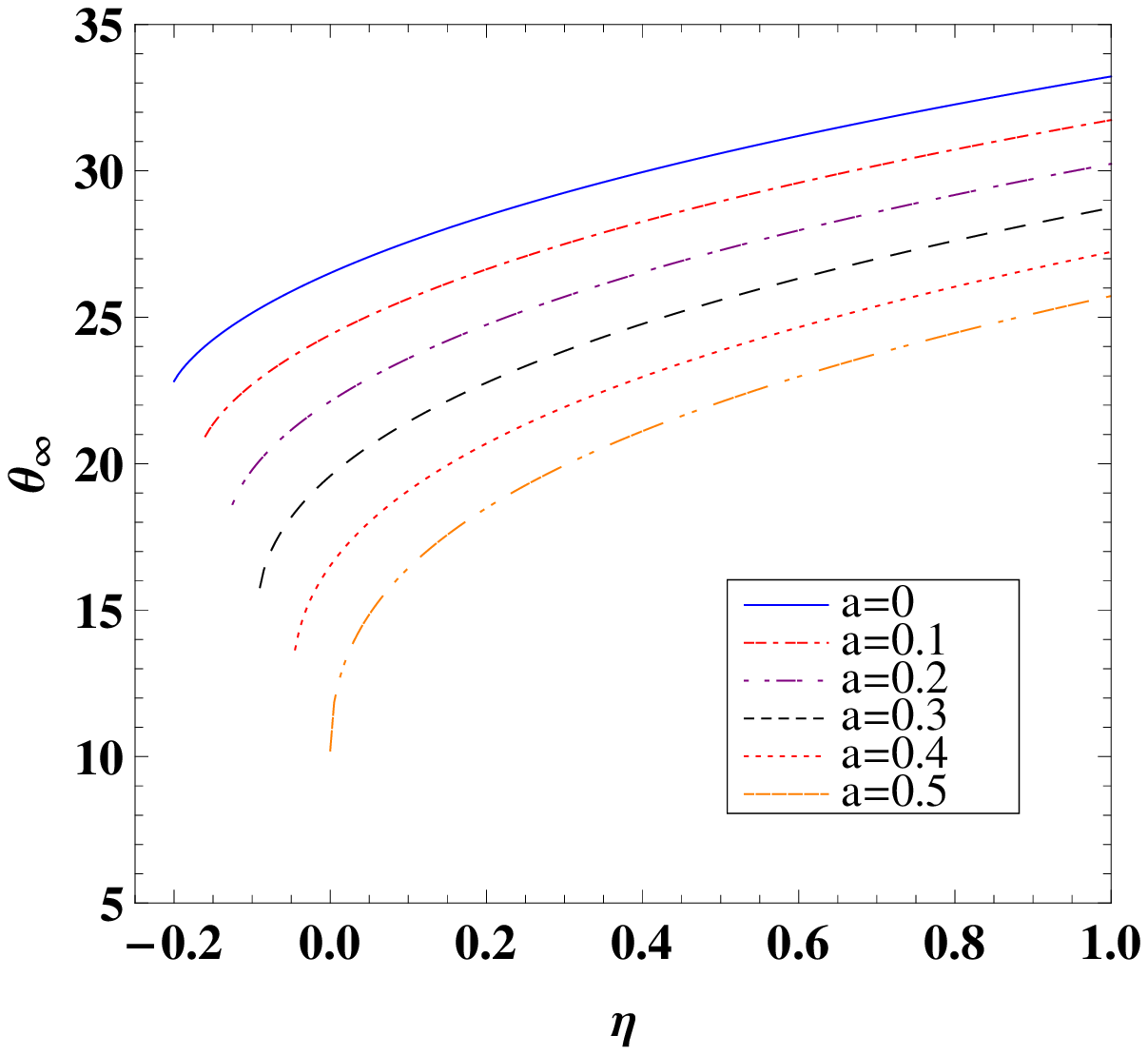}
\;\;\;\;\includegraphics[width=5cm]{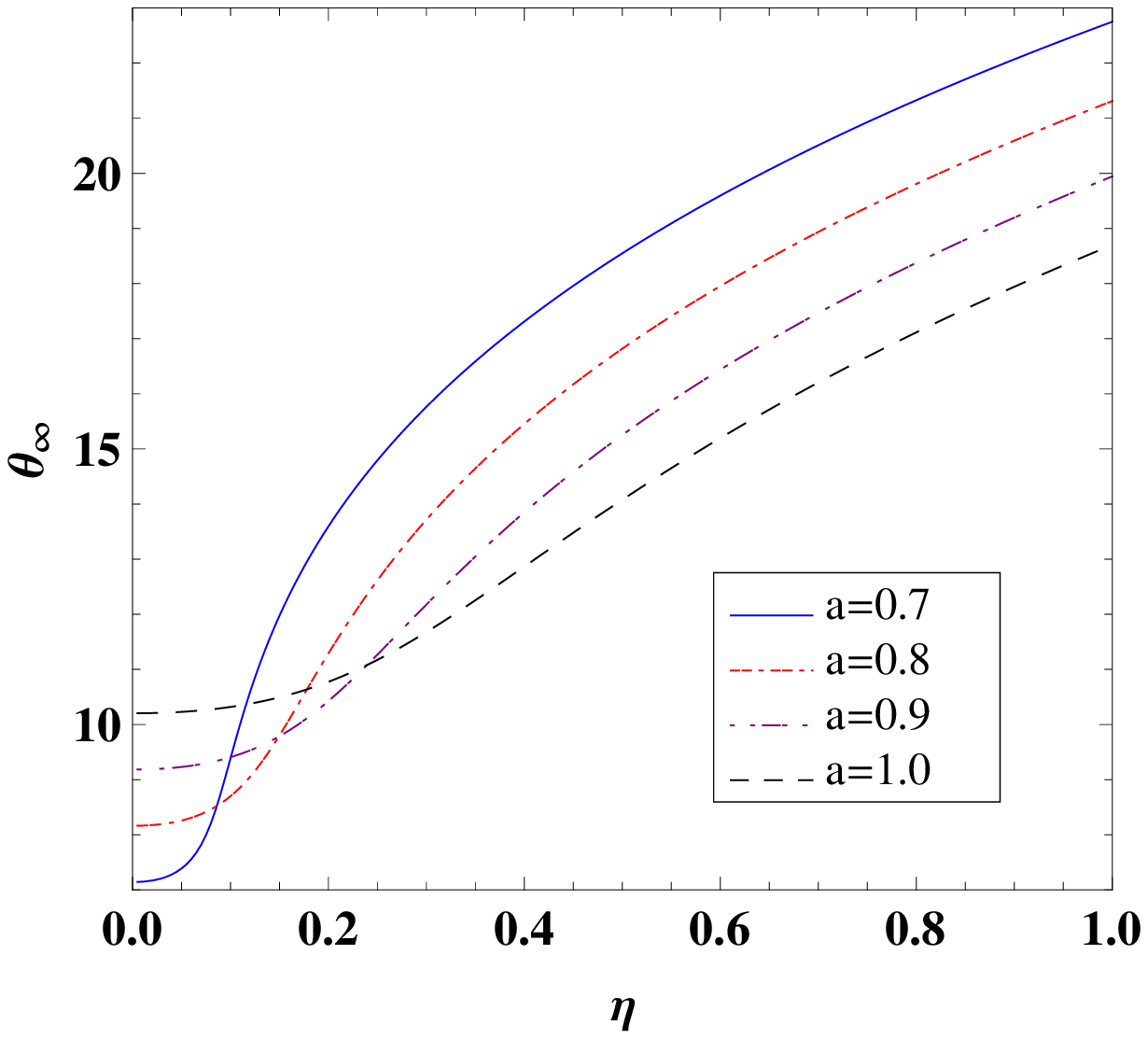}
\caption{Variety of the
innermost relativistic image $\theta_{\infty}$ with the deformed
parameter $\eta$ for different $a$. Here, we set $2M=1$.}
\end{center}
\end{figure}
\begin{figure}
\begin{center}
\includegraphics[width=5cm]{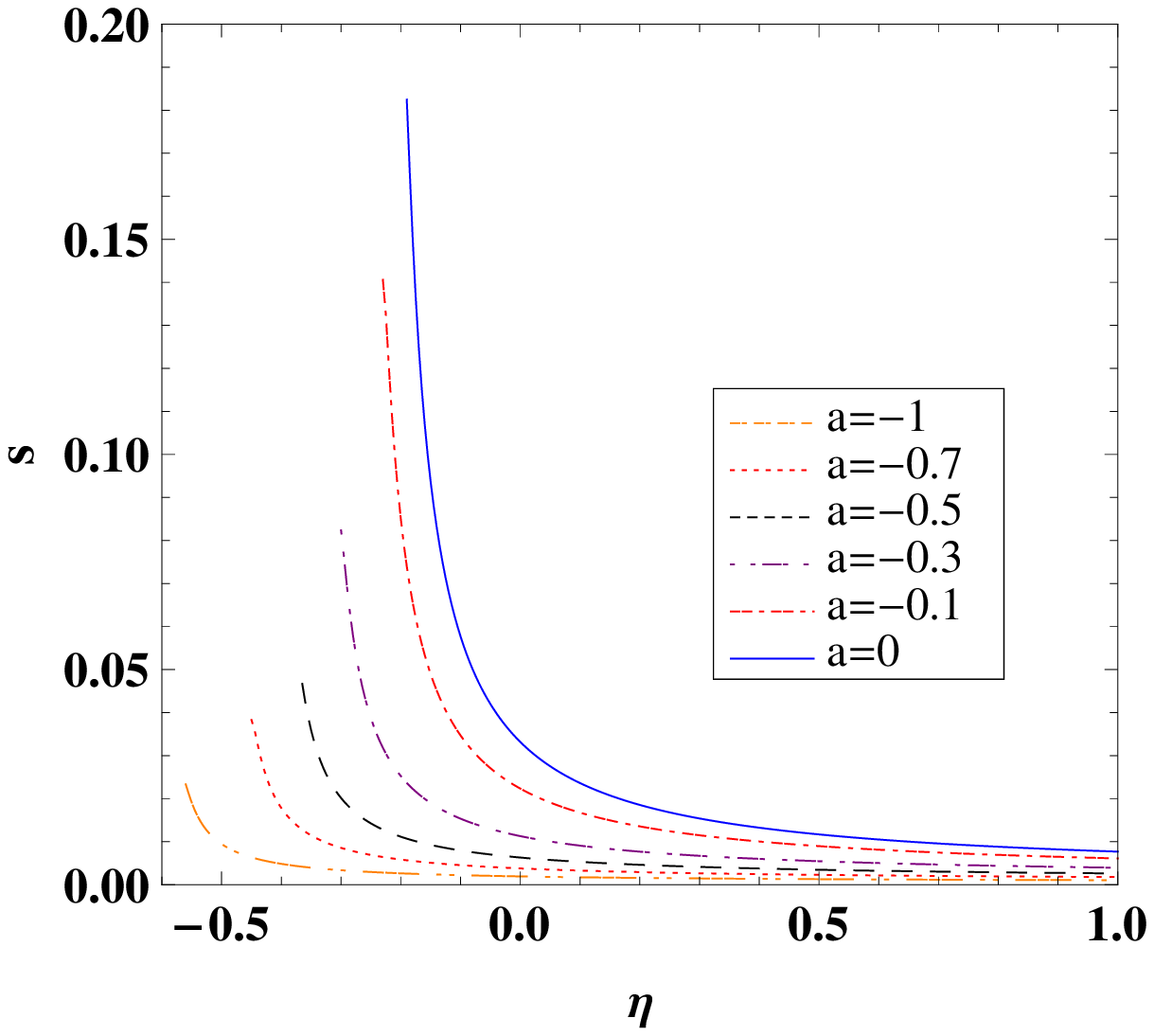}
\;\;\;\;\includegraphics[width=5cm]{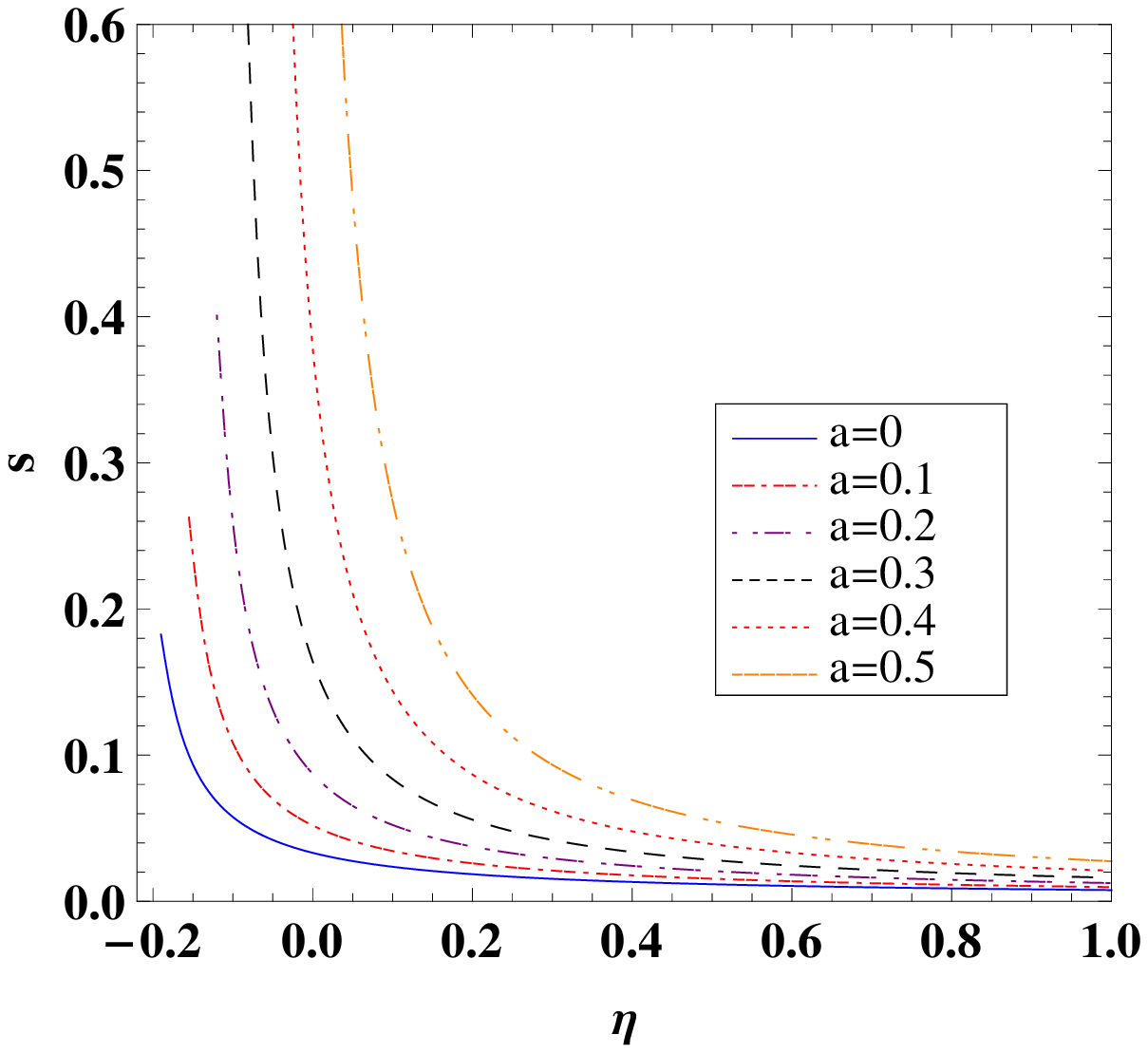}
\;\;\;\;\includegraphics[width=5cm]{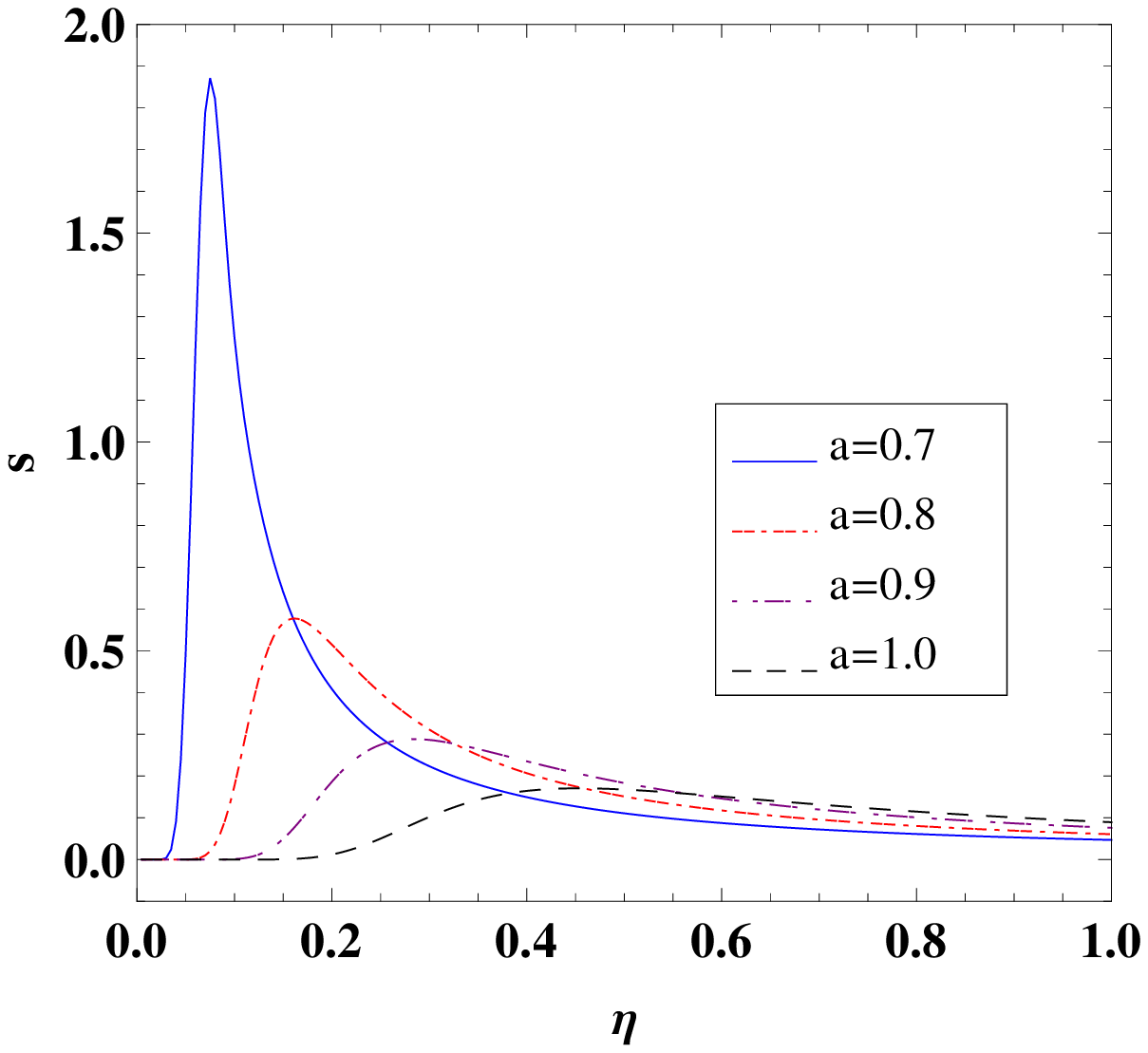}
\caption{Variety of the
angular separation $s$ with the deformed parameter $\eta$ for
different $a$. Here, we set $2M=1$.}
\end{center}
\end{figure}
\begin{figure}
\begin{center}
\includegraphics[width=5cm]{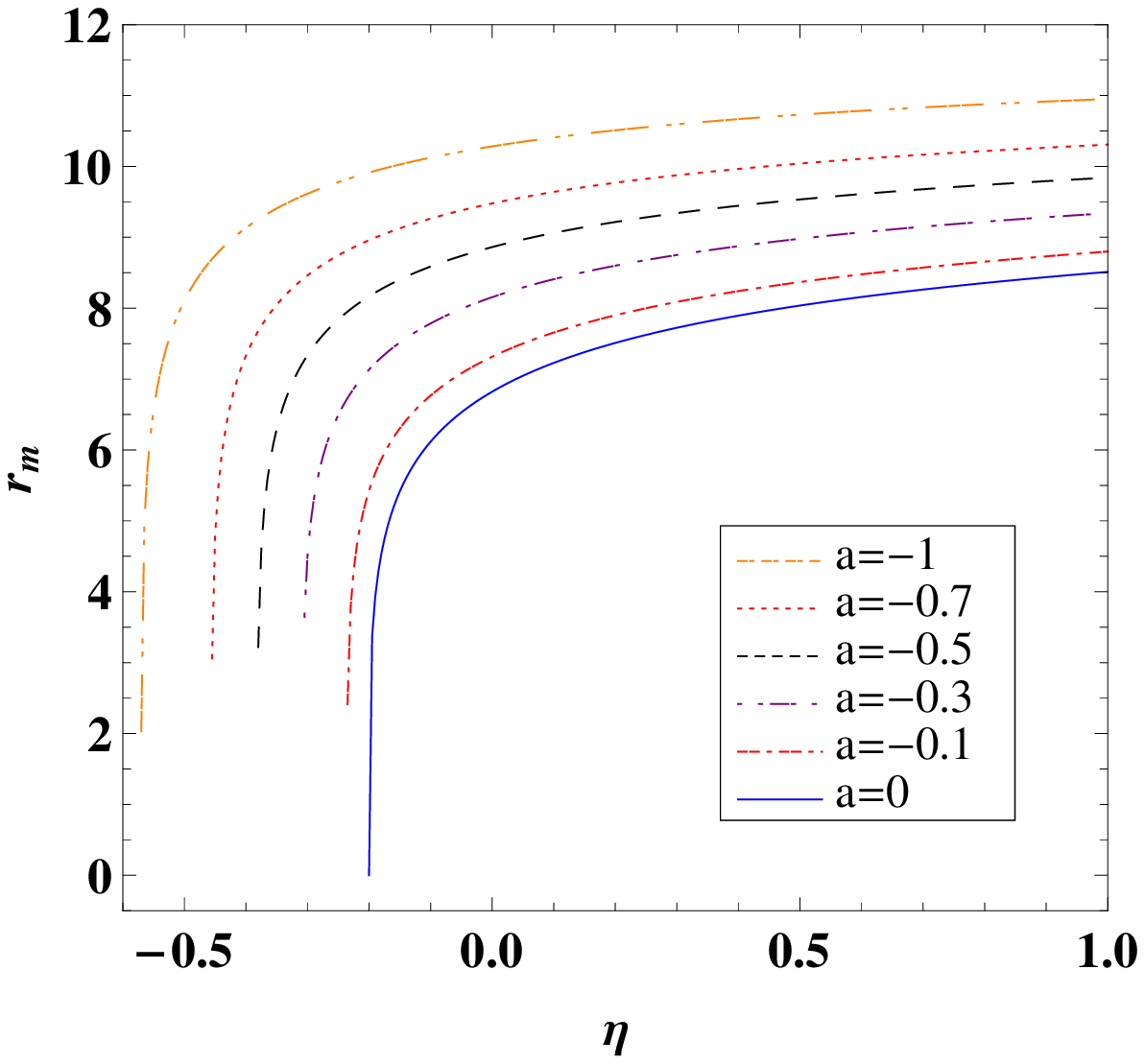}\;\;\;\;
\includegraphics[width=5cm]{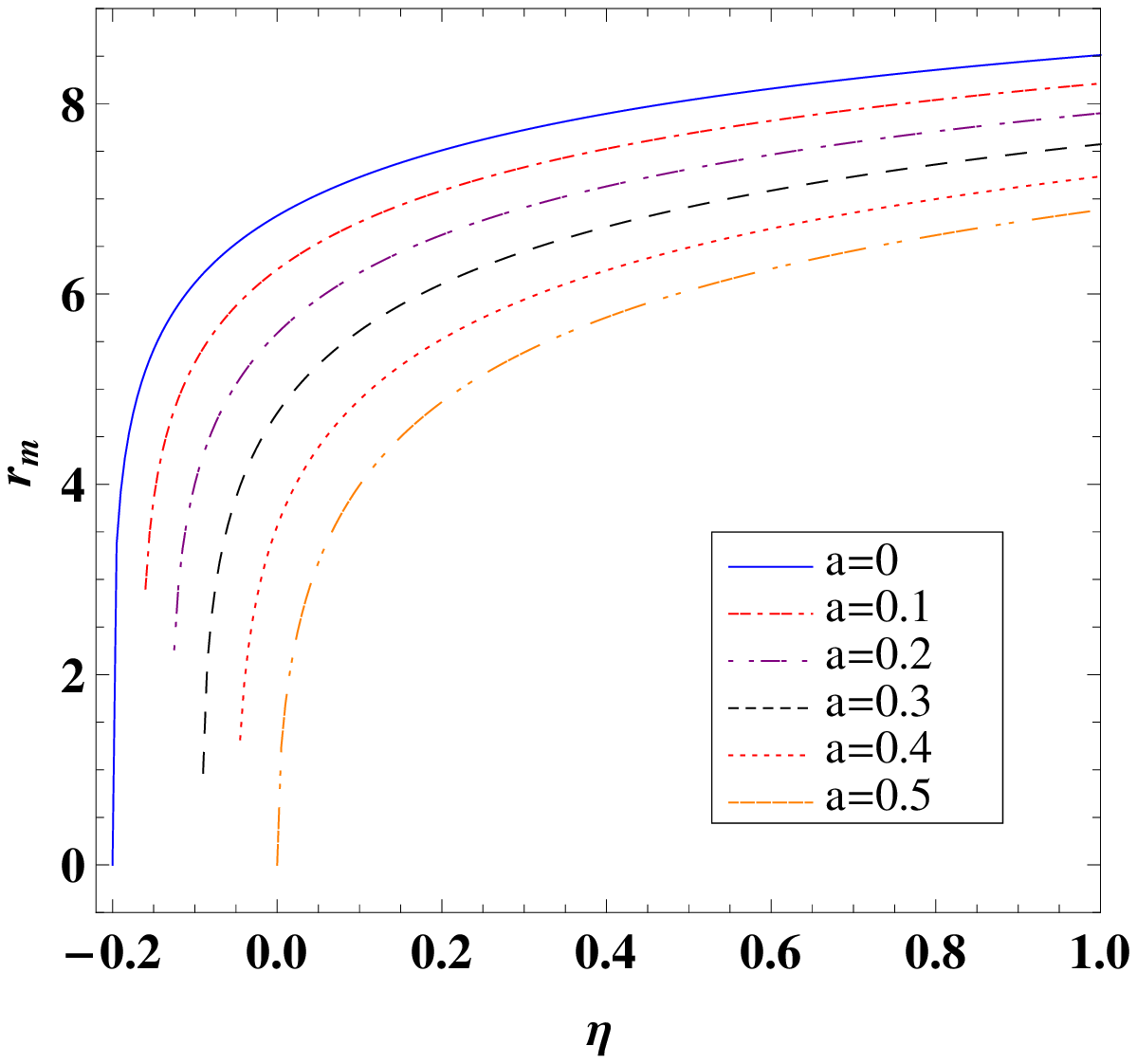}\;\;\;\;
\includegraphics[width=5cm]{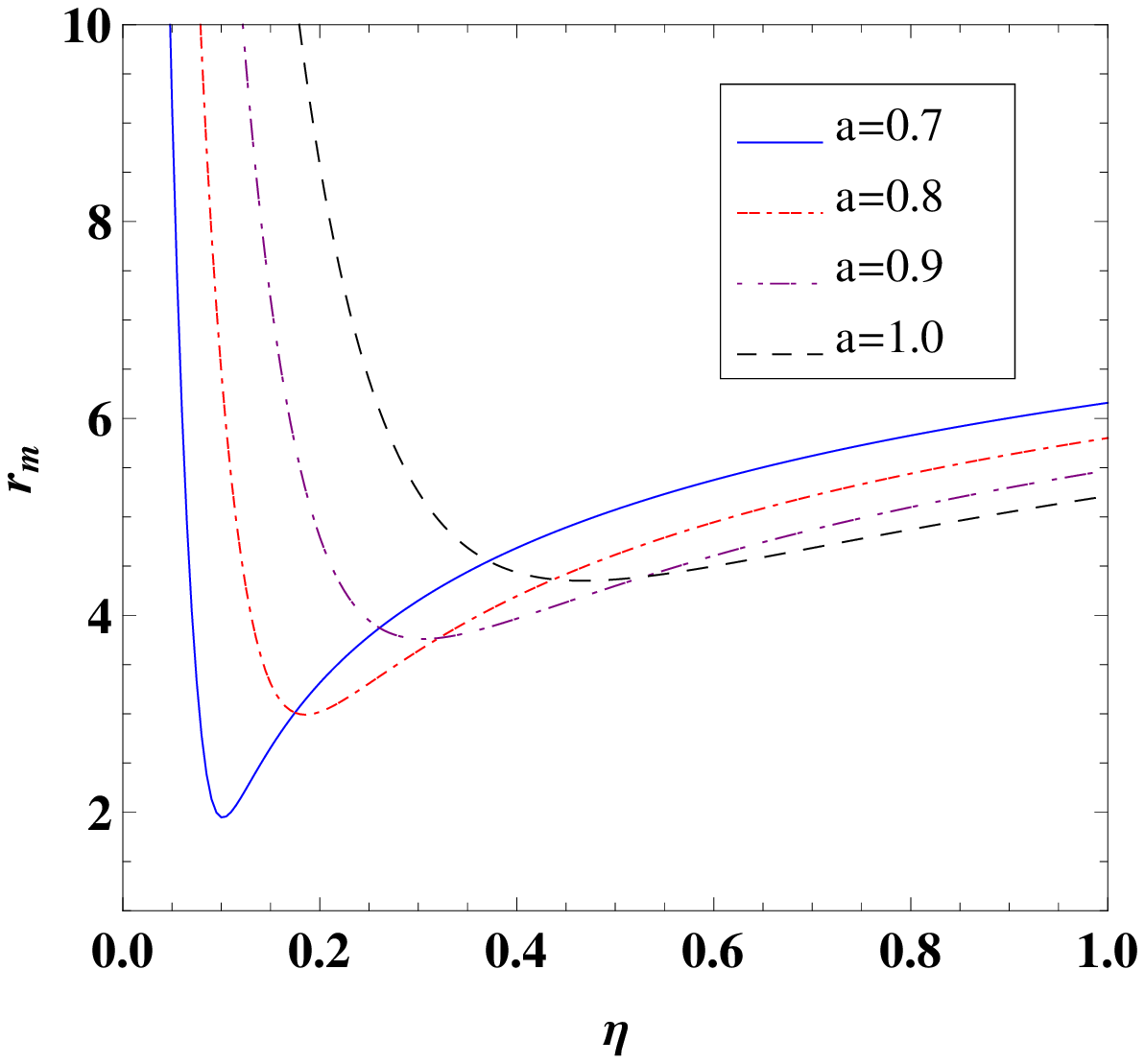}
\caption{Variety of the
relative magnitudes $r_m$ with the deformed parameter $\eta$ for
different $a$. Here, we set $2M=1$.}
\end{center}
\end{figure}
\begin{eqnarray}
u_{ps}=D_{OL}\theta_{\infty}.\label{uhs1}
\end{eqnarray}
In order to estimate the coefficients $\bar{a}$ and $\bar{b}$, one
has to separate at least the outermost image from all the others.
As in Refs.\cite{VB1,VB2}, we consider here a perfect situation where only the outermost image $\theta_1$ is separated as a single image and all the remaining ones are packed together at $\theta_{\infty}$. In this perfect situation, the angular separation $s$ and the relative
magnitudes $r_m$ between the first image and other ones
can be simplified further as
\cite{VB1,VB2,Gyulchev1}
\begin{eqnarray}
s&=&\theta_1-\theta_{\infty}=\theta_{\infty}e^{\frac{\bar{b}-2\pi}{\bar{a}}},\nonumber\\
r_m&=&2.5\log{\frac{\mu_1}{\sum^{\infty}_{n=2}\mu_n}}
=\frac{5\pi}{\bar{a}}\log{e},\label{ss1}
\end{eqnarray}
Through measuring $s$, $\theta_{\infty}$ and $r_m$ from observation experiments, one can obtain the coefficients $\bar{a}$, $\bar{b}$ in strong deflection limit and the minimum impact parameter $u_{ps}$. Comparing these values from observation with those predicted by theoretical models, we can extract information about the parameters of compact object.

For Milk Way Galaxy, the mass of the central object of is estimated recently
to be $4.4\times 10^6M_{\odot}$ \cite{Genzel1} and its distance is
around $8.5kpc$, which leads to the ratio of the mass to the distance
$M/D_{OL} \approx2.4734\times10^{-11}$.  With this data,  we present the numerical value for the angular position of the relativistic images $\theta_{\infty}$, the angular separation $s$ and the relative magnitudes $r_m$ are plotted in Figs.(11), (12) and (13). For the cases with $a<0.5$,
one can obtain that the angular position of the relativistic images $\theta_{\infty}$ increases with the parameters $\eta$ and decreases with the rotation parameters $a$. It is converse to that in the Johannsen-Psaltis rotating non-Kerr spacetime. The angular separation $s$ between $\theta_{1}$ and $\theta_{\infty}$ decreases with the parameters $\eta$ and increases with the rotation parameters $a$.  However, in the cases with $a>0.5$, we find that $\theta_{\infty}$ is an increasing function of $a$ as $\eta$ is near the limit value $\eta_{min}$. The curve of $s$ with the deformation parameter $\eta$ has a peak for different $a$. With increase of $a$, the value of peak drops down and its position moves along the right. The change of $r_{m}$ with the parameters $\eta$ and $a$ is converse to that of the angular separation $s$ in both cases.

Finally, we will consider the time delays between the relativistic images, which is another important kind of observables in strong gravitational lensing \cite{VBt,Tim2,Tim1,Yi1,Yi101}. From the null geodesics equation (\ref{cedi}), one has
\begin{eqnarray}
\frac{dt}{dx}&=&\frac{\sqrt{B(x)}[C(x)-JD(x)]}{\sqrt{C(x)-2JD(x)-J^2A(x)}\sqrt{D(x)^2+A(x)C(x)}}.
\end{eqnarray}
The time for the photon traveling from the source to the observer can be expressed as \cite{VBt}
\begin{eqnarray}
T(x_0)&=&2\int^{\infty}_{x_0}\bigg|\frac{dt}{dx}\bigg| dx-\int^{\infty}_{D_{OL}}\bigg|\frac{dt}{dx}\bigg| dx-\int^{\infty}_{D_{LS}}\bigg|\frac{dt}{dx}\bigg| dx.\label{T0}
\end{eqnarray}
Considering that observer and source are very far from
the black hole, the last two terms in Eq.(\ref{T0}) can be neglected and then the time delay between two photons travelling on different trajectories can be simplified as
\begin{eqnarray}
T_1-T_2&=&2\int^{\infty}_{x_0,1}\bigg|\frac{dt}{dx}(x,x_{0,1})\bigg| dx-2\int^{\infty}_{x_{0,2}}\bigg|\frac{dt}{dx}(x,x_{0,2})\bigg| dx.\label{T10}
\end{eqnarray}
Supposing $x_{0,1}<x_{0,2}$, one can find
\begin{eqnarray}
T_1-T_2&=&\tilde{T}(x_{0,1})-\tilde{T}(x_{0,2})+2\int^{x_{0,2}}_{x_{0,1}}
\frac{\tilde{P}(x,x_{0,1})}{\sqrt{A_{0,1}}} dx+2\int^{\infty}_{x_{0,2}}\bigg[\frac{\tilde{P}(x,x_{0,1})}{\sqrt{A_{0,1}}}-
\frac{\tilde{P}(x,x_{0,2})}{\sqrt{A_{0,2}}}\bigg]dx,\label{T20}
\end{eqnarray}
with
\begin{eqnarray}
\tilde{P}(x,x_{0})&=&\frac{\sqrt{B(x)A(x_0)}[C(x)-JD(x)]}{\sqrt{C(x)}
\sqrt{D(x)^2+A(x)C(x)}},\nonumber\\
\tilde{T}(x_{0,1})&=&\int^{1}_{0}\tilde{R}(z,x_0)f(z,x_0)dz,\nonumber\\
\tilde{R}(z,x_0)&=&\frac{2x^2}{x_0\sqrt{C(z)}}\frac{\sqrt{B(z)|A(x_0)|}
[C(z)-JD(z)]}{\sqrt{D^2(z)+A(z)C(z)}}\bigg(1-\frac{1}{\sqrt{A(x_0)}f(z,x_0)}\bigg).
\end{eqnarray}
After some similar operations in the calculation of the deflection, one can obtain \cite{VBt}
\begin{eqnarray}
\tilde{T}(u)=-\tilde{a}\log{\bigg(\frac{u}{u_{ps}}-1\bigg)}+\tilde{b}+\mathcal{O}(u-u_{ps}), \label{tf1}
\end{eqnarray}
with
\begin{eqnarray}
&\tilde{a}&=\frac{\tilde{R}(0,x_{ps})}{\sqrt{q(x_{ps})}}, \nonumber\\
&\tilde{b}&= -\pi+\tilde{b}_R+\tilde{a}\log{\bigg\{\frac{2q(x_{ps})C(x_{ps})}
{u_{ps}A(x_{ps})[D(x_{ps})+JA(x_{ps})]}\bigg\}}, \nonumber\\
&b_R&=\int^{1}_{0}[\tilde{R}(z,x_{ps})f(z,x_{ps})-\tilde{R}(0,x_{ps})f_0(z,x_{ps})]dz,\label{cot1}
\end{eqnarray}
which diverges logarithmically in the strong-field limit for the cases with the marginally circular photon orbit.
Assuming the source, the lens and the observer are aligned almost in a line, the time delay between a $n$-loop and a $m$-loop relativistic image can be approximated as
\begin{eqnarray}
\Delta T_{n,m}=\Delta T^0_{n,m}+\Delta T^1_{n,m}, \label{tfs1}
\end{eqnarray}
where
\begin{eqnarray}
\Delta T^0_{n,m}&=&2\pi(n-m)\frac{\tilde{a}}{\bar{a}},\nonumber\\
\Delta T^1_{n,m}&=&2\sqrt{\frac{B(x_{ps})}{A(x_{ps})}}
\sqrt{\frac{x^2_{ps}u_{ps}A(x_{ps})[D(x_{ps})+JA(x_{ps})]}{2q(x_{ps})C(x_{ps})}}
e^{\frac{\bar{b}}{2\bar{a}}}
\bigg[e^{-\frac{m\pi}{\bar{a}}}-e^{-\frac{n\pi}{\bar{a}}}\bigg]. \label{tfs2}
\end{eqnarray}
\begin{figure}
\begin{center}
\includegraphics[width=5cm]{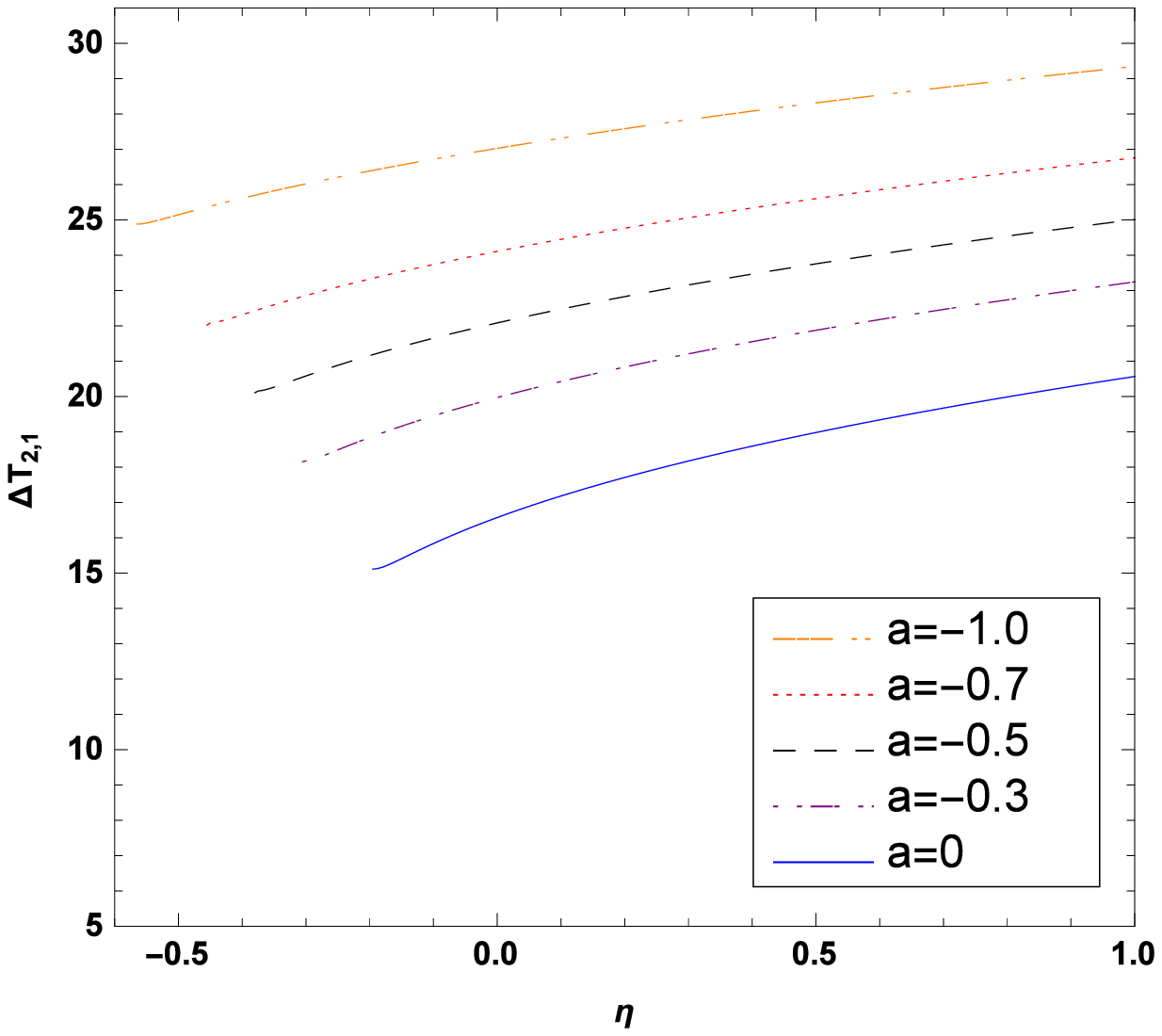}\;\;\;\;
\includegraphics[width=5cm]{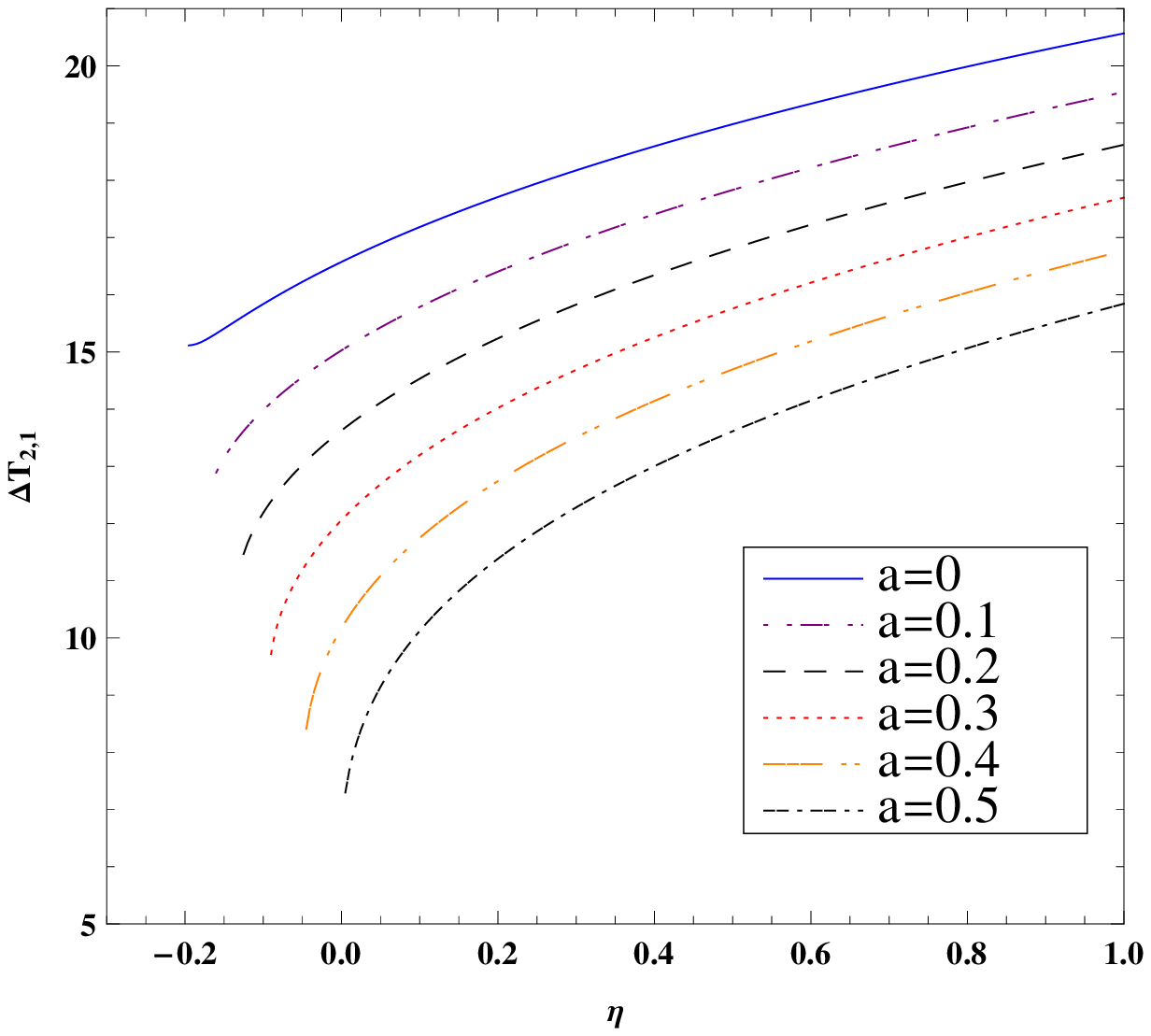}\;\;\;\;
\includegraphics[width=5cm]{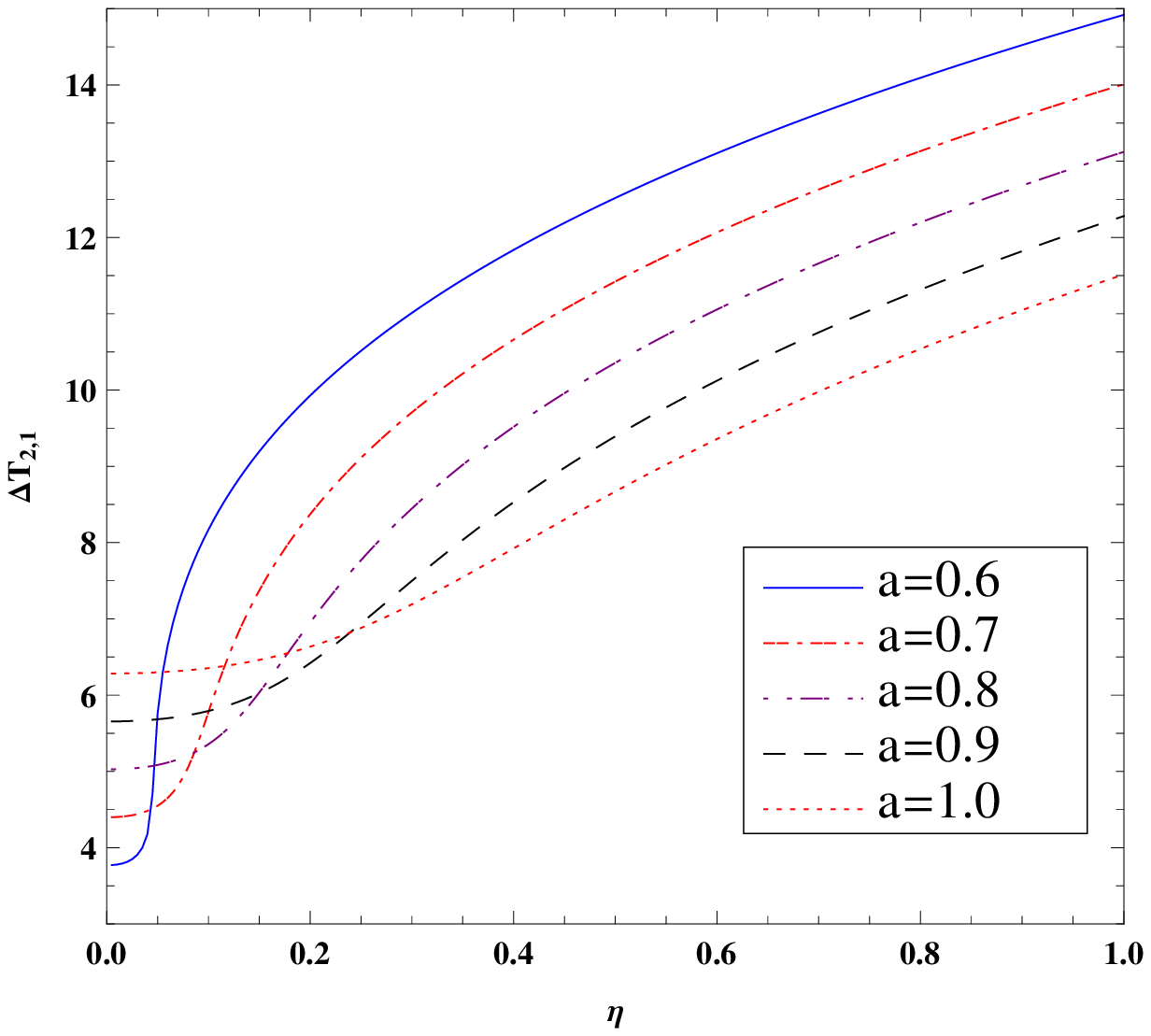}
\caption{Variety of the time delays between the relativistic images with the deformed parameter $\eta$ for different $a$ in a Konoplya-Zhidenko rotating non-Kerr spacetime. Here, we set $2M=1$, $n=2$ and $m=1$.}
\end{center}
\end{figure}
In Fig.(14), we present the time delay between the first relativistic image and the second one in a Konoplya-Zhidenko rotating non-Kerr spacetime. It is shown that the time delay $\Delta T_{2,1}$ decreases with the rotation parameter $a$ except in the case with $a>0.5$ and the value $\eta$ near the limit $\eta_{min}$ in which $\Delta T_{2,1}$ is an increasing function of $a$.
 With increase of the deformation parameter $\eta$, it first decreases and then increases for the non-positive $a$ and increases monotonously for the positive $a$.
\begin{figure}
\begin{center}
\includegraphics[width=6cm]{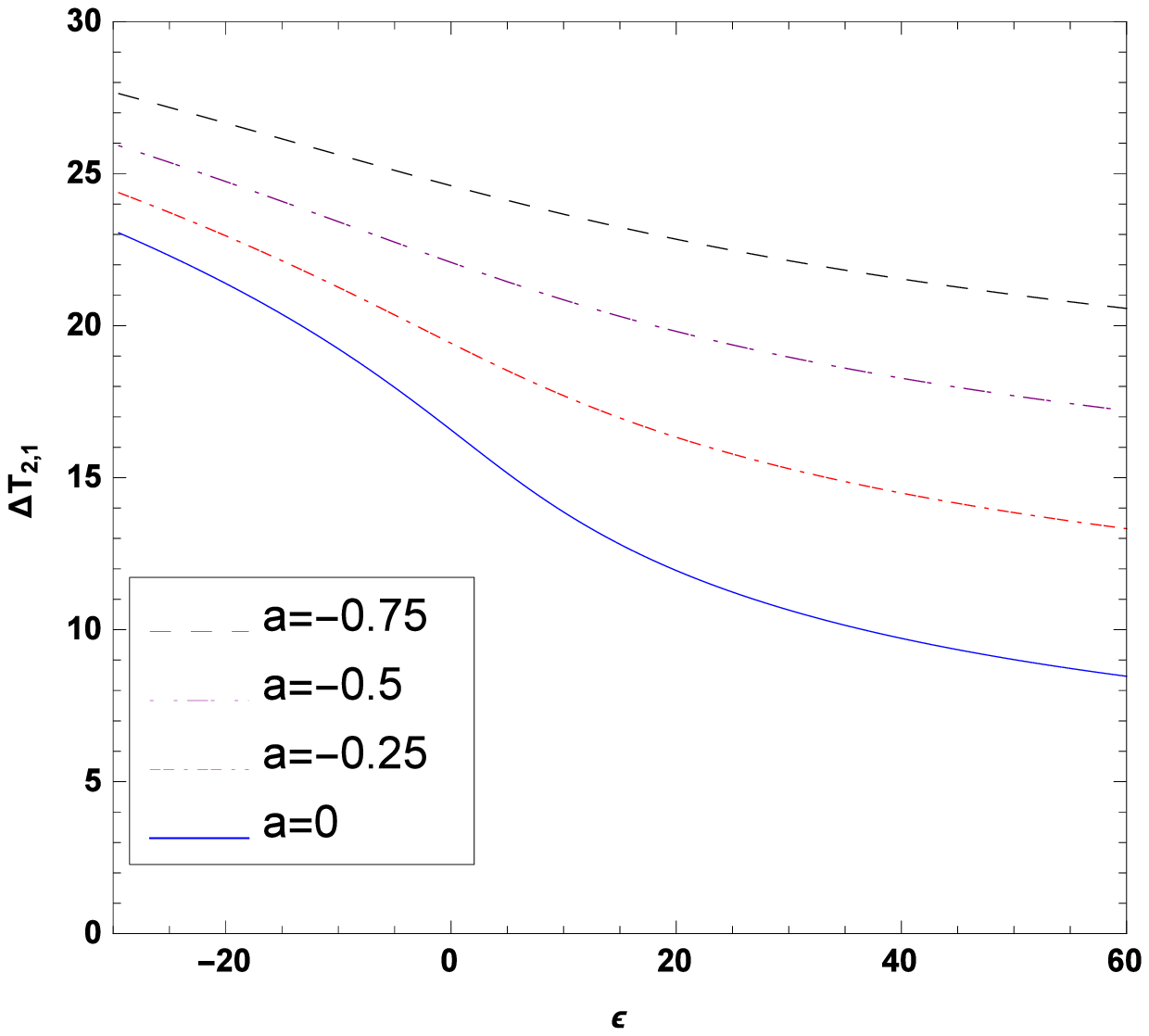}\;\;\;\;
\includegraphics[width=6cm]{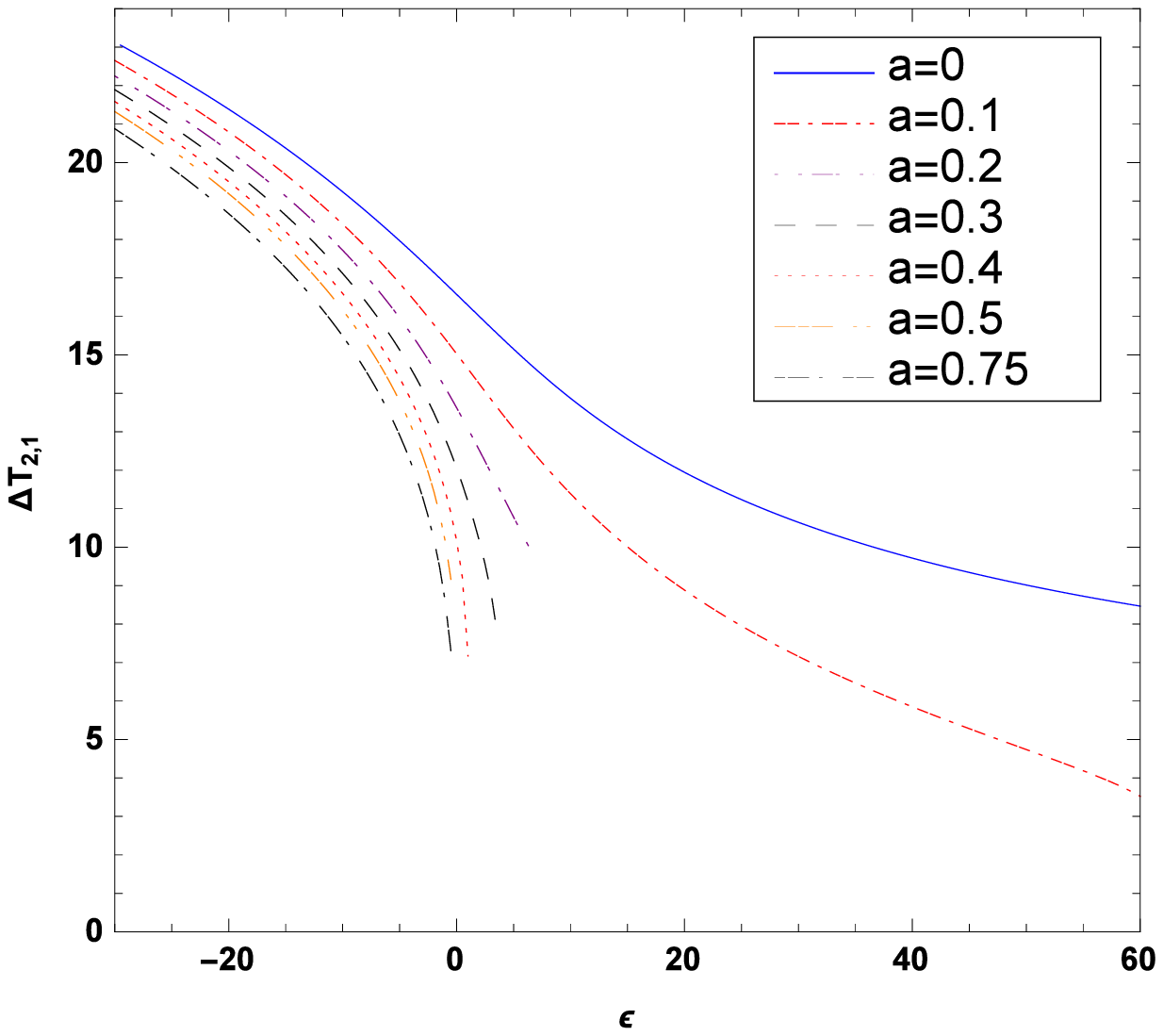}
\caption{Variety of the time delays between the relativistic images with the deformed parameter $\epsilon$ for different $a$ in a Johannsen-Psaltis  rotating non-Kerr spacetime. Here, we set $2M=1$, $n=2$ and $m=1$.}
\end{center}
\end{figure}
In order to make a comparison, in Fig.(15), we also present the time delays between the first relativistic image and the second one in a Johannsen-Psaltis  rotating non-Kerr spacetime, which tells us that the time delays decreases with the deformation parameter $\epsilon$  and the rotation parameter $a$. It means that the change of the time delays between the relativistic images with the deformation parameter in the Konoplya-Zhidenko rotating non-Kerr spacetime has quality difference from that in the Johannsen-Psaltis spacetime.

\section{summary}

In this paper we have investigated the propagation of photon in a Konoplya-Zhidenko rotating non-Kerr spacetime with an extra deformation parameter. We find that the deformed parameter together with
the rotation parameter imprint in the marginally circular photon
orbit, the deflection angle, the coefficients in strong field
lensing, the observational gravitational lensing variables and the time delay between two relativistic images. The condition of existence of horizons is not inconsistent with that of the marginally circular photon orbit, which is different from those in the Johannsen-Psaltis  rotating non-Kerr spacetime. In the cases without horizon, the spacetime is classified by the deformation parameter and the rotation parameter as two kinds: (i) the singularity is completely naked and (ii) the singularity is covered by the marginally circular orbit. This is very similar to that in the Janis-Newman-Winicour spacetimes.
The deflection angle of the light ray near WNS diverges logarithmically in the strong-field limit because of the existence of the marginally circular orbit radius, but the deflection angle of the light ray closing very to the SNS is a certain finite value $\alpha_s$.  The sign of $\alpha_s$ depends on the rotation parameter $a$ and the deformation parameter $\eta$. It means that in the Konoplya-Zhidenko rotating  non-Kerr spacetime the
behavior of the deflection angle near the singularity differs from those in the Janis-Newman-Winicour spacetime with the naked singularity and  in the Johannsen-Psaltis rotating non-Kerr spacetime.

For all values of $a$,  the relativistic images are farer from the optical axis for a larger deformed parameter. With increase of the deformation parameter, the separability $s$ decreases and the relative magnitudes $r_{m}$ increases as $a<0.5$. However, in the cases with $a>0.5$, there exist a peak in the curve of $s$ with the deformation parameter $\eta$ for different $a$. With increase of $a$, the value of peak drops down and its position moves along the right. The change of $r_{m}$ with the parameters $\eta$ and $a$ is converse to that of the angular separation $s$ in this case.
The time delay $\Delta T_{2,1}$ decreases with the rotation parameter $a$. With increase of the deformation parameter $\eta$, $\Delta T_{2,1}$ first decreases and then increases for the non-positive $a$ and increases monotonously for the positive $a$. These properties of the relativistic images differ from those in the Johannsen-Psaltis  rotating non-Kerr spacetime, which could  provide a possibility to check the no-hair theorem and to distinguish varieties of alternative theories of gravity in the future astronomical observations.

\section{\bf Acknowledgments}
We would like to thank Prof. Alexander Zhidenko for his/her useful comments.
This work was partially supported by the National Natural
Science Foundation of China under Grant No.11275065,  No. 11475061,
the construct program of the National Key Discipline, and the Open
Project Program of State Key Laboratory of Theoretical Physics,
Institute of Theoretical Physics, Chinese Academy of Sciences, China
(No.Y5KF161CJ1).

\end{document}